\documentclass[preprint,12pt]{elsarticle}

\usepackage{amssymb}

\usepackage[hidelinks]{hyperref}

\usepackage[detect-mode,detect-weight]{siunitx}
\usepackage{multirow}

\usepackage{lineno}
 \usepackage{xcolor}
 \usepackage{bm}
\newcommand{\bstar}{\ensuremath{\beta^*}}
\newcommand{\lstar}{\ensuremath{L^*}}

\journal{arXiv}

\begin{document}

\begin{frontmatter}



\title{FCC-hh Experimental Insertion Region Design\tnoteref{t1}
}


\author[1]{Jos\'{e} L. Abelleira}
\author[2]{Robert B. Appleby}
\author[3]{Sergey Arsenyev}
\author[4]{Javier Barranco}
\author[3]{Michael Benedikt}
\author[3]{Maria Ilaria Besana}
\author[5]{Oscar Blanco Garc\'{i}a}
\author[5]{Manuela Boscolo}
\author[6]{David Boutin}
\author[3]{Xavier Buffat}
\author[3]{Helmut Burkhardt}
\author[3]{Francesco Cerutti}
\author[6]{Antoine Chanc\'{e}}
\author[7]{Francesco Collamati}
\author[1]{Emilia Cruz-Alaniz}
\author[6]{Barbara Dalena}
\author[3]{Michael Hofer}
\author[3]{Barbara L. Humann}
\author[3]{Angelo Infantino}
\author[3]{Jacqueline Keintzel}
\author[3]{Andy Langner}
\author[3]{Marian L\"{u}ckhof}
\author[3]{Roman Martin}
\author[4]{Tatiana Pieloni}
\author[2]{Haroon Rafique}
\author[3]{Werner Riegler}
\author[1]{L\'{e}on Van Riesen-Haupt}
\author[3]{Daniel Schulte}
\author[1]{Andrei Seryi}
\author[4]{Claudia Tambasco}
\author[3]{Rogelio Tom\'{a}s\corref{cor1}}\ead{rogelio.tomas@cern.ch}\cortext[cor1]{Corresponding author}
\author[3]{Frank Zimmermann}

\address[1]{John Adams Institute, University of Oxford, Oxford OX1 3RH, United Kingdom}
\address[2]{University of Manchester, M13 9PL Manchester, United Kingdom}
\address[3]{CERN, CH 1211 Geneva 23, Switzerland}
\address[4]{EPFL, CH-1015 Lausanne, Switzerland}
\address[5]{INFN-LNF, Via Fermi 40, 00044 Frascati, Italy}
\address[6]{CEA, IRFU, SACM, Centre de Saclay, F-91191 Gif-sur-Yvette, France}
\address[7]{INFN-Rome, Piazzale Aldo Moro 2, 00185, Rome}

\tnotetext[t1]{The research presented in this document is part of the European Circular Energy-Frontier Collider Study (EuroCirCol) project which has received funding from the European Union’s Horizon 2020 research and innovation programme under grant No 654305.}

\begin{abstract}
The Future Circular Collider study is exploring possible designs of circular colliders for the post-LHC era, as recommended by the European Strategy Group for High Energy Physics. One such option is FCC-hh, a proton-proton collider with a centre-of-mass energy of 100~TeV. The experimental insertion regions are key areas defining the performance of the collider. This paper presents the first insertion region designs with a complete
assessment of the main challenges, as collision debris with two orders of magnitude larger power
than current colliders, beam-beam interactions in long insertions, dynamic aperture for optics
with peak $\beta$ functions one order of magnitude above current colliders, photon background from synchrotron radiation and cross talk between the insertion regions.
An alternative design avoiding the use of crab cavities with a small impact on performance is also
presented.

\end{abstract}


\end{frontmatter}


\section{Overview}
FCC-hh will provide proton-proton collisions at a center-of-mass energy of \SI{100}{TeV}, a factor~7 higher than the LHC. The goal for the integrated luminosity is set to \SI{20}{ab^{-1}} in each high luminosity experiment. This ambitious goal can be reached by an operational scenario with 10 years of operation using the less ambitious parameters (Baseline option) followed by 15 years of operation at the  Ultimate parameters. Table~\ref{tab:rma:FCC_parameters} shows the two parameter sets for the high luminosity Insertion Regions (IRs) and compares them with the respective parameters of LHC and  High Luminosity LHC (HL-LHC). The most notable difference between Baseline and Ultimate are the goals for the $\beta$ functions at the Interaction Point (IP), $\bstar$, leading to a significant increase in instantaneous luminosity at Ultimate optics. 

\begin{figure}
	\centering
	\includegraphics*[trim={0cm 0cm 0cm 0cm},clip=true, width=0.7\textwidth ]{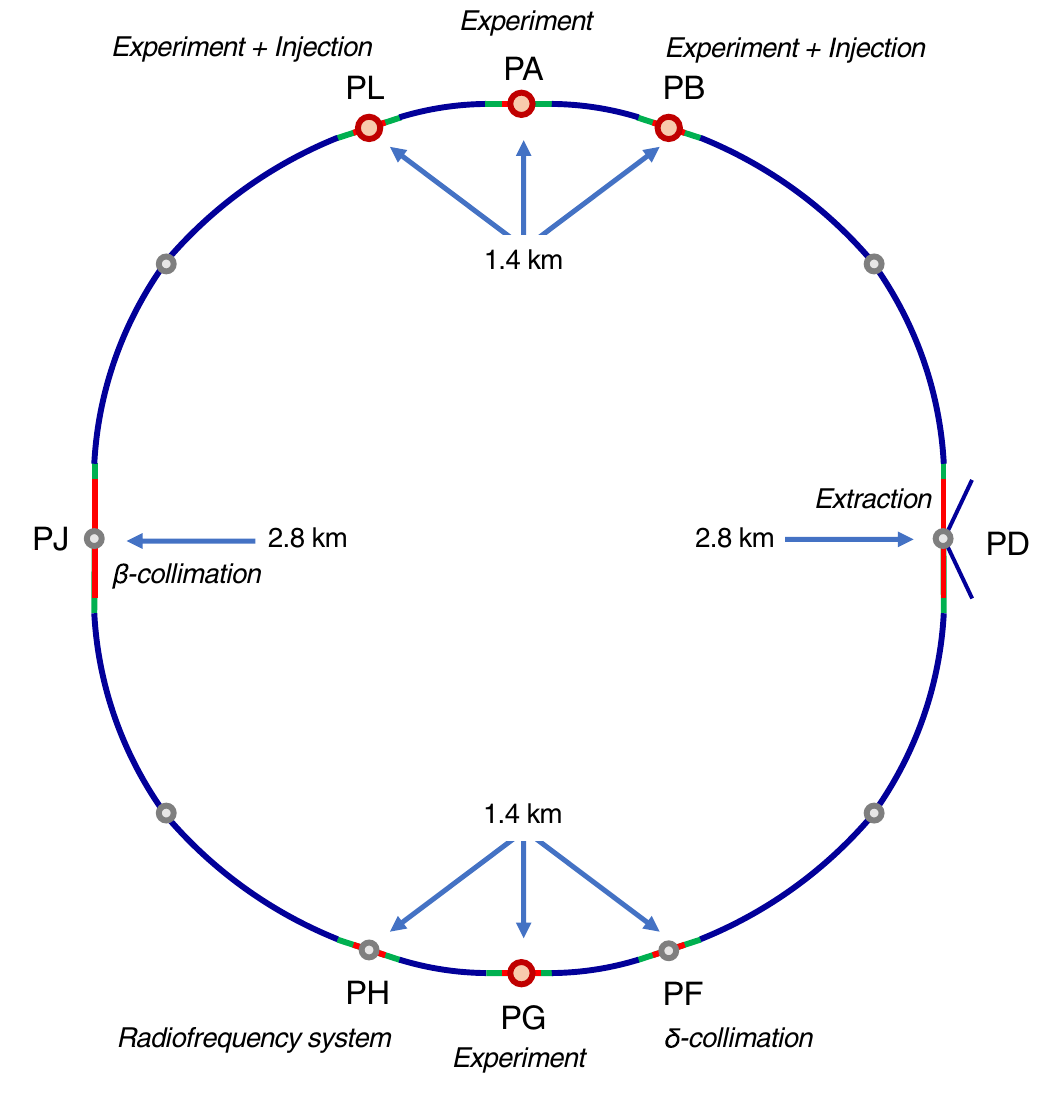}
	\caption{\label{layout}
	FCC-hh layout with points A-L labelled as PA-PL.}
\end{figure}
FCC-hh layout is shown in Fig.~\ref{layout}.
The high luminosity experiments are located in IRs
around the IPs at points A and G (PA and PG in in Fig.~\ref{layout}), while another two experiments 
are placed in IRs around point L and point B (PL and PB in in Fig.~\ref{layout}), which also contain the injection areas.
Due to the high center-of-mass energy and the high luminosity the total power released at the IP towards each side increases from \SI{1}{kW} in the LHC or \SI{4.75}{kW} in HL-LHC up to \SI{260}{kW} in FCC-hh at ultimate parameters. Most of this power will be absorbed in the detector but debris particles emitted at small angles will travel down the beam pipe and impact in the insertion region magnets, possibly causing quenches and degrading the material. Consequently, the radiation load from collision debris has been identified as a key issue of the final focus system early in the design phase of the IRs~\cite{bib:rma:PhysRevAccelBeams.20.081005}. Unifying adequate protection of the triplet magnets with a high luminosity performance has been the driving factor of the IR layout. Nb$_{3}$Sn technology has been chosen for the IR magnets for his superior performance
in terms of magnetic field and operational temperature margins. The possible flux jumps
in these magnets should be investigated in the future~\cite{bib:fluxjusmps}.

\begin{table}
  \begin{center}
  \caption{\label{tab:rma:FCC_parameters} Key parameters of FCC-hh compared to LHC and HL-LHC.}
  \begin{tabular}{ |m{4.8cm}|c|c|cc| } 
	\hline
	& \bf LHC & \bf HL-LHC &  \multicolumn{2}{c|}{\bf FCC-hh} \\
	&		&		&	\bf Baseline	&	\bf Ultimate	\\ 	\hline
	Center-of-mass energy [\si{\tera eV}] 		&	14		&	14		&	\multicolumn{2}{c|}{100}	\\ \hline
	Injection energy [\si{\tera eV}] 			&	0.45	&	0.45	&	\multicolumn{2}{c|}{3.3}	\\ \hline
	Ring circumference [\si{\kilo m}] 			&	26.7	&	26.7	&	\multicolumn{2}{c|}{97.75}	\\ \hline
	Arc dipole field [\si{T}] 					&	8.33	&	8.33	&	\multicolumn{2}{c|}{16}	\\ \hline
	Number of IPs								&	2+2		&	2+2		&	\multicolumn{2}{c|}{2+2}	\\ \hline
	Number of bunches per beam	$n_{\text{b}}$	&	2808	&	2748	&	\multicolumn{2}{c|}{10600 (53000)}	\\ \hline
	Beam current [\si{A}]						&	0.58	&	1.11	&	\multicolumn{2}{c|}{0.5}	\\ \hline
	Peak luminosity/IP [\si{\num{e34} \centi m^{-2}s^{-1}}]	&	1		&	5	&	5	&	30	\\ \hline
	Events/crossing								&	27	&	135	&	170	& 	1020 (204)	\\ \hline
	Stored beam energy	[\si{\giga J}]			&	0.4	&	0.7	&	\multicolumn{2}{c|}{8.4}	\\ \hline
	Synchrotron power per beam [\si{\mega W}] 	&	0.0036		&	0.0073	&	\multicolumn{2}{c|}{2.4}	\\ \hline
	Arc synchrotron radiation [\si{W/m/beam}]	&	0.18	&	0.35	&	\multicolumn{2}{c|}{28.4}	\\ \hline
	IP beta function $\bstar$ [\si{m}]			&	0.4	&	0.15	&	1.1	&	0.3	\\ \hline
	Bunch spacing [\si{\nano s}]				&	25	&	25	&	\multicolumn{2}{c|}{25 (5)}	\\ \hline
	Initial norm. rms emittance $\epsilon_{\text{n}}$ [\si{\micro m}]	&	3.75	&	2.5	&	\multicolumn{2}{c|}{2.2 (0.45)}	\\ \hline
	Initial bunch population $N_{\text{b}}$[\si{\num{e11}}]	&	1.15	&	2.2	&	\multicolumn{2}{c|}{1.0 (0.2)}	\\ \hline
	Transv. emittance damping time [\si{h}]		&	25.8	&	25.8	&	\multicolumn{2}{c|}{1.1}	\\ \hline
	RMS bunch length [\si{\centi m}]		&	\multicolumn{2}{c|}{7.55}	&	\multicolumn{2}{c|}{8}	\\ \hline
	RMS IP beam size [\si{\micro m}]		&	16.7	&	7.1	&	6.8 & 3.5	\\ \hline
	Full crossing angle $\theta$ [\si{\micro rad}]		&	285	&	590	&	104 & 200	\\ \hline
  \end{tabular}
  \end{center}
\end{table}

In terms of  chromaticity correction it has been estimated that the sextupoles in the arcs are able to correct around 557 units of chromaticity. The natural chromaticity for the case with $\beta^*=\SI{30}{cm}$ is below this value, and therefore the chromaticity can be corrected. However this is not the case for beyond ultimate optics. While the aperture of the final focus system can accommodate a $\bstar$ of almost 20~cm for the nominal crossing angle of Table~\ref{tab:rma:FCC_parameters}, the strength of the sextupoles necessary to correct the chromaticity is above the achievable maximum. If operation beyond ultimate $\bstar$ is desirable an achromatic telescopic squeezing scheme as foreseen for the HL-LHC~\cite{bib:fartoukh_2013_ATS} could be used to increase the chromatic correction efficiency of the arc sextupoles.

Although the high mass of protons usually keeps the synchrotron radiation produced in hadron colliders low, the high beam energy of FCC-hh gives rise to the concern that the photon background in the experimental regions might grow to notable levels. Hence, a closer investigation of the synchrotron radiation was necessary in order to quantify the impact. The simulation of the photon background concluded that the synchrotron radiation is not expected to be an issue for the experiments.

Debris from proton collision at the interactions points may create background in the other detectors. Protons with an energy close to the nominal beam energy travelling far in the beam pipe before being intercepted, as well as muons passing through the rock between two experiments are of particular concern. Tracking studies of the protons and an analysis of the muon range in rock were performed, concluding that the cross talk between experiments is negligible.

The high luminosity IR design relies on the availability of crab cavities to compensate the luminosity loss due to the crossing angle needed to keep beam-beam long range effects under control as described in~\cite{barrancoFCCweek2018}. As this technology is currently being tested in proton accelerators for the first time it is desirable to have an alternative that avoids crab cavities. Flat beam optics are a good candidate for this as small beamsizes can be achieved in the non-crossing plane to increase the luminosity whilst not having to increase crossing angle as much as when the beamsize is reduced in both planes, therefore reducing the luminosity loss from the geometric overlap of the colliding bunches. Corresponding optics have been developed, using an alternative triplet layout.

In addition to the two high luminosity insertion regions situated around points A and G, FCC-hh features two low luminosity insertion regions around points B and L, much like the LHC. In absence of a physics case -- and consequently luminosity goals or space constraints -- for these two experimental regions, an initial design is proposed that can reach an integrated luminosity of \SI{500}{fb^{-1}}. An alternative for the low luminosity IRs is FCC-eh \cite[Sec. 2.8]{bib:rma:shortCDR}, a lepton-hadron collider with insertion region scaled up from the LHeC~\cite{bib:rma:PhysRevAccelBeams.18.111001}.

A filling scheme with 5~ns bunch spacing is considered to mitigate the event pile-up in the detectors. The corresponding parameters are shown in Table~\ref{tab:rma:FCC_parameters} in parenthesis. This option has not been addressed in the IR design
although it reduces aperture needs thanks to the lower emittance and could pose operational difficulties to keep beams
in collision as the beam size reduces below \SI{1}{\micro \metre} during the physics fill due to synchrotron radiation damping.

\begin{figure}
	\centering
	\includegraphics*[trim={1cm 2cm 1cm 2cm},clip=true, width=\textwidth ]{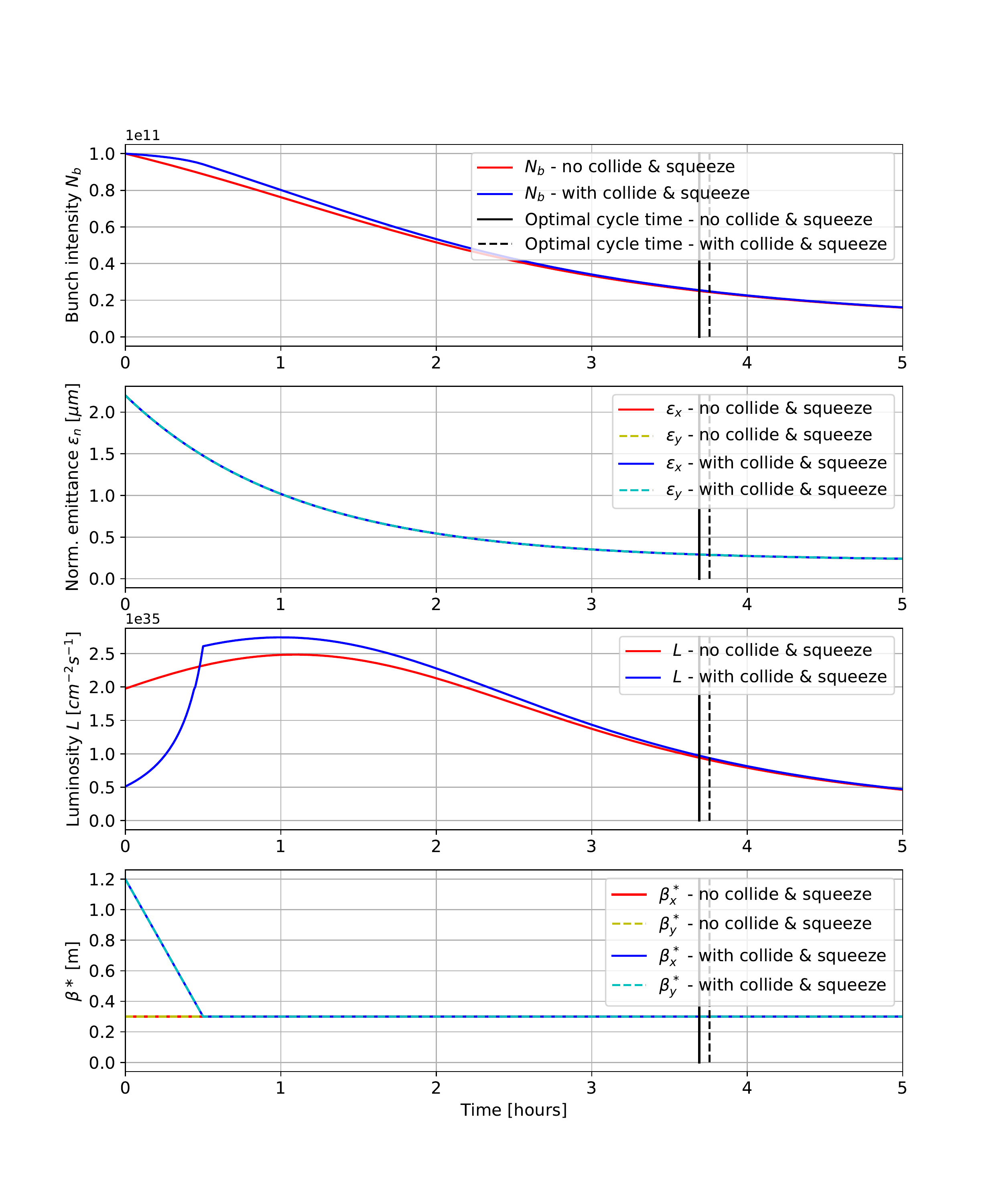}
	\caption{\label{collideandsqueeze}
	Beam parameters during FCC-hh physics fills with and without collide \& squeeze.}
\end{figure}

Simulations of collective effects have determined a change of the operational mode that will now assume a collide and squeeze approach~\cite{bib:buffat:evian2012squeezing}. 
This represents an easy mitigation without a significant penalty in integrated luminosity. Beams will collide at larger $\bstar$, around 1.2~m, and continue the $\bstar$ squeeze to the minimum beta while colliding. This will avoid the reduction of the stability area due to long-range beam-beam effects acting against the stabilizing effect of the octupole magnets 
and will provide enough margin in stability as required by the Run~2 experimental evidences of the LHC. Figure~\ref{collideandsqueeze} shows an example of how FCC-hh physics fill could look like with a collide \& squeeze scenario with $\beta^{*}$ decreasing linearly from \SI{1.2}{m} to \SI{0.3}{m} in 30~minutes compared to collision without collide \& squeeze. It should be noted that the average luminosity production is not reduced significantly in this scenario since the optics squeeze is happening during collision, shortening the turn-around time (not pictured).
The simulations in Fig.~\ref{collideandsqueeze} neglect emittance growth from luminosity burn-off,
recently evaluated in~\cite{bib:emitlumi}.

\section{System Layout and Optics}
\label{sec:optics}

This section  describes the layout and the optics designs
of the high and low luminosity EIRs.
An alternative optics design for the high luminosity EIR
without crab cavities is also presented.

\subsection{Baseline Design of the High Luminosity EIRs}

\begin{figure}
	\centering
	\includegraphics[width=\textwidth]{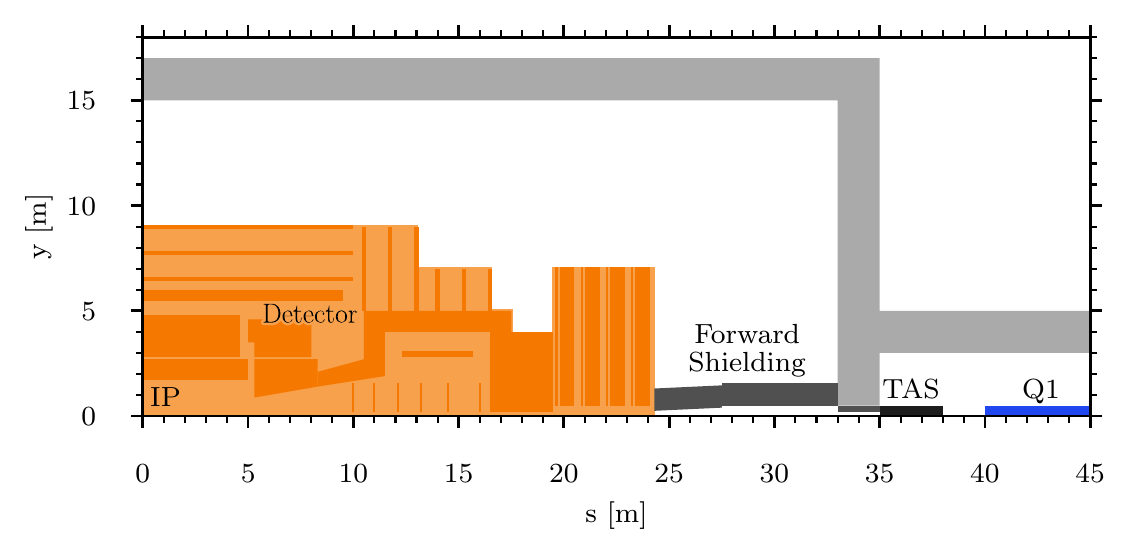}
	\caption{\label{fig:rma:detector_region_layout} Detector and insertion region layout leading to the $\lstar = \SI{40}{m}$ lattice. The IP is located at $(0,0)$.}
\end{figure}

Early studies of the final focus system layout concluded that the main contributor to the minimum $\bstar$ is the overall length of the triplet, while the drift between the IP and first quadrupole , $\lstar$, plays a minor role \cite[Sec. III D]{bib:rma:PhysRevAccelBeams.20.081005}. This led to a clear strategy to minimize $\bstar$ with significant amounts of shielding reducing the free aperture of the final focus magnets: to choose the smallest $\lstar$ that does not restrict the detector design and to increase triplet length until dynamic aperture or chromaticity become obstacles. In this strategy the machine-detector interface plays a key role as it defines $\lstar$. A sketch of the detector region layout is shown in Fig.~\ref{fig:rma:detector_region_layout}. While the detector has a total length of about \SI{50}{m}, extending to \SI{25}{m} on either side of the IP, the opening scenario requires a total cavern length of \SI{66}{m}. During operation, the gap between detector and cavern wall will be occupied by the forward shielding that protects the detector from secondaries back-scattered from the
passive absorber for charged particles (TAS), a \SI{3}{m} long copper absorber that protects the final focus magnets from collision debris. The aperture in the \SI{2}{m} thick wall between cavern and tunnel is equipped with a cast iron absorber to complete the forward shielding. The TAS is located \SI{35}{m} to \SI{38}{m} from the IP. With an additional space of \SI{2}{m} reserved for vacuum equipment and for the end of the magnet cryostat, first quadrupole of the final focus triplet starts at $\lstar = \SI{40}{m}$.

The beam pipe at the IP is made of \SI{0.8}{mm} thick beryllium and has an inner radius of \SI{20}{mm}. This pipe extends to $\pm \SI{8}{m}$ to either side of the IP and is followed by a beryllium cone with an opening angle of \SI{2.5}{mrad} corresponding to a pseudorapidity of $\eta = 6$. From \SI{16}{m} from the IP on, the inner radius of the aluminium beam pipe is constant at \SI{40}{mm}, this is necessary for the opening of the detector.

\begin{figure}
	\centering
	\includegraphics*[width=\textwidth]{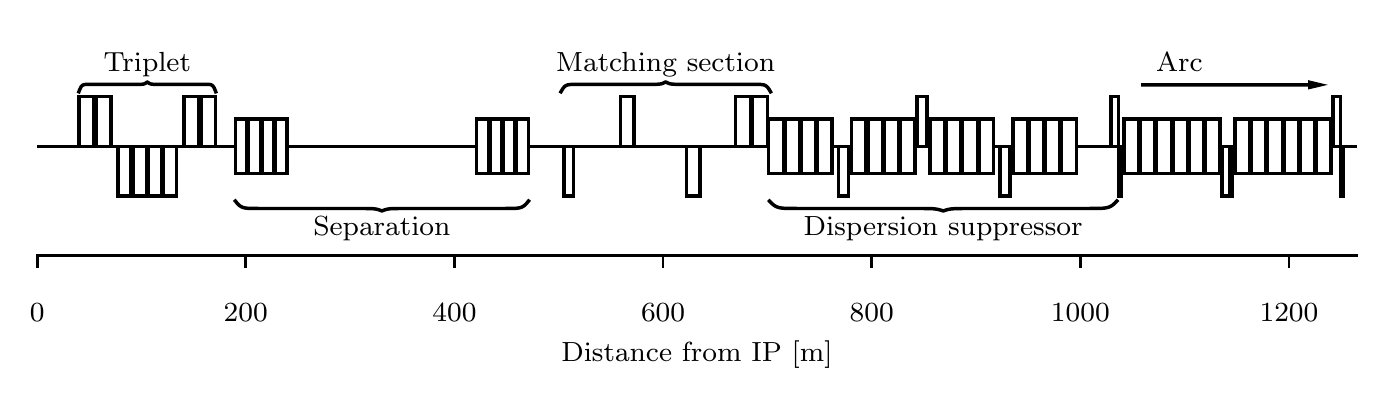}
	\caption{\label{fig:rma:FCC_hh_IR_layout} Layout of the high luminosity interaction region. The layout is antisymmetric around the IP at $(0,0)$.}
\end{figure}

\begin{figure}
	\centering
	\includegraphics*[width=0.8\textwidth]{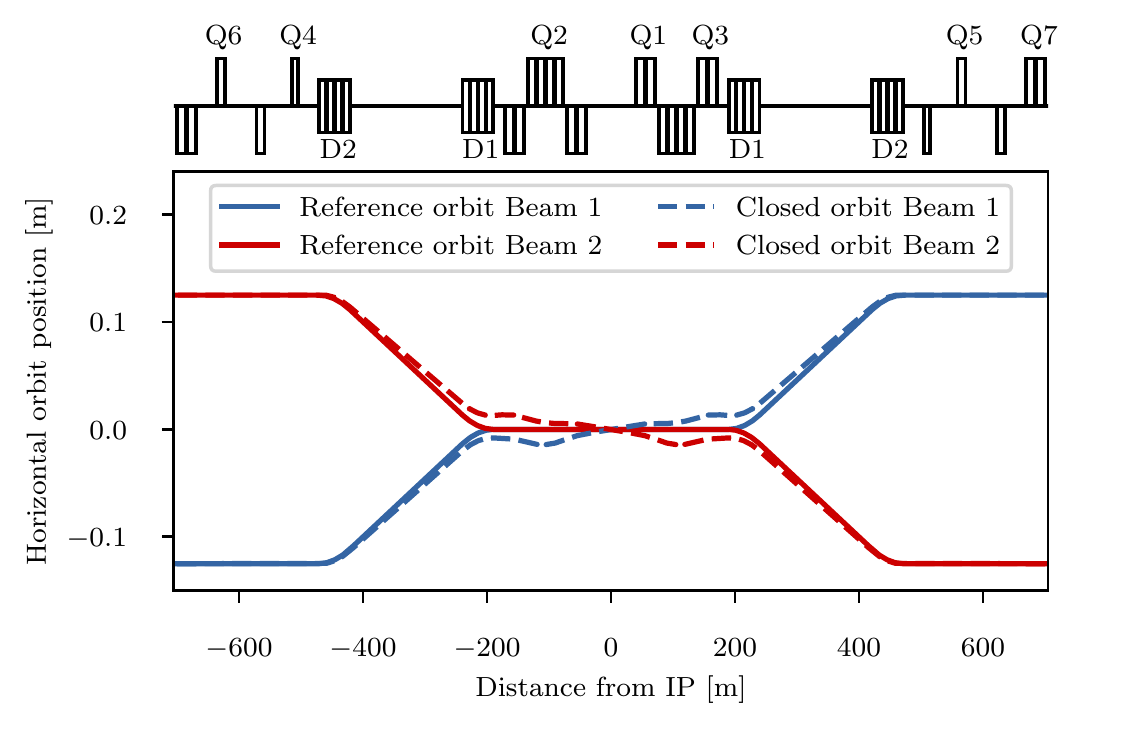}
	\caption{\label{fig:rma:FCC_hh_IP_orbits} Reference orbits (solid lines) and closed orbits with crossing angles (dashed lines) in the interaction region.}
\end{figure}

The interaction region layout of FCC-hh follows the same principles as the LHC and HL-LHC interaction regions. The layout is shown in Fig.~\ref{fig:rma:FCC_hh_IR_layout}.
Starting at the interaction point, the strongly focused and highly divergent beams pass a drift space with the length $\lstar$ chosen to accommodate the detector. Following this drift space, a final focus system comprised of three large aperture quadrupoles (hence called the triplet) focuses the beams in both the horizontal and vertical planes. The triplet consists of single aperture magnets that host both beams. The triplets on both sides of the IP are powered antisymmetrically. This has the advantage that the triplet region is optically identical for both beams. Behind the triplet, a shared aperture dipole D1 separates the two beams. After a drift, the double bore dipole D2 bends the separated beams onto parallel orbits again. The resulting reference orbits are shown in Fig.~\ref{fig:rma:FCC_hh_IP_orbits}. Also depicted are orbit excursions that let the two beams collide with a crossing angle in order to avoid parasitic collisions outside the detector area.
Four more quadrupoles Q4-Q7 make up the following matching section that occupies the rest of the straight section. The straight section is connected to the arcs by a two cell dispersion suppressor. To provide enough degrees of freedom to match all required beam parameters from the IP to the arcs, the four matching section quadrupoles, the three individually powered quadrupoles of the dispersion suppressor Q8-Q10 as well as three tuning quadrupoles in the first arc cell QT11-QT13 are used for the matching procedure.

\subsubsection{Final Focus Triplet}
\label{sec:Final_focus_triplet}
\begin{figure}
	\centering
	\includegraphics*[width=0.8\textwidth]{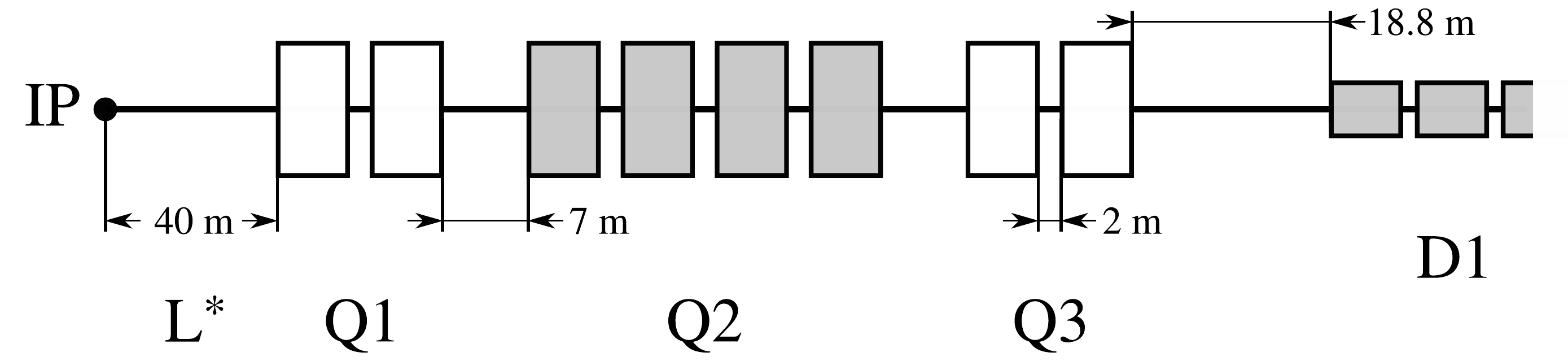}
	\caption{\label{fig:rma:triplet_drifts} Layout of the final focus triplet.}
\end{figure}

The final focus design strategy calls for a long triplet in order to achieve small $\bstar$ values. In practice, not only chromaticity and dynamic aperture were limiting factors for the triplet length, but also the total length of the straight section that determines the arc side focal length of the final focus system, as well as the strength of the Q7 quadrupole. Furthermore the lengths of individual magnets must be equal or below \SI{14.3}{m} in order to be compatible with a cryostat length of \SI{15}{m}. The relative lengths of Q1, Q2 and Q3 were adopted from HL-LHC. As suggested in \cite{bib:rma:deMaria:1064704}, Q1 was chosen to have a smaller aperture and higher gradient than Q2 and Q3 in order to minimize $\bstar$. The specification for the triplet quadrupoles are listed in Table~\ref{tab:rma:hl_eir_magnet_specifications} and the layout of the final focus triplet shown in Fig.~\ref{fig:rma:triplet_drifts}. Q1 and Q3 are made up of two submagnets each with length of \SI{14.3}{m}. For the interconnects a drift space of \SI{2}{m} is reserved between the submagnets. 
The drift between Q1 and Q2 as well as Q2 and Q3 is longer at \SI{7}{m} and must house orbit correctors, BPMs and vacuum equipment. Q2 consists of four \SI{12.5}{m} long submagnets. This not only allows for a similar length ratio as in the HL-LHC but also to place orbit correctors in the cryostat of the outermost Q2 magnets. Behind Q3, \SI{18.8}{m} of space are reserved for higher order multipole correctors to compensate triplet field errors. 

A \SI{35}{mm} thick inner shielding of the tungsten alloy INERMET180 protects the triplet magnets from collision debris. Furthermore the free aperture is reduced by a gap for the liquid helium for cooling, the Kapton insulator, a beam screen, a gap for the insulation of the beam screen as well as the cold bore that scaled with of the coil aperture radius. The individual radial thickness of these layers are detailed in Tab.~\ref{tab:Beam_Screen_Components} and have been modeled as simple layers.

\begin{table}[h]
	\centering	

	\begin{tabular}{|l|c|}
        \hline
		Component & Radial Thickness \\
		\hline
		Cold bore & 5.44 \% of coil radius \\
		Liquid helium cooling & 1.5 mm \\
		Kapton insulation & 0.5 mm \\
		Beam screen & 2.05 mm \\
		Beam screen insulation & 2 mm\\
		\hline
	\end{tabular}
    \caption{Radial thickness of various components installed between quadrupole coil and beam in quadrupole triplet.}
	\label{tab:Beam_Screen_Components}
\end{table}

\begin{figure}[tbh!]
	\centering
	\includegraphics*[width=\textwidth]{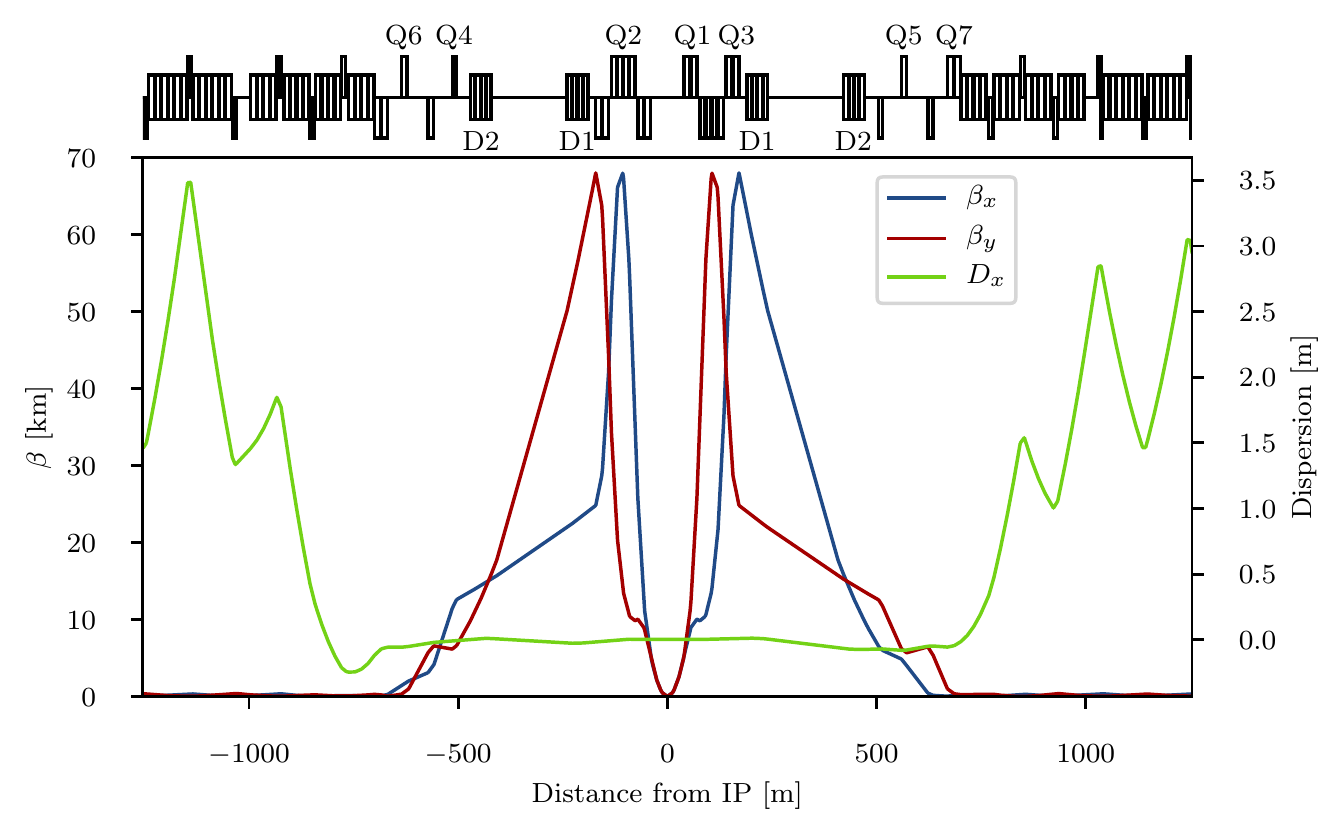}
	\caption{\label{fig:rma:IR_beta_0300} Optics of the high luminosity EIR with $\bstar = \SI{0.3}{m}$.}
\end{figure}
\begin{figure}
	\centering
	\includegraphics*[width=0.7\textwidth]{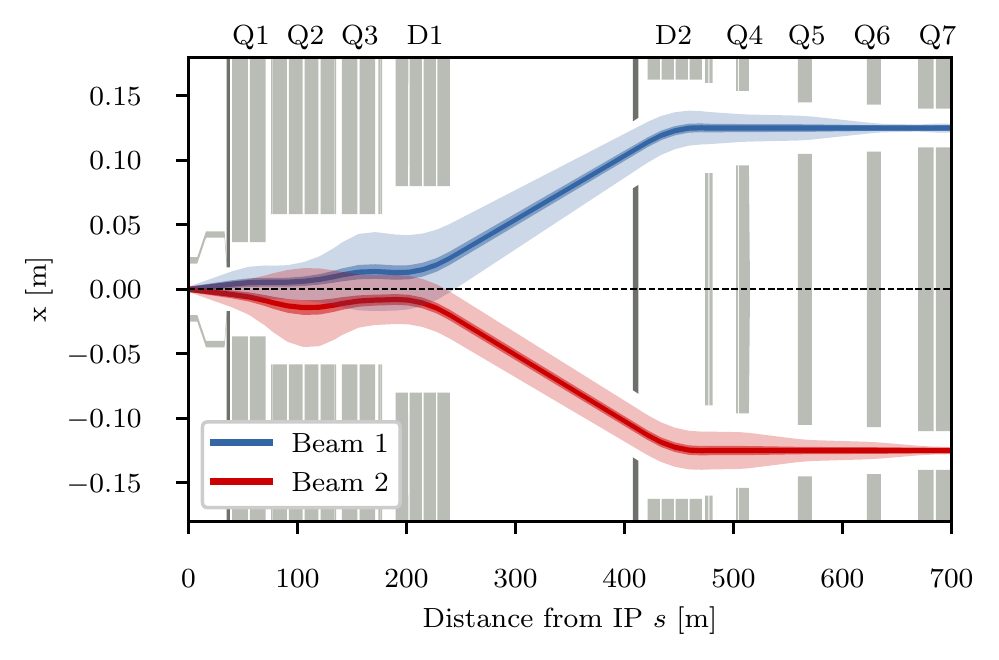}
	\caption{\label{fig:rma:IPA_apertures_0300} Overall layout of the insertion region between the IP and Q7. For each beam, the closed orbit, the \SI{2}{$\sigma$} envelope and the \SI{15.5}{$\sigma$} envelope for the ultimate $\bstar$ of \SI{0.3}{m} are shown. The beam sizes include a $\beta$ beating of \SI{10}{\%} and a closed orbit uncertainty of \SI{2}{mm}. Magnet apertures and the detector region beam pipe are illustrated with light gray while absorbers are shown in dark gray. The large aperture triplet magnets leave significant aperture margins.}
\end{figure}
\begin{figure}
	\centering
	\includegraphics*[width=\textwidth]{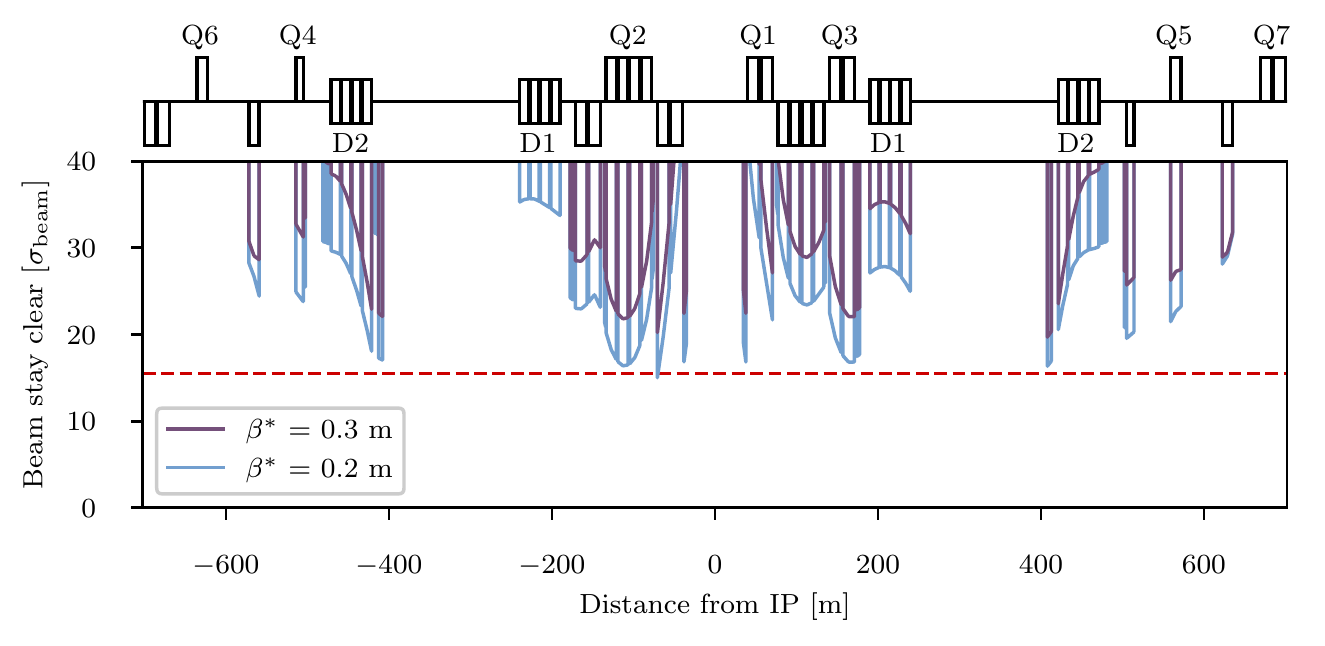}
	\caption{\label{fig:rma:IPA_beamstayclear_0300_0200} Beam stay clear of the high luminosity EIR for horizontal crossing and ultimate as well as beyond ultimate optics. For $\bstar = \SI{0.2}{m}$ the beam stay clear is just below the minimum of \SI{15.5}{$\sigma$} in the left Q1, suggesting a slightly larger $\bstar$ can be accommodated.}
\end{figure}

Despite this significant reduction of the free aperture, the triplet can accommodate a beam with lower than ultimate $\bstar$. Figure~\ref{fig:rma:IR_beta_0300} shows the $\beta$ functions and horizontal dispersion in the Experimental Insertion Region (EIR) and Fig.~\ref{fig:rma:IPA_apertures_0300} the corresponding aperture usage. Although aperture and alignment tolerances are not included in Fig.~\ref{fig:rma:IPA_apertures_0300}, the beam stay clear depicted in Fig.~\ref{fig:rma:IPA_beamstayclear_0300_0200} clearly shows that the ultimate optics have a significant margin in terms of aperture. In fact, optics with almost $\bstar = \SI{0.2}{m}$ can be achieved, although the chromaticity correction will not suffice with the current arc layout.

\subsection{Alternative Triplet and Flat Optics of the High Luminosity EIRs}

In parallel to the final focus triplet described in Section~\ref{sec:Final_focus_triplet}, efforts were made to design an alternative triplet \cite{bib:jab:alternative}. This alternative was designed using an algorithm that systematically scans the design parameter space to find the shortest possible triplet that has sufficient beam stay clear and shielding~\cite{bib:lva:vanRiesen-Haupt:IPAC2017-TUPVA043}. In a first approximation, the code scans through the entire design parameter space to estimate the beam stay clear using the thin lens approximation. It then does a more precise scan using the MAD-X aperture module in a smaller area identified by the approximation.

The design was worked on iteratively with energy deposition studies to determine the right amount of shielding required to protect the triplet from the collision debris. In a first iteration, the optimisation code was used to find the shortest triplet with 1.5~cm of tungsten shielding. This triplet was integrated into the baseline EIR and energy deposition studies were performed to estimate the amount of shielding needed. Next, the triplet was optimised again with the new shielding estimate and again integrated and tested. This process was repeated several times until a triplet was found that is as short as possible whilst still having sufficient beam stay clear and shielding.

In the course of this optimisation it was found that peaks in energy deposition could be minimized if all triplet quadrupoles had similar coil radii~\cite{bib:abe:Abelleira:IPAC2018-MOPMK003,bib:abe:vanRiesen-Haupt:IPAC2018-MOPMK007}. Therefore, the optimisation code was modified to find triplets made of quadrupoles of equal radii. In order to fulfil the technical requirements, the quadrupoles of the ideal solution had to be split into sub-magnets that were no longer than 15~m. The resulting triplet consisted of seven 15~m sub-magnets with equal radii and similar gradients -- the details of the magnets in this triplet are shown in Table~\ref{tab:lva:MagnetsAlternative}. Like the baseline triplet, the main quadrupoles in the alternative triplet are separated by 7~m drifts to leave space for correctors and instrumentation, whilst the sub-magnets only need 2~m separation to leave room for connectors.

\begin{table}
   \centering
   \caption{Properties of quadrupole groups in alternative triplet.}
   \begin{tabular}{|l|c|c|c|}
              \hline
       \textbf{Parameter} & \multicolumn{3}{c|}{\textbf{Quadrupole}}  \\
         & Q1 & Q2 & Q3  \\
            \hline
       Sub-Magnets & 2 & 3 & 2 \\
              \hline
       Sub-Magnet Length [m] & 15 & 15 & 15 \\
              \hline
       Coil Radius [mm] & 96.5 & 96.5 & 96.5 \\
              \hline
       Gradient [T/m] & 106 & 112 & 99 \\
              \hline
       Shielding [mm] & 44.2 & 33.2 & 24.2 \\
       \hline
   \end{tabular}
   \label{tab:lva:MagnetsAlternative}
\end{table}

\begin{figure}
	\centering
	\includegraphics[width=0.7\textwidth]{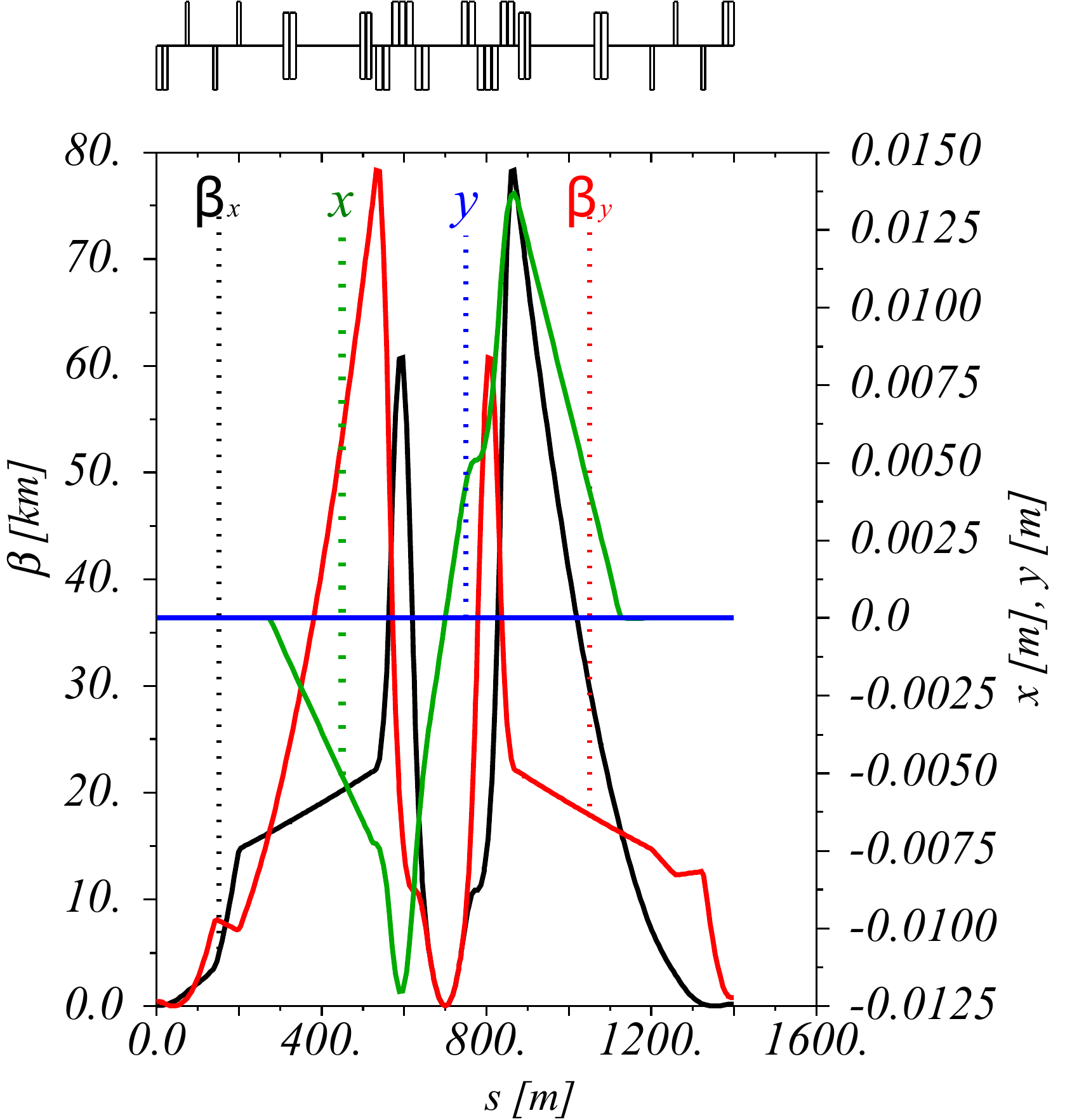}
	\caption{$\beta$ functions and orbit for EIR optics with alternative triplet and $\beta_{x,y}^*=$0.3~m.}
		\label{fig:lva:OpticsAlternative}
\end{figure}

The triplet was integrated into the same EIR as in Section~\ref{sec:Final_focus_triplet}, leaving the same 18.8~m drift between Q3 and the first separation dipole for the correction package. The matching quadrupoles in the EIR were used to match the Twiss functions to the arc. The resulting optics in the triplet are shown in Fig.~\ref{fig:lva:OpticsAlternative}, which also shows the beam orbit for a \SI{200}{ \micro rad} crossing in the horizontal plane.

Once the triplet was integrated the beam stay clear in the individual quadrupoles was reassessed and the shielding was increased wherever possible. This lead to an increase in shielding in Q2 and Q1 by 9~mm and 20~mm respectively. These increases are possible because the $\beta$ functions and orbit are smaller near the front of the triplet, hence leaving more space for potential shielding. This distribution in shielding is advantageous since most of the collision debris will hit the magnets closer to the IP. The exact amounts of shielding are also shown in Table~\ref{tab:lva:MagnetsAlternative}. 

\begin{figure}
	\centering
	\includegraphics[width=0.7\textwidth]{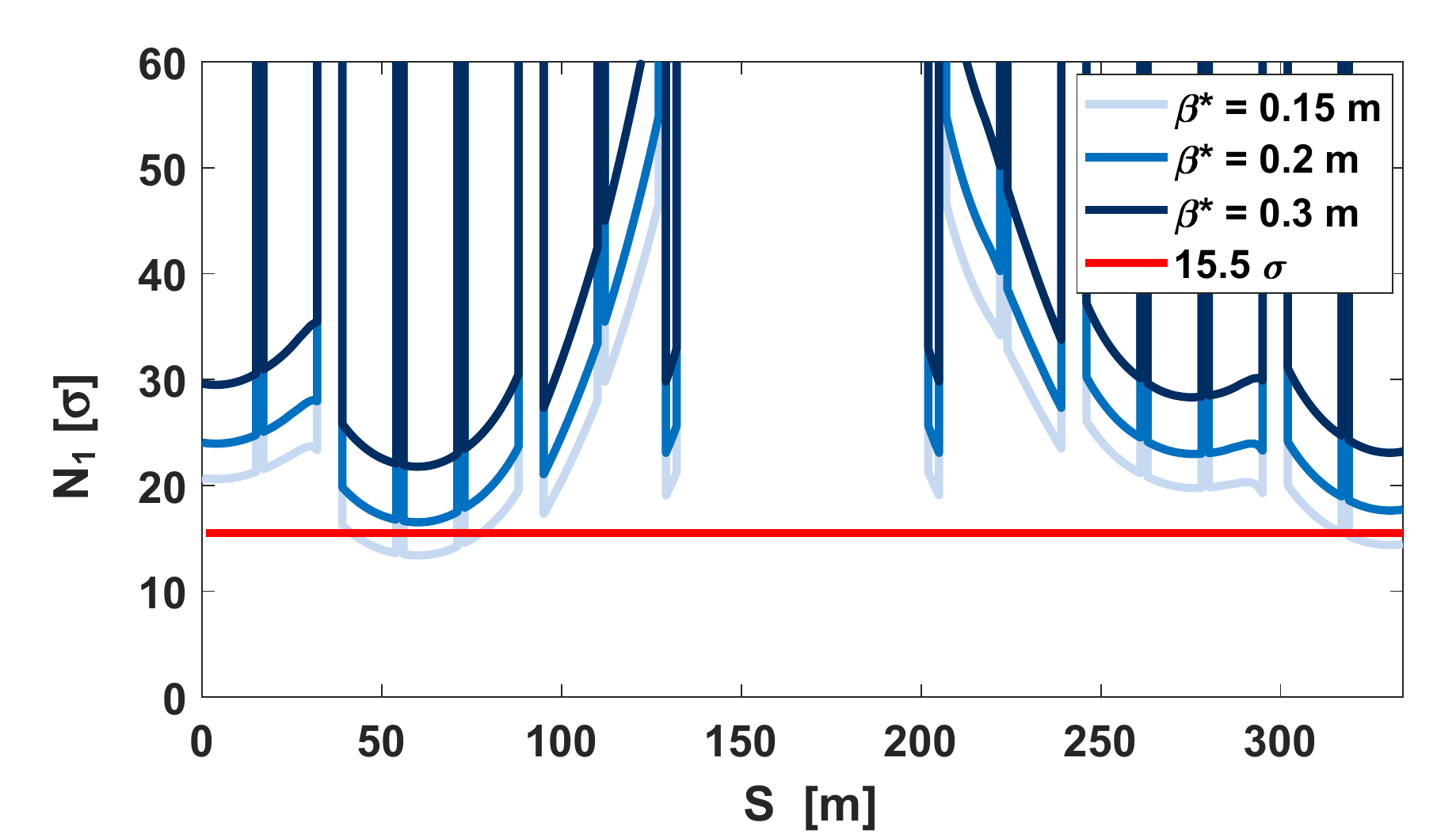}
	\caption{Plot showing BSC for $\beta^* =$ 0.15 m, 0.2 m and 0.3~m for the alternative triplet.}
	\label{fig:lva:ReachAlternative}
\end{figure}

Whilst the ultimate collision optics aims for a $\beta^*_{x,y}$ of 0.3~m, the shielding was designed to leave $15.5~\sigma$ for an optics with a $\beta^*_{x,y}$ of 0.2~m to provide a luminosity handle. The aperture studies were performed using the same technical specifications for the cooling, cold bore and beam screen as outlined in Section~\ref{sec:Final_focus_triplet} and the results are shown in Fig.~\ref{fig:lva:ReachAlternative}. As one can see from Fig.~\ref{fig:lva:ReachAlternative}, the alternative triplet can comfortably reach a $\beta^*$ of 0.3~m and even 0.2~m. Figure~\ref{fig:lva:ReachAlternative} also shows the beam stay clear for a case with $\beta^* = $ 0.15~m, whilst this is lower than the required it may still be a viable option should the beam current be low enough to change the collimator settings accordingly. A low beam current may be one of the reasons why the $\beta^*$ would need to be decreased in the first place to compensate for the loss in luminosity.

The alternative triplet can also be used for a flat optics, which can be exploited to compensate for the luminosity loss in case crab cavities are not feasible\cite{bib:abe:Abelleira:IPAC2018-MOPMK003, bib:tpi:Tomas:charmonix2014}. Without crab cavities the average integrated luminosity with round optics would decrease from $\SI{8.9}{fb^{-1}}$ to $\SI{6.3}{fb^{-1}}$ per day, however, the proposed flat beam optics could still achieve a luminosity of $\SI{7.2}{fb^{-1}}$ per day, without crab cavities.
This can be achieved without changing the gradients in the triplet but re-matching the $\beta$ functions using the matching section quadrupoles. In this initial study we assumed a $\rm 1.2~m \times 0.15~m$ flat optics. Detailed studies should be conducted.
Table~\ref{Tab:jab:comparison} shows a comparison between the main parameters of the round and flat optics.

The normalized separation for the flat optics is set at a value 30~\% higher than for round optics. This is achieved by an increase in the crossing angle.
The reason comes from beam-beam studies, that requires an increase in beam to beam separations at the long-range encounters in the case of flat optics to maintain the dynamic aperture similar to the equivalent round optics case. This is due to two factors: an uncompensated tune shift that can be in average corrected for and a different dimension of the detuning with amplitude. Preliminary results can be found in \cite{bib:jab:FCC_18_beam-beam}. For our final beta ratio of 8 an 80 \% larger normalized beam to beam separation is needed. However thanks to the collide and squeeze operation this need comes only when the beam emittances have shrunken significantly and the beam intensities have been reduced by 10\%. We therefore assume that only a 30\% larger normalized separation is required for this study case.

\begin{table}
	\centering	
	\caption{Parameters of the different optics for the alternative triplet.}
	 \label{Tab:jab:comparison}
	\begin{tabular}{|l|c|c|}
		\hline
		\textbf{Parameter} & \textbf{Round} & \textbf{Flat} \\
		\hline
		$\beta_x^*$ [m] & 0.3 & 1.2  \\ 
				\hline
		$\beta_y^*$ [m] & 0.3 & 0.15  \\ 
				\hline
	    Full crossing angle [$\mu$rad] & 200 &  130\\ 
	    		\hline
	    Beam-beam separation [$\sigma$] & 17 & 22  \\ 
		\hline
	\end{tabular}
\end{table}

\begin{figure}
	\centering
   \includegraphics[width=0.7\columnwidth]{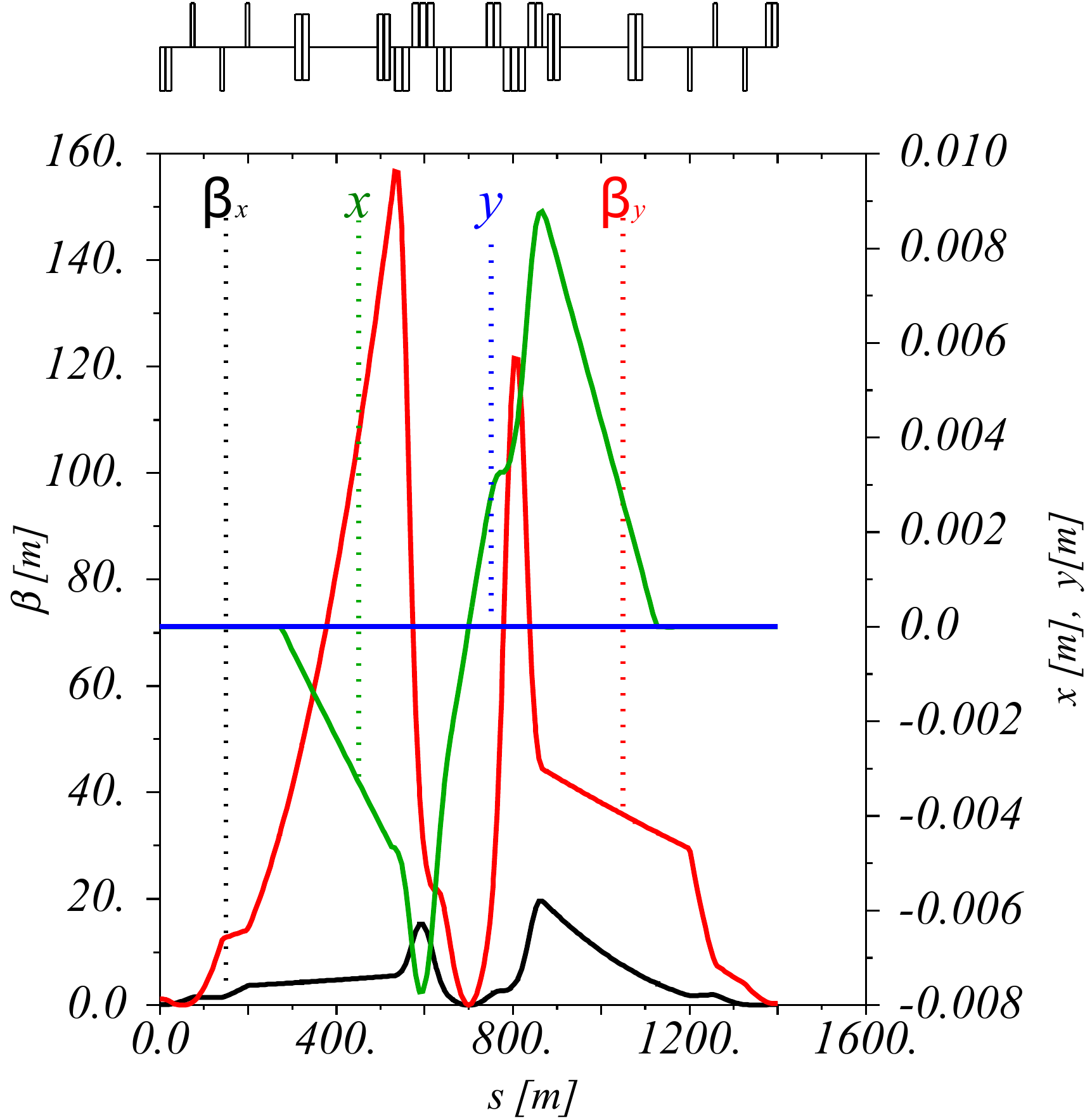}
	\caption{$\beta$ Functions and orbit for $\rm 1.2 \times 0.15~m$ flat EIR collision optics with alternative triplet and $\beta_{x}^*=$1.2~m, $\beta_{y}^*=$0.15~m.}
		\label{fig:jab:Optics40Flat}
\end{figure}

\begin{figure}[ht!]
	\centering
	\includegraphics[width=0.7\columnwidth]{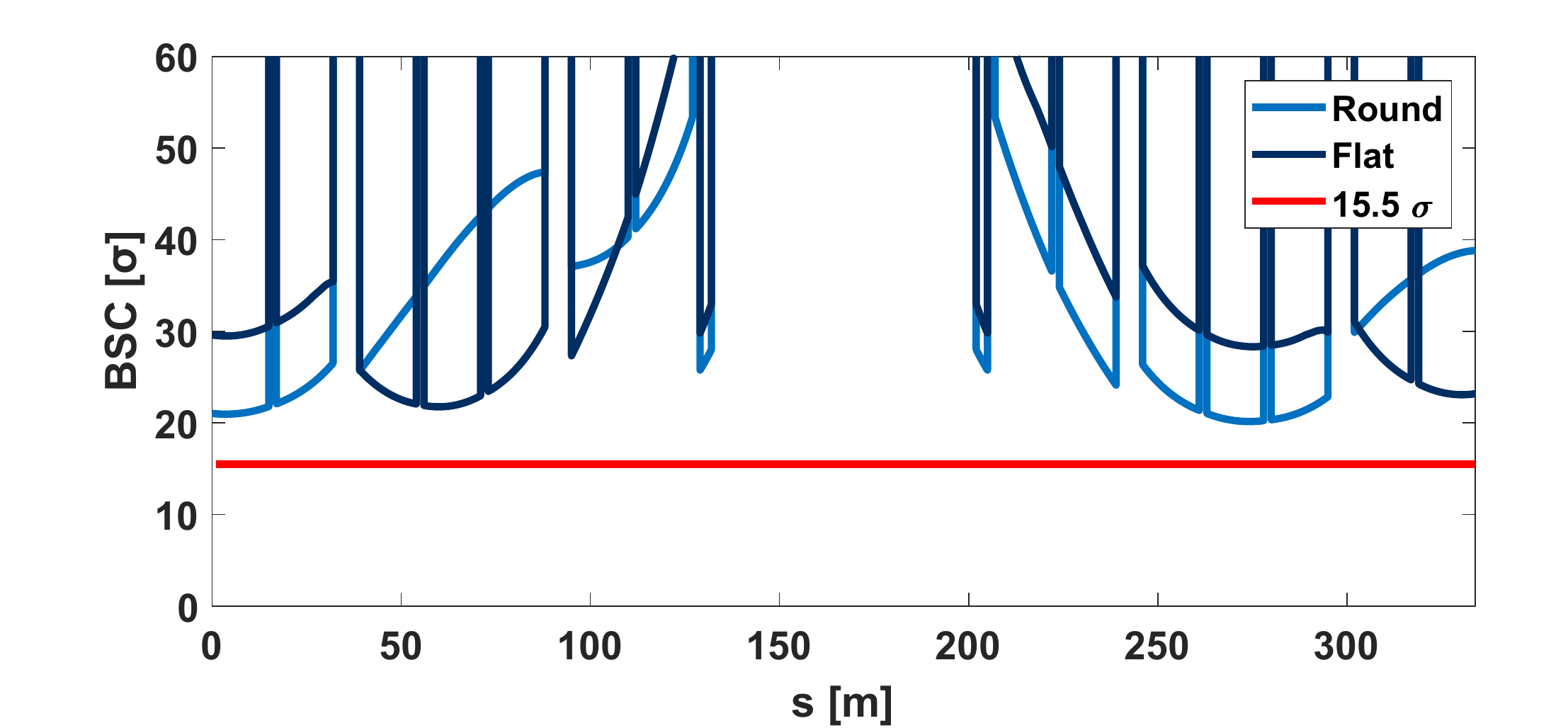}
	\caption{Plot showing BSC in triplet for flat and round optics.}
	\label{fig:jab:Reach40Flat}
\end{figure}

Figure~\ref{fig:jab:Optics40Flat} shows the EIR optics for the flat option.
The corresponding BSC are shown in Fig.~\ref{fig:jab:Reach40Flat}. The flat optics still provides sufficient BSC. 

\subsection{Low luminosity EIRs}
\label{sec:low_luminosity_ir}

\begin{figure}
   \centering
	\includegraphics*[width=\textwidth]{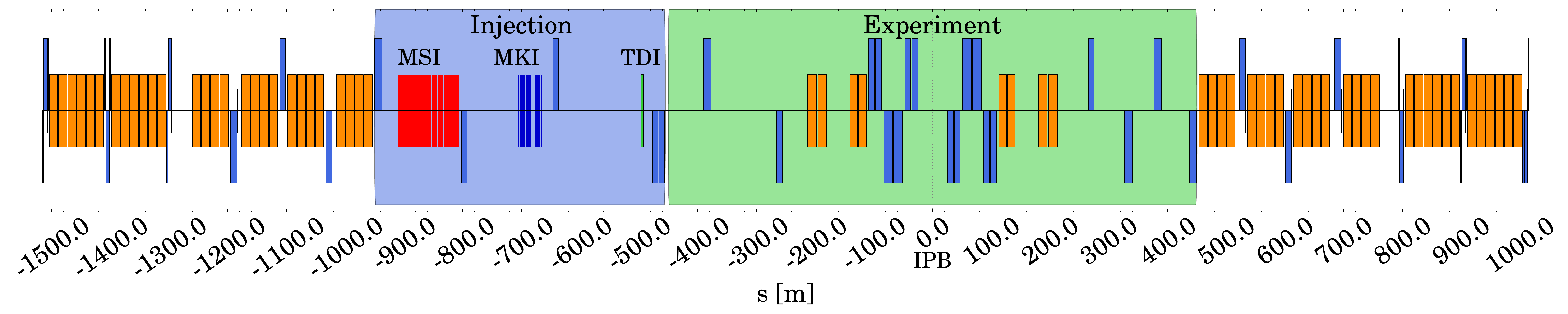}
   \caption{Layout of the low luminosity insertion in point B.}
   \label{fig:mho:low_lumi_ipb_layout}
\end{figure}

In addition to the high luminosity IRs, located around points A and G, the FCC-hh will also host two low luminosity experimental insertions in points B and L. Similar to the LHC, in these insertion also the beams from the injector chain will be injected upstream of the experiments. 
Due to initial injection hardware considerations, the length of half cells containing such hardware has been set to \SI{150}{m}. Contrary to the LHC injection/experimental insertions, two additional half cells were added after the injection cells. These give the possibility to add more injection protection elements to protect both the superconducting magnets and experiment from misinjected beam. Due to this, the interaction point is not located in the center of the straight section but rather \SI{250}{m} further downstream.
Unlike the high luminosity experimental insertions, currently no required performance is established for these low luminosity experiments and subsequently no target $\beta^*$ can be specified. Similarly, due to the lack of a detector design and required cavern length,  $L^*$ has been tentatively set to \SI{25}{m}.
The layout of this combined injection/experimental insertion for point B based on these considerations is presented in Fig.~\ref{fig:mho:low_lumi_ipb_layout}.

The final focus triplet left and right of the interaction point consists of three quadrupoles. Each one of these quadrupoles is split into two submagnets to keep the magnet length below \SI{15}{m}. 
Each one of the Q1 and Q3 submagnets has a length of \SI{10}{m} wheres each Q2 submagnet is \SI{15}{m} long. Between each submagnet, a drift space with a length of \SI{2}{m} has been reserved for the interconnects. All triplet quadrupoles have the same coil aperture of \SI{64}{mm}. The specifications of the triplet quadrupoles can be found in Tab.~\ref{tab:rma:ll_eir_magnet_specifications}.
This aperture is further reduced, because of the presence inside the coil of a liquid helium layer, a Kapton insulator layer, and the stainless steel cold bore, using the same specifications as described in Tab.~\ref{tab:Beam_Screen_Components}. A \SI{10}{mm} thick tungsten (INERMET180) shielding is finally put inside the cold bore to mitigate the radiation in the superconducting coils. The available radial aperture for the beam is therefore reduced to \SI{18.25}{mm}.
In order to keep the separation section after the triplet as short as possible a superconducting solution was chosen. 
Using two \SI{12.5}{m} long shared aperture separation dipoles D1 with a field strength of \SI{12}{T} and two \SI{15}{m} long double aperture recombination dipoles D2 with a field strength of \SI{10}{T} the length of this section can be keep under \SI{100}{m}. 
Due to the aforementioned considerations on the injection hardware and the added additional cells, the matching sections on the left and right hand side of the insertion do not have the same length. 
On the non-injection side of the insertion, four matching quadrupoles make up the matching section which is \SI{235}{m} long. 
The \SI{735}{m} long matching section on the injection side of the insertion consist six matching quadrupoles. Between Q8 and Q9 the injection septum (MSI) is located which deflects the injected beam in the vertical plane. In the following half cell between Q7 and Q8 the injection kicker system (MKI) is installed which provides a horizontal kick to put the injected beam on the closed orbit. The quadrupole Q8 between the MSI and the MKI was chosen to be horizontally defocussing to provide an additional horizontal kick which helps in reducing the required kick strength of the injection kickers. A \SI{4}{m} long absorber (TDI) to protect superconducting magnets further downstream from misinjected beam is installed in the cell between Q6 and Q7. Each of these half cells is \SI{150}{m} long.

\begin{figure}
   \centering
	\includegraphics*[width=\textwidth]{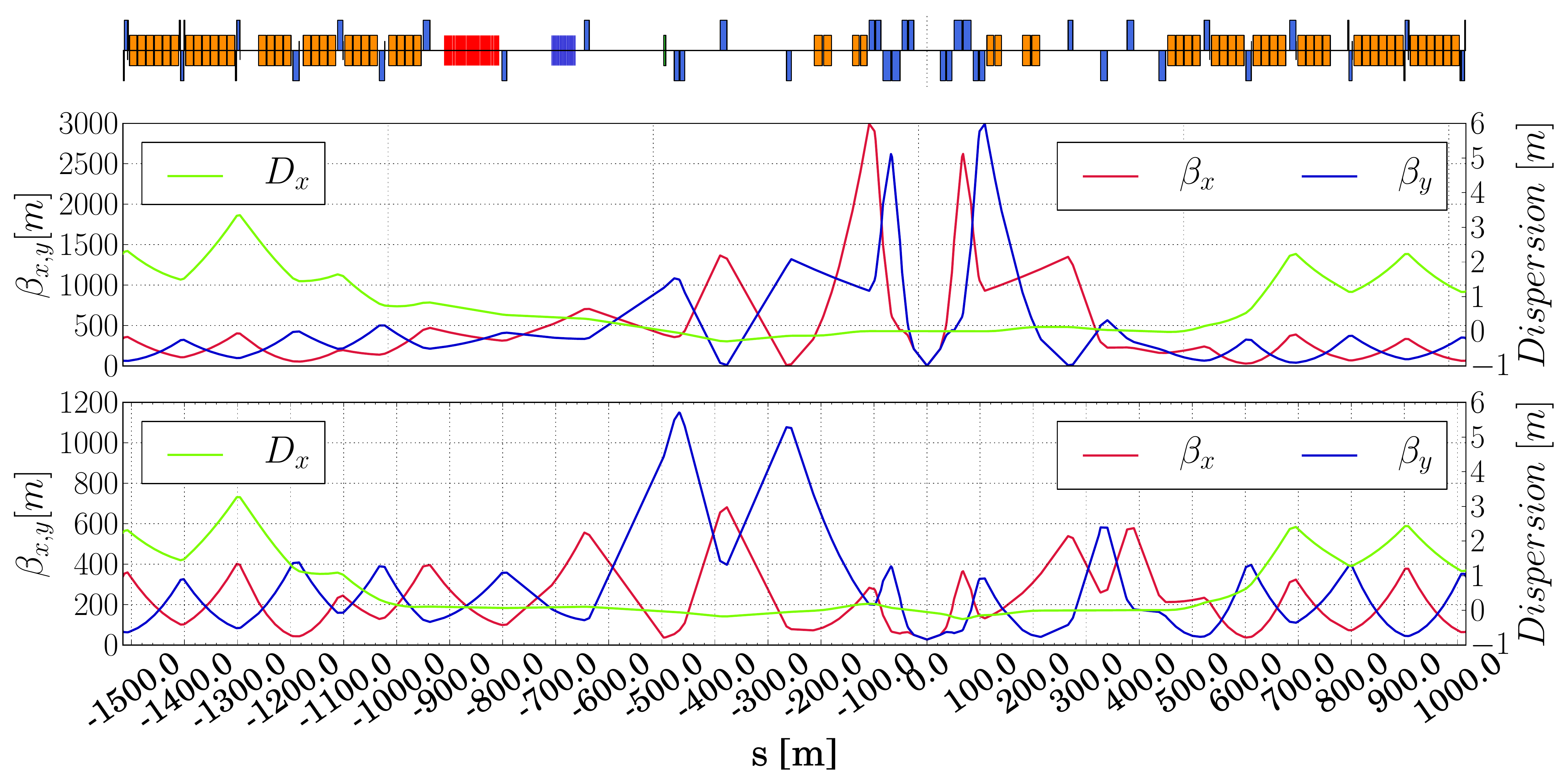}
   \caption{Collision optics (top) and injection optics (bottom) for the low luminosity insertion in point~B.}
   \label{fig:mho:low_lumi_ipb_optics}
\end{figure}

\begin{figure}[htb!]
   \centering
	\includegraphics*[width=\textwidth]{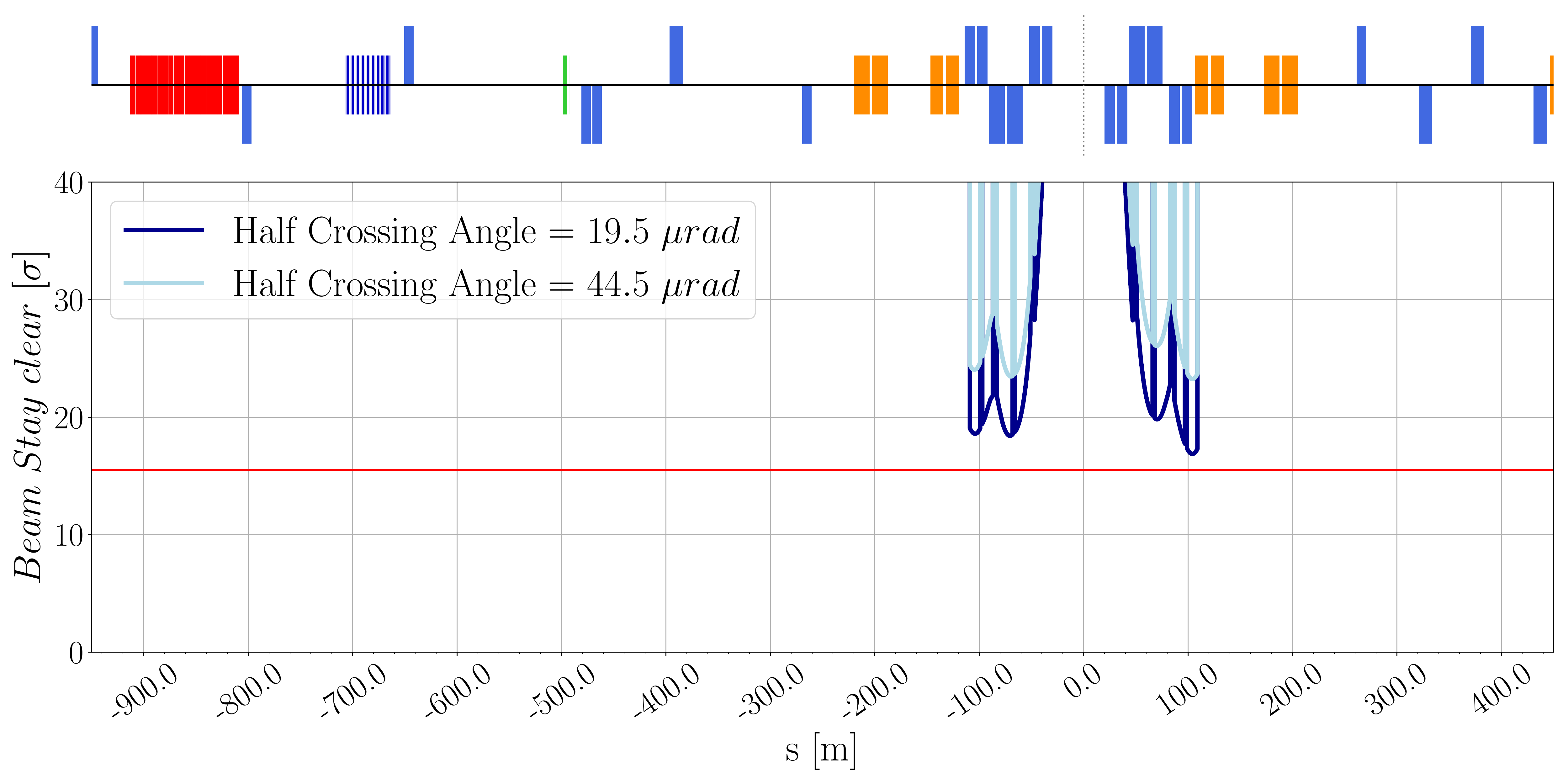}
   \caption{Aperture in the low luminosity insertion in point~B corresponding to a half crossing angle of~\SI{19.5}{ \micro rad} and for the maximum possible half crossing angle of~\SI{44.5}{ \micro rad}. The minimum allowed beam stay clear of $15.5~\sigma$ is indicated in red.}
   \label{fig:mho:low_lumi_ipb_xing}
\end{figure}

Both the optics for collision energy as well as for an injection energy are presented in Fig.~\ref{fig:mho:low_lumi_ipb_optics}.
At collision energy a minimum $\beta^*$ of \SI{3}{m} has been matched. The crossing angle for these insertion has obtained from scaling the normalized separation of the high luminosity insertion~\cite{bib:ach:IPAC2016-TUPMW020} with the reduced number of long range encounters. This leads to a normalized separation of $5.25~\sigma$, corresponding to a half crossing angle of~\SI{19.5}{ \micro rad}. With this crossing angle the beam stay clear in the triplet is well above the minimum allowed beam stay clear of $15.5~\sigma$ and could be further increased, which is illustrated in Fig.~\ref{fig:mho:low_lumi_ipb_xing}.
However, a crossing angle of \SI{180}{\micro rad} at full intensity as detailed in Section~\ref{sec:Beam-Beam crossing angle} would lead to a beam stay clear below limit of $15.5~\sigma$. To comply with both constraints, the $\beta^*$ at begin of collision then has to be set to \SI{19}{m}. The minimum $\beta^*$ of \SI{3}{m} could then be reached after 1.5 hours, assuming a reduction of the separation to $20~\sigma$ at this point. During collision, the beams are always colliding with transverse offset to keep the head-on beam-beam tune shift from these two experiments below $1-2 \times 10^{-3}$.  

At injection energy, the crossing angle is limited by the triplet aperture to a beam separation of $7~\sigma$. Further studies are required to assess the viability of this separation. If the separation proves to be insufficient, the shielding in the triplet could be reduced, in turn potentially decreasing also the achievable integrated luminosity.

For the injection optics various constraints had to be taken into account to provide optimum injection protection efficiency. The horizontal phase advance between the MKI and the TDI should be $90^\circ$ to ensure any kicker failure translating into an orbit offset at the TDI. The beta functions at the TDI were matched to the largest possible values to increase the beam size which in turn reduces the peak energy density in case of an injection kicker malfunction. Furthermore, the dispersion function in the straight section is kept below \SI{30}{cm}.

\begin{figure}
   \centering
	\includegraphics*[width=\textwidth]{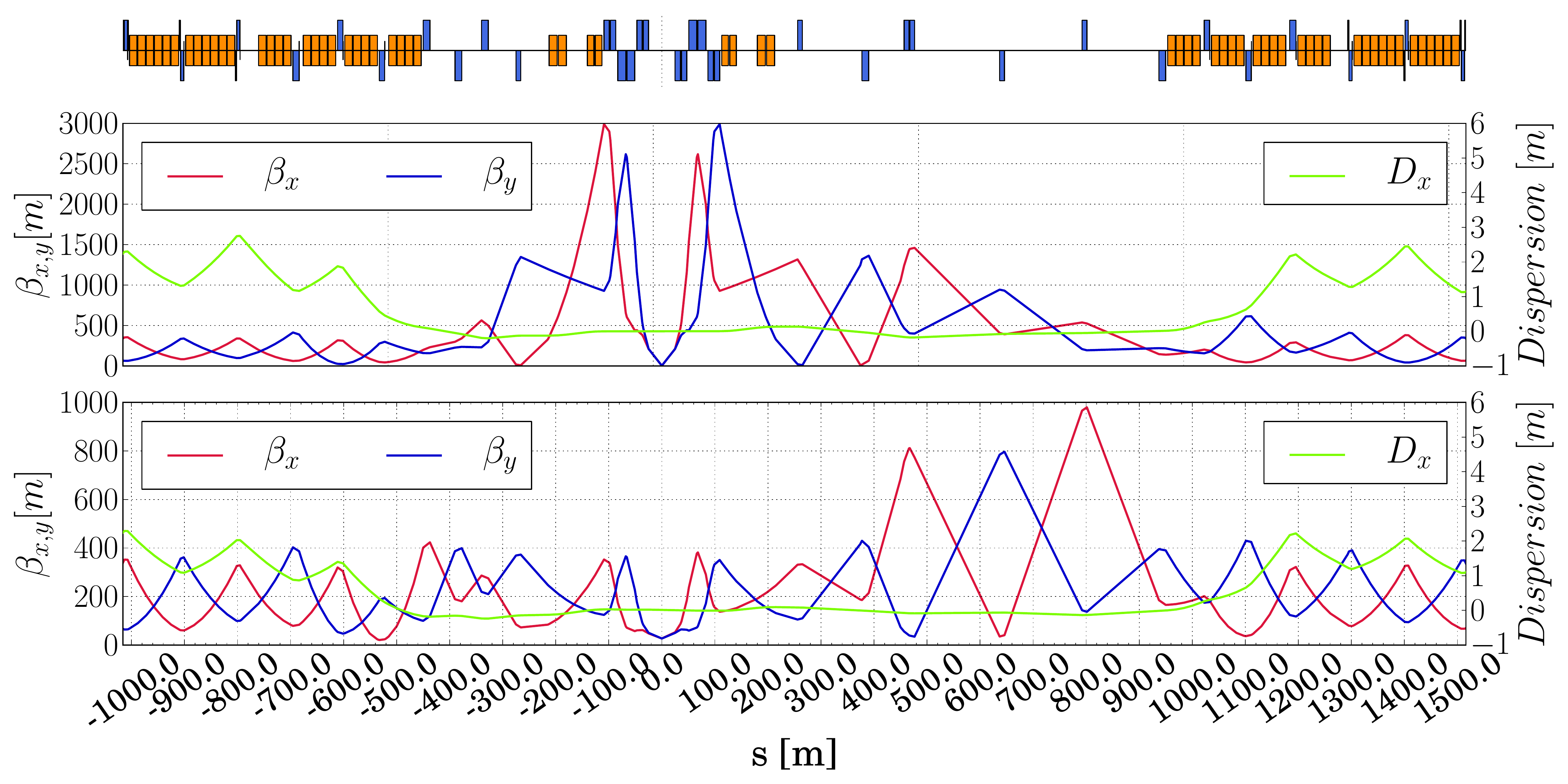}
   \caption{Collision optics (top) and injection optics (bottom) for the low luminosity insertion in point~L.}
   \label{fig:mho:low_lumi_ipl_optics}
\end{figure}

As the injection of beam 2 will take place in point L the low luminosity insertion for this straight section is mirrored with respect to s. Here the interaction point is located \SI{250}{m} upstream of the middle of the straight section for beam 1. Both the collision optics with a $\beta^*$ of \SI{3}{m} and injection optics with a $\beta^*$ of \SI{27}{m} are illustrated in Fig.~\ref{fig:mho:low_lumi_ipl_optics}.

\subsection{Crab Cavities}
In the long shared aperture section around the IP, the two counter rotating beams must be separated by an orbit bump in order to avoid parasitic beam-beam encounters that occur every \SI{3.75} {m} left and right of each IP. The two beams only cross each other at the IP with a crossing angle $\theta$. The crossing angle determines the separation of the beam in the shared aperture and thus the long range beam-beam separations. The minimum crossing angle was determined by beam-beam studies to be \SI{200}{\micro rad} for the ultimate $\bstar$ of \SI{0.3}{m} and then scaled for other optics to provide the same normalized separation of $\approx$ \SI{17}{$\sigma$} for the maximum bunch intensity. Table~\ref{tab:rma:crossing_angle_vs_lumi_reduction} list the crossing angle for a set of collision optics together with the luminosity reduction factor caused by the reduced geometric overlap of the bunches at the IP due to the crossing angle. For ultimate optics and beyond, FCC-hh is not able to provide even half of the luminosity head-on collisions would provide. It is clear that the luminosity reduction in the high luminosity EIRs must be compensated by crab cavities.
\begin{table}
  \begin{center}
  \caption{\label{tab:rma:crossing_angle_vs_lumi_reduction} Crossing angle and luminosity reduction due to crossing angle for different collision optics  for an emittance of $\epsilon_{n} = \SI{2.2}{\micro \metre}$.}
  \begin{tabular}{ |m{3.5cm}|c|c|c| } 
	\hline
	  \bf Optics version & \bf $\bstar$ [\si{m}] & \bf Full crossing  & \bf Luminosity \\ 
	& & \bf angle $\theta$ [\si{\micro rad}] & \bf reduction factor \\ \hline
	baseline & 1.1 & 104 & 0.85 \\ \hline
	ultimate & 0.3 & 200 & 0.40 \\ \hline
	beyond ultimate & 0.2 & 245 & 0.28 \\ \hline
  \end{tabular}
  \end{center}
\end{table}

\begin{figure}
  \begin{center}
    \includegraphics*[width=0.45\textwidth]{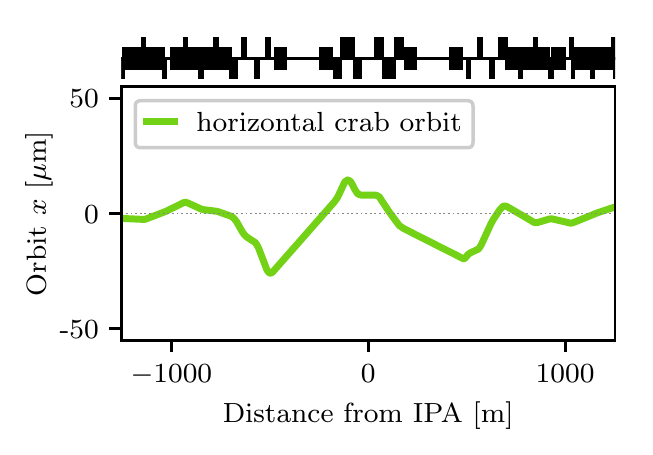}
    \includegraphics*[width=0.45\textwidth]{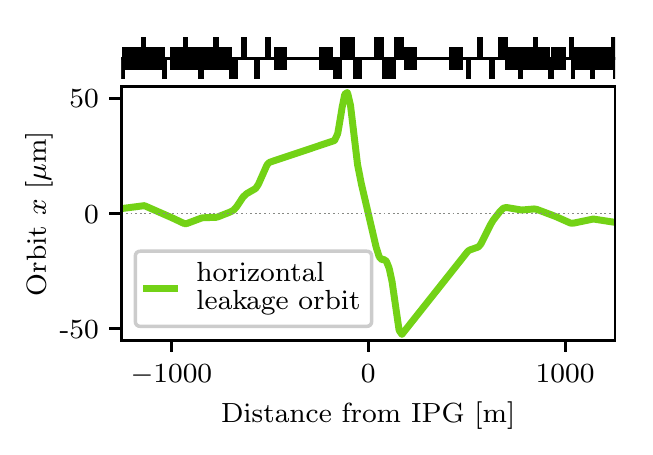}
    \includegraphics*[width=0.45\textwidth]{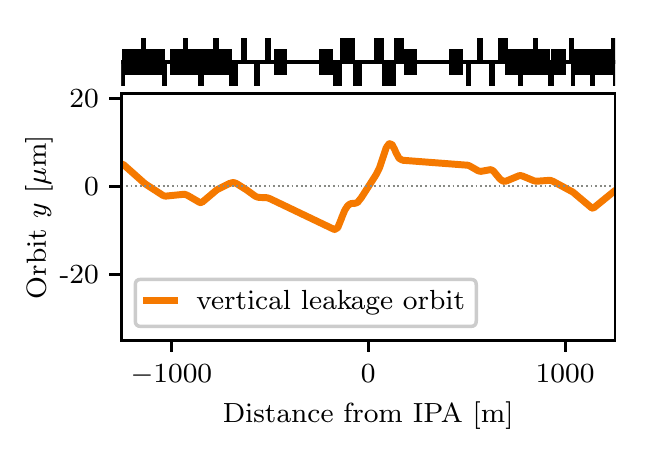}
    \includegraphics*[width=0.45\textwidth]{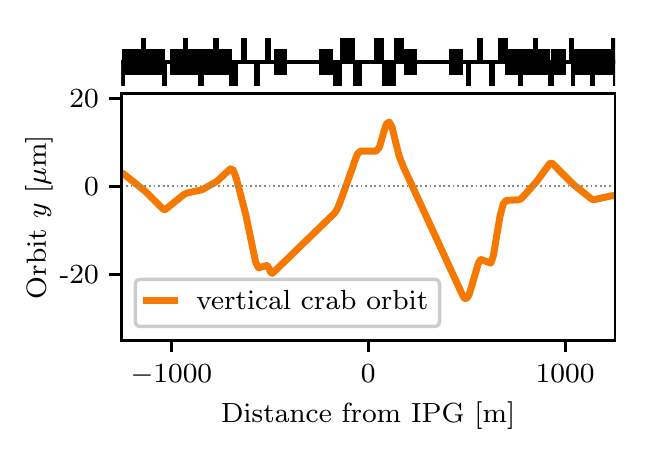}
  \end{center}
  \caption{Crab orbits for $\bstar = \SI{0.3}{m}$ and orbit leakage into the other high luminosity EIR.}
  \label{fig:rma:crab_orbits}
\end{figure}

Initial studies with crab cavities show that a crab voltage of \SI{13.4}{MV} per beam on either side of each high luminosity IP is needed to provide full crabbing in ultimate optics, corresponding to \SI{107.2}{MV} in total. Half of this voltage must be horizontally deflecting in one EIR, the other half vertically deflecting in the other EIR. For optics beyond ultimate parameters, the crab voltage increases up to $8\times \SI{18.1}{MV}$. Orbit leakage of the crab orbit into the arcs varied strongly during the evolution of the lattice. In the latest lattice version it appears to be small, causing only small orbit aberrations in the other IPs. More detailed studies should be performed to get a better control of the orbit leakage in the future. The crab orbits and orbit leakage into the other high luminosity EIR are shown in Fig.~\ref{fig:rma:crab_orbits} for ultimate optics.

\section{Dynamic Aperture Studies}
 The  insertion regions presented in the previous sections have been integrated into the FCC-hh lattice and Dynamic Aperture (DA) studies are run to validate the particle stability. Below simulations without and with beam-beam interaction are presented.

\subsection{Dynamic Aperture with Triplet Errors}
This section covers DA studies run at collision energy to evaluate the impact of errors on the magnets and analyze possible correction schemes to achieve the minimum DA necessary to ensure the stability of the beam. The studies covered in this section do not include beam-beam effects, and therefore, based on the experience of the LHC, the target DA was set to be 10$\sigma$. DA studies including beam-beam effects, with a corresponding lower target DA are covered in Section~\ref{sec:Beam-Beam crossing angle}. 

Given the large beta functions and integrated length of the quadrupoles of the final focus triplet of the high luminosity EIRs, the effects of systematic and random non-linear errors in the magnets had a severe impact on the stability of the beam. Therefore, DA studies at collision energy with errors on the triplet and crossing angle on proved to be challenging. Several corrections were implemented to compensate for the reduction on DA, but it was found that at collision energy two corrections were particularly important to achieve a DA above the target: the first one is to optimize the phase between the main IRs (IRA and IRG), which comprises running simulations to find which phase provided the best compensation between the errors of both IRs and therefore the higher DA; and the second one the implementation of non-linear correctors in the IR, to minimize the resonance driving terms arising from the errors in the triplet. 

\begin{figure}
	\centering
	\includegraphics*[width=0.7\textwidth]{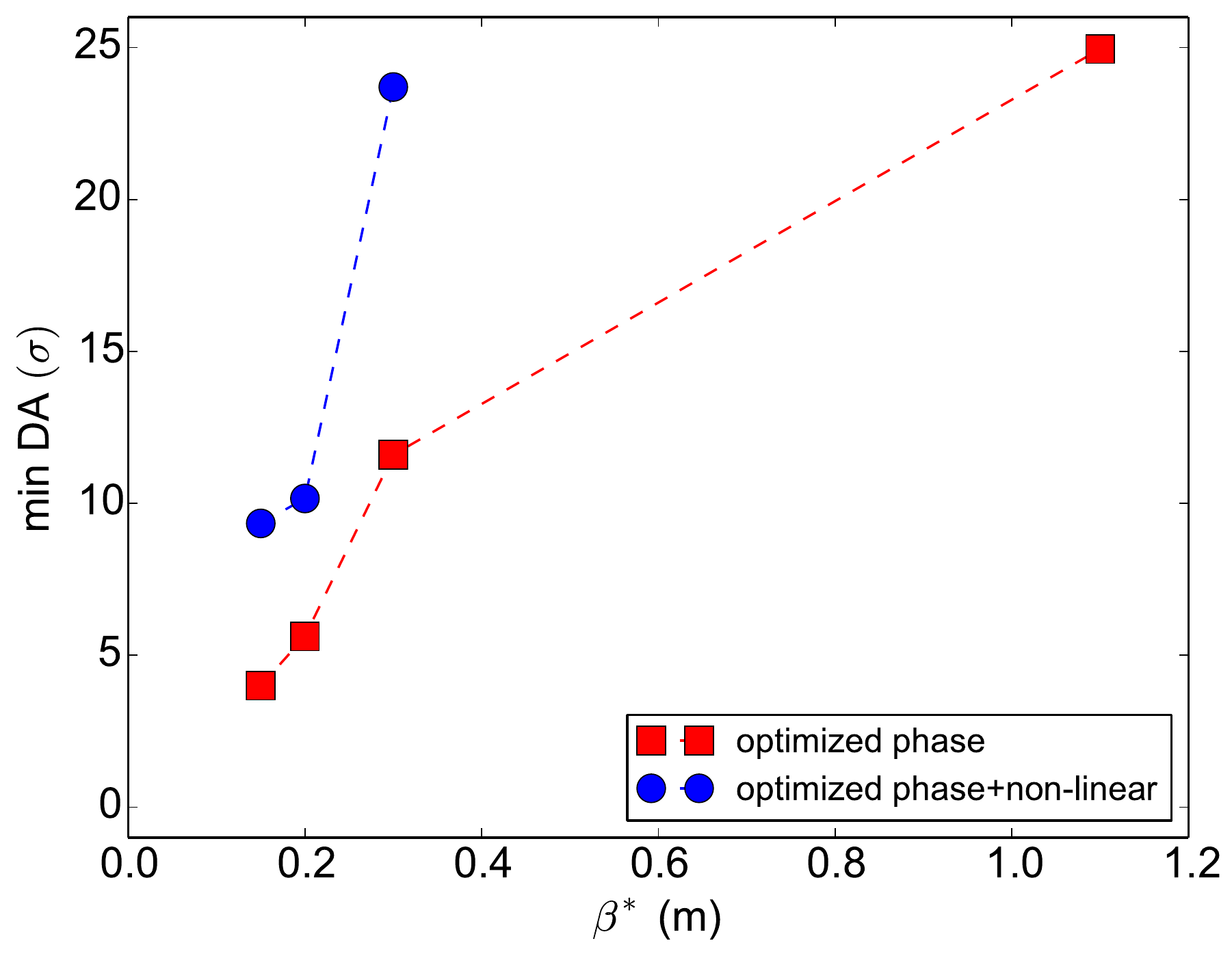}
	\caption{\label{fig:eca:davsbetaIR}Minimum DA over 60 seeds versus $\beta^*$ with and without non-linear correctors.}
\end{figure}

The minimum DA vs $\beta^*$ at collision energy with errors on both the arcs and the triplets, and with and without non-linear correctors is shown on Fig.~\ref{fig:eca:davsbetaIR}. 
A significant increase is observed on all cases when adding the non-linear correctors, except for the case with $\beta^*$=1.1~m whose DA is already large without non-linear correctors. The ultimate case with $\beta^*$=0.3~m shows a DA above the target of 10 $\sigma$ even without non-linear correctors, by optimizing the phase advance between the main IRs and other corrections; however, the use of non-linear correctors is still recommended in case other errors affect the DA. In the presence of beam-beam effects different optimized phases are required and detailed optimization should be foreseen. The figure also shows dynamic apertures for optics with $\beta^*$ below the ultimate 0.3~m. The use of non-linear correctors becomes essential for these cases.

\subsection{Beam-Beam effects and crossing angle}
\label{sec:Beam-Beam crossing angle}
\begin{figure}
	\centering
	\includegraphics*[width=0.49\textwidth]{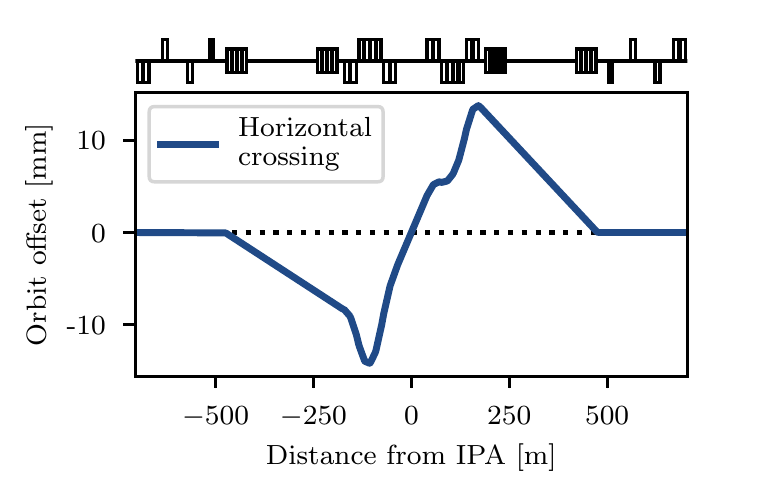}
	\includegraphics*[width=0.49\textwidth]{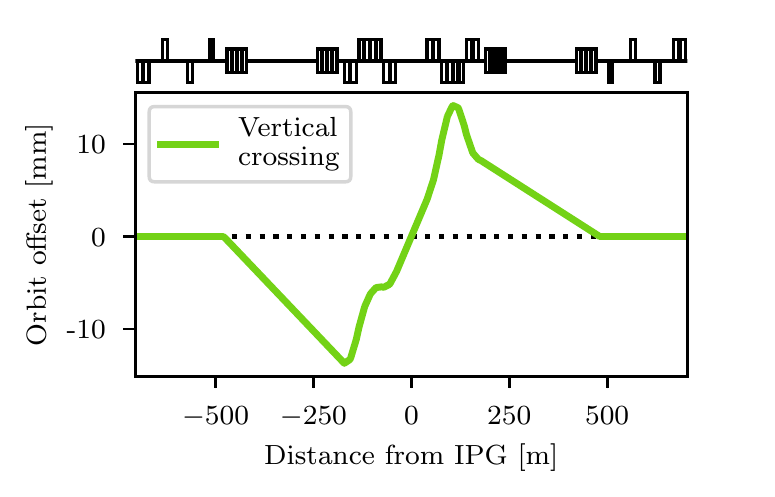}
	\caption{\label{fig:rma:FCC_hh_IPA_0300_orbit_bump} Orbit bump for a \SI{200}{\micro rad} crossing angle required at $\bstar = \SI{0.3}{m}$ in the horizontal plane for IPA and in the vertical plane for IPG.}
\end{figure}

\begin{figure}
    \centering
    \includegraphics*[width=0.8\textwidth]{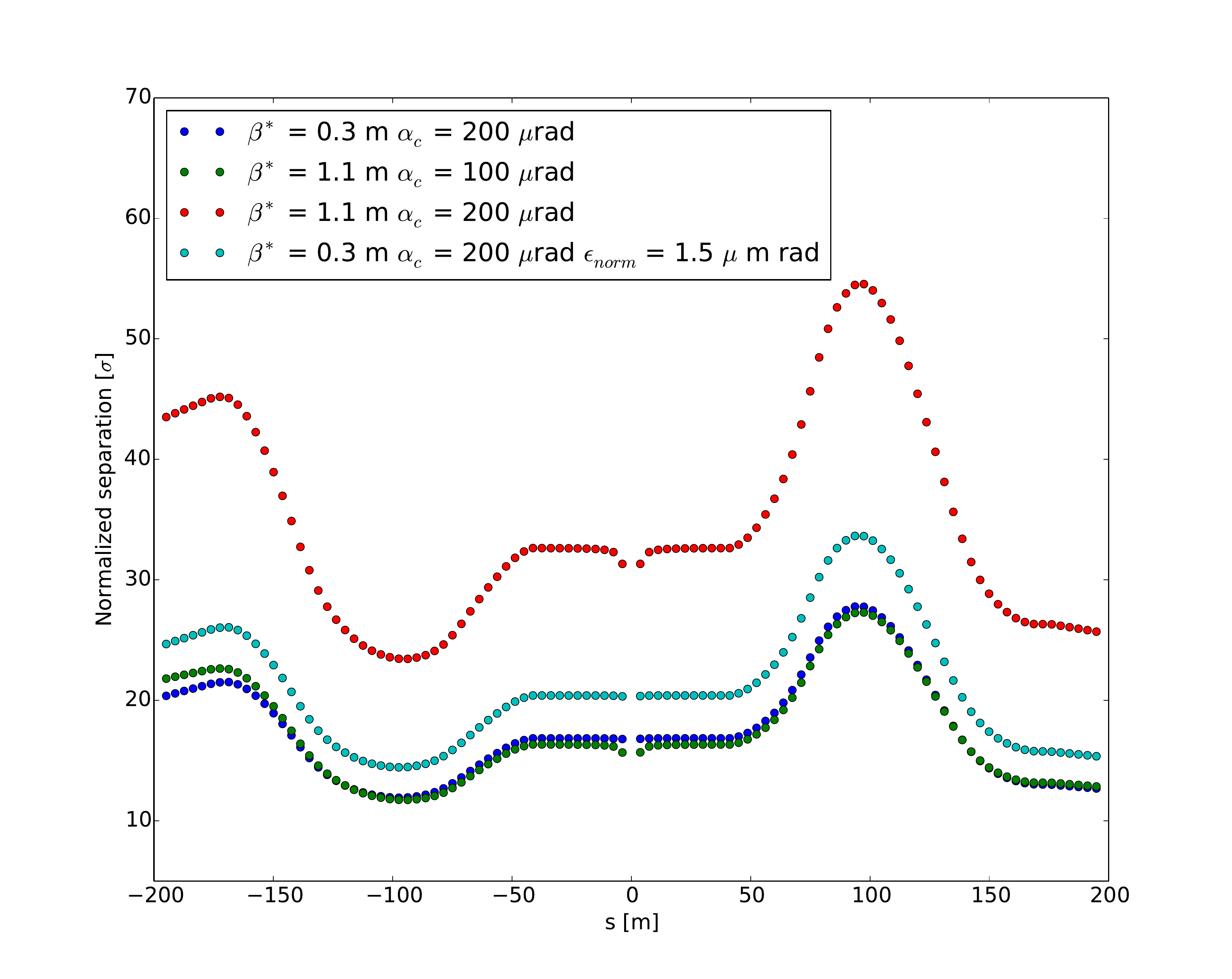}
    \caption{Beam-beam long range separations in units of the transverse beam size in the IR for the Ultimate scenario with $\bstar = \SI{0.3}{m}$ with a crossing angle $\theta =\SI{200}{\micro rad}$ and the nominal normalized emittance $\epsilon_n=\SI{2.2}{\micro m}$ (blue dots) and for a reduced normalized emittance of $\epsilon_n=\SI{1.5}{\micro m}$ (light blue dots). The beam-beam long range for the Baseline scenario for  $\bstar = \SI{1.1}{m}$ are also plotted with a crossing angle $\theta=\SI{200}{\micro rad}$ (red dots) and with a reduced crossing angle $\theta=\SI{100}{\micro rad}$ (green dots). For this last cases the nominal normalized emittance of $\epsilon_n=\SI{2.2}{\micro m}$ has been considered.}
    \label{fig:sepLR}
\end{figure}

The beam-beam interaction can limit the performance of a particle collider. In fact, the beam-beam interaction can induce particle losses, resulting in a reduction of the beam lifetimes and can create a high background for physics experiments. 
In addition, the beam-beam interaction can be responsible for an elevated heat and radiation load on the collimation system, can induce emittance blow-up and can cause coherent beam instabilities with a consequent reduction of the luminosity reach.
\begin{figure}
    \centering
    \includegraphics*[width=0.7\textwidth]{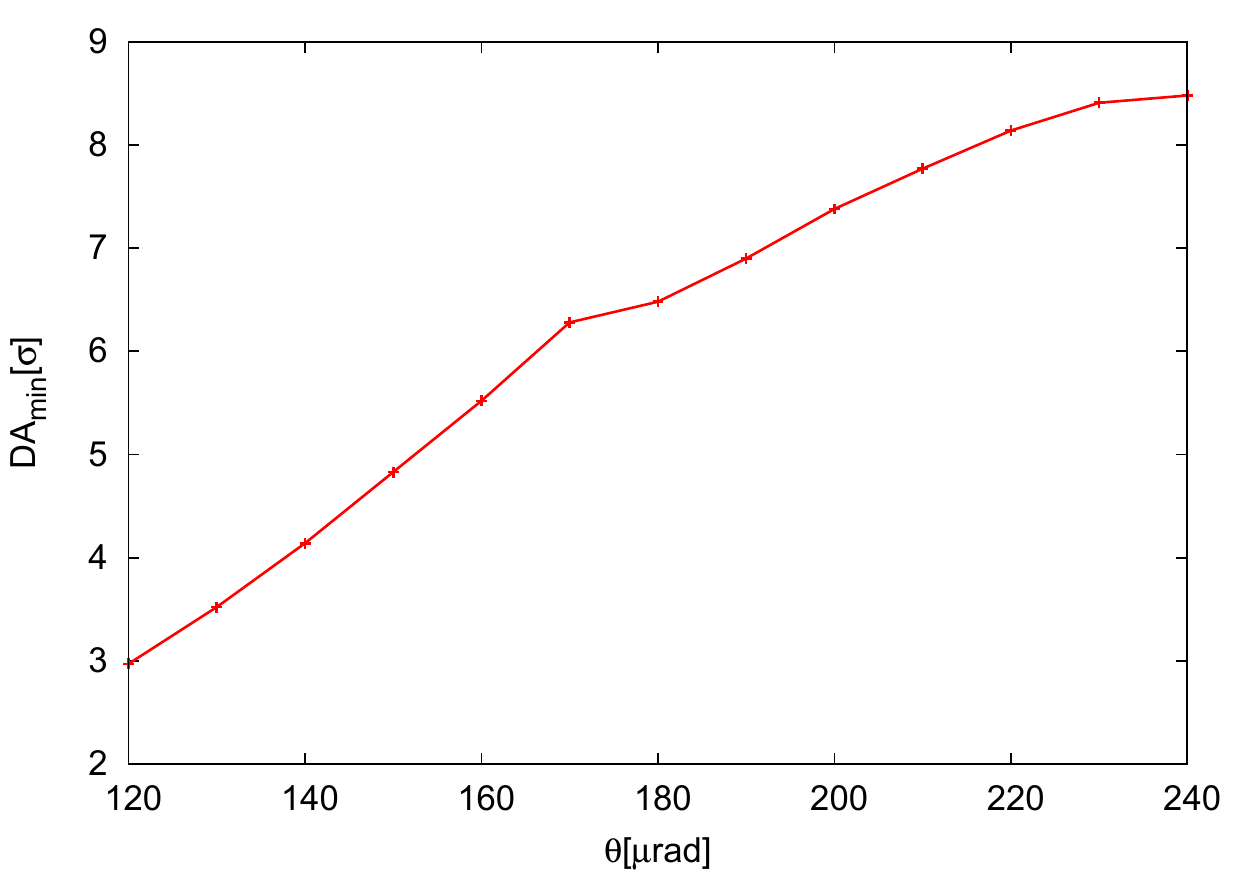}
    \caption{Minimum Dynamic Aperture in the presence of beam-beam interactions as a function of the crossing angle at the interaction points A and G for the Ultimate collisions optics with $\bstar = \SI{0.3}{m}$ and $\lstar = \SI{40}{m}$~\cite{barrancoFCCweek2018}.}
    \label{fig:DAbeambeam}
\end{figure}
The design of FCC-hh is based on the LHC beam-beam theoretical studies and experience~\cite{Herr:2014bca, Pieloni1968515, buffat2017094, pielonithesis}. The beams collide head-on in two high luminosity interaction points (IPA and IPG). According to the filling scheme used, the bunches experience different number of head-on and long range collisions generating two different families of bunches~\cite{herrbbeffects}. The so-called Nominal bunches are located in the middle of a train while the PACMAN bunches are located in the head or in the tails of the train. Due to empty slots at the interaction points the PACMAN bunches experience fewer long range interactions than Nominal bunches, leading to different beam-beam effects. As for the LHC and the HL-LHC, an alternating crossing scheme is chosen for the two high luminosity experiments in IPA and IPG, in order to passively compensate for tune and chromaticity shifts for PACMAN bunches~\cite{Herr:604005}. In these studies we have assume the beam crossing with a finite horizontal angle for IPA and a vertical one in IPG as shown in Fig.~\ref{fig:rma:FCC_hh_IPA_0300_orbit_bump}. Different schemes have also explored and seem  feasible with reduced beam-beam long-range effects~\cite{barrancoEuroCirc} but will require further studies.
Two additional, lower luminosity experiments are located in IPB and IPL. Assuming that the four experiments operate in proton-proton collisions with 25~ns bunch spacing, 352 long range encounters are expected compared to the 120 long range beam-beam interactions of the LHC.
Detailed beam-beam studies have been carried out by means of weak-strong as well as strong-strong models by using the SixTrack~\cite{bib:SixTrack, schmidt94} and COMBI~\cite{combiweb, combitatiana, combiWT} codes. The SixTrack code has been used for the computation of the area of stable motion in real space, the DA. A detailed lattice description and the LHC experimental data have been employed for the benchmark of the SixTrack code with and without beam-beam effects~\cite{crouchthesis, PhysRevSTAB.15.024001}.
The COMBI code makes use of a self-consistent treatment, including a simplified lattice description, and provides the evolution of macroscopic beam parameters, such as, the beam intensity and the emittance together with the Landau damping of coherent beam instabilities~\cite{pielonithesis, PhysRevSTAB.17.111002}. For the DA studies presented here, the approach used is similar to the LHC and HL-LHC design studies~\cite{Grote:691988, Luo:692074}. The LHC observations have shown that below a simulated DA of 4~$\rm{\sigma}$ a reduction of beam lifetime starts to appear~\cite{Herr:2014bca}. As described in~\cite{crouchthesis, ipaccrouch} a strong correlation exists  between the beam intensity lifetime and the simulated DA for different beam configurations with and without beam-beam interactions. The conservative approach to ensure the good performance of the machine is to target a DA with beam-beam as large as the collimation gap, which is 7.2~$\rm{\sigma}$ for FCC-hh.
As it is shown in Fig.~\ref{fig:DAbeambeam}, a DA of 7.2~$\rm{\sigma}$ is ensured with a crossing angle $\theta=\SI{200}{\micro rad}$ in IPA and IPG for the nominal normalized emittance of $\epsilon_n=\SI{2.2}{\micro m}$ and at the Ultimate $\bstar$ of $ \SI{0.3}{m}$.  The corresponding orbit bumps at the two interaction points are shown in Fig.~\ref{fig:rma:FCC_hh_IPA_0300_orbit_bump} and the beam-beam long range separations in units of the transverse beam size are shown in Fig.~\ref{fig:sepLR}. For the Ultimate scenario with $\bstar = \SI{0.3}{m}$ and the chosen crossing angle of \SI{200}{\micro rad} (the blue dots), the long range separation at the first encounter is 17~$\rm{\sigma}$. At this separation, the value of DA (7.2 $\rm{\sigma}$) is well above 4 $\rm{\sigma}$ providing sufficient margin to avoid additional particle losses on the collimation system due to beam-beam diffusive mechanisms (including a relative momentum deviation of $10^{-4}$). 
In addition, margins are also left for high chromaticity operations (up to 20 units) if required for mitigation of coherent beam instabilities, for operating in the presence of multipolar lattice errors~\cite{garciaIpac} or to also collide in IPB and IPL. 
For the Baseline scenario with collisions at $\bstar = \SI{1.1}{m}$, the long range beam-beam separation is well above 30 $\rm{\sigma}$  (the red dots). 

\begin{figure}
  \centering
  \includegraphics*[width=0.7\textwidth]{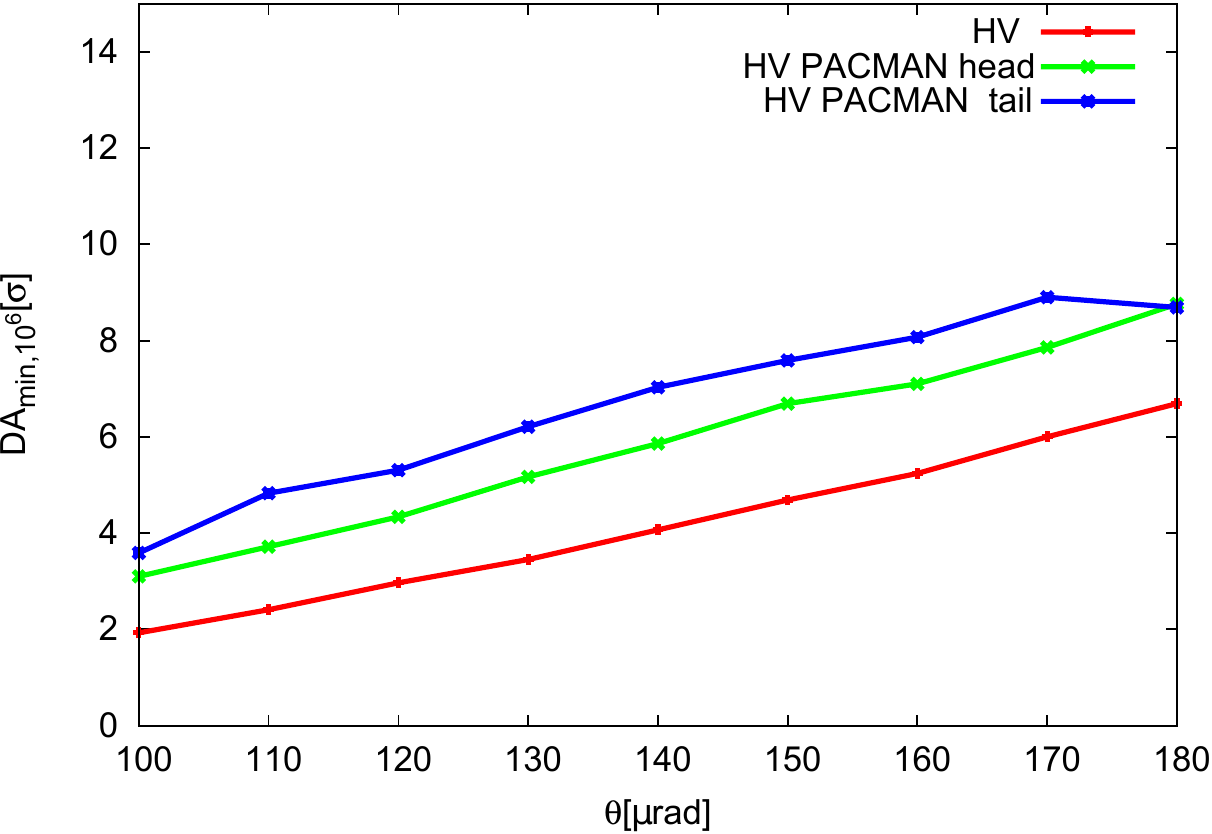}
  \caption{Minimum Dynamic Aperture, evaluated over $10^6$ turns, as a function of the crossing angle in IPA and IPG for H-V alternating crossing scheme for Nominal bunches (the red line) and for PACMAN bunches in the head of a train (the green line) and in the tail of a train (the blue line)~\cite{garciaIpac}.}
  \label{fig:daHVPacman}
\end{figure}

The DA as a function of the crossing angle in IPA and IPG for PACMAN bunches is shown in Fig.~\ref{fig:daHVPacman} (the blue and the green lines) for the H-V alternating crossing scheme. The red line corresponds to Nominal bunches. As visible, the DA for PACMAN bunches is always above the DA for Nominal bunches. The PACMAN effects of tune and chromaticity shifts are negligible assuming the passive compensation  with alternating crossing planes in IPA and IPG~\cite{barrancoEuroCirc}.

\begin{figure}
    \centering
    \includegraphics*[width=0.65\textwidth]{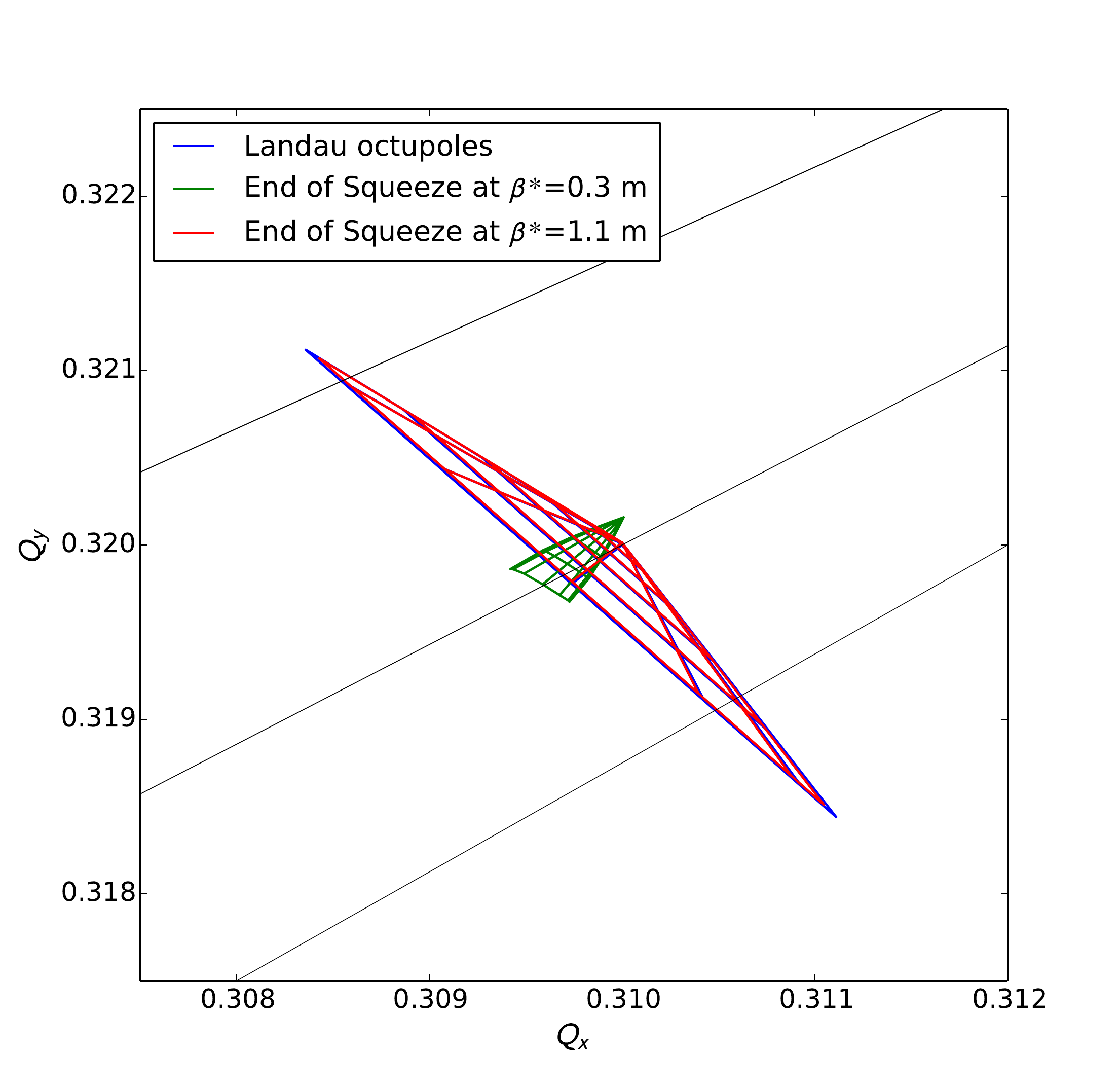}
    \caption{Two-dimensional tune footprints for particles up to 6 $\rm{\sigma}$ amplitude at the end of the beta squeeze including long-range beam-beam interactions and Landau octupoles powered with negative polarity, for the Ultimate scenario (the green color) and for the Baseline scenario (the red color). For comparison the case with only Landau octupoles powered with negative polarity (maximum strength) is also shown (the blue color). Note that the blue footprint does not depend on $\beta^*$ as Landau octupoles are placed in the arc.}
    \label{fig:footeos}
\end{figure}

The two-dimensional expected tune footprints for particles up to 6 $\rm{\sigma}$ amplitude are shown in Fig.~\ref{fig:footeos} for the two values of $\bstar$ ($\bstar = \SI{1.1}{m}$ and $\bstar = \SI{0.3}{m}$) with and without the effects of the long-range beam-beam effects. Without long-range beam-beam effects the tune spread in frequency is generated by the Landau octupoles (the blue color) at maximum strength and powered with negative polarity as described in~\cite{Gareyte}. The tune spread is then reduced by the long-range beam-beam effects depending on the interactions strength. In  Fig.~\ref{fig:footeos} the Ultimate scenario with  $\beta^*=\SI{0.3}{\meter}$ , corresponding to a minimum beam to beam separation at the long-range encounters of 17  $\rm{\sigma}$ (the green color) and the the Baseline scenario with $\beta^*=\SI{1.1}{\meter}$  (the red color) at relaxed beam to beam separations of 32 $\rm{\sigma}$, are shown. Despite the tune spread reduction due to the beam-beam long-range compensating the octupole spread, such configuration is preferred because it maximizes the DA~\cite{JShiEPAC}.
The strategy proposed for the FCC is to collide head-on  at the two main IPs before the long-range interactions reduce significantly the tune spread provided by the Landau octupole system. Such reduction occurs during the betatron squeeze and collisions should be foreseen around 1.1 meters  $\beta^*$, value at which no reduction on the tune footprint is visible (the red color in Fig.~\ref{fig:footeos}) compared to the pure Landau octupoles tune spread (the blue color in Fig.~\ref{fig:footeos}).
If no coherent instabilities are observed, a reduction of the crossing angle at $\bstar = \SI{1.1}{m}$ is possible down to a minimum value of $\SI{100}{\micro rad}$. In fact, for this value of the crossing angle a DA of $\approx$ 7 $\rm{\sigma}$ is still preserved since the beam-beam long range separations (the green dots in Fig.~\ref{fig:sepLR}) are the same as the Ultimate case with a crossing angle of $\SI{200}{\micro rad}$ (the blue dots in Fig.~\ref{fig:sepLR}).

The beam parameters of the "Collide \& Squeeze" scheme, together with the luminosity evolution, are shown in Fig.~\ref{collideandsqueeze} as function of time. When the Ultimate $\bstar = \SI{0.3}{m}$ is reached the normalized emittance is reduced to $\epsilon_n=\SI{1.5}{\micro m}$ due to synchrotron radiation, as shown in Fig.~\ref{collideandsqueeze}. The corresponding beam-beam long range separations are also shown in Figure~\ref{fig:sepLR}.
As expected the long range beam-beam separation at the first encounter is larger w.r.t. the Ultimate case and it is about 20 $\rm{\sigma}$ (the light blue dots).

\begin{figure}[ht!]
    \centering
    \includegraphics*[width=0.65\textwidth]{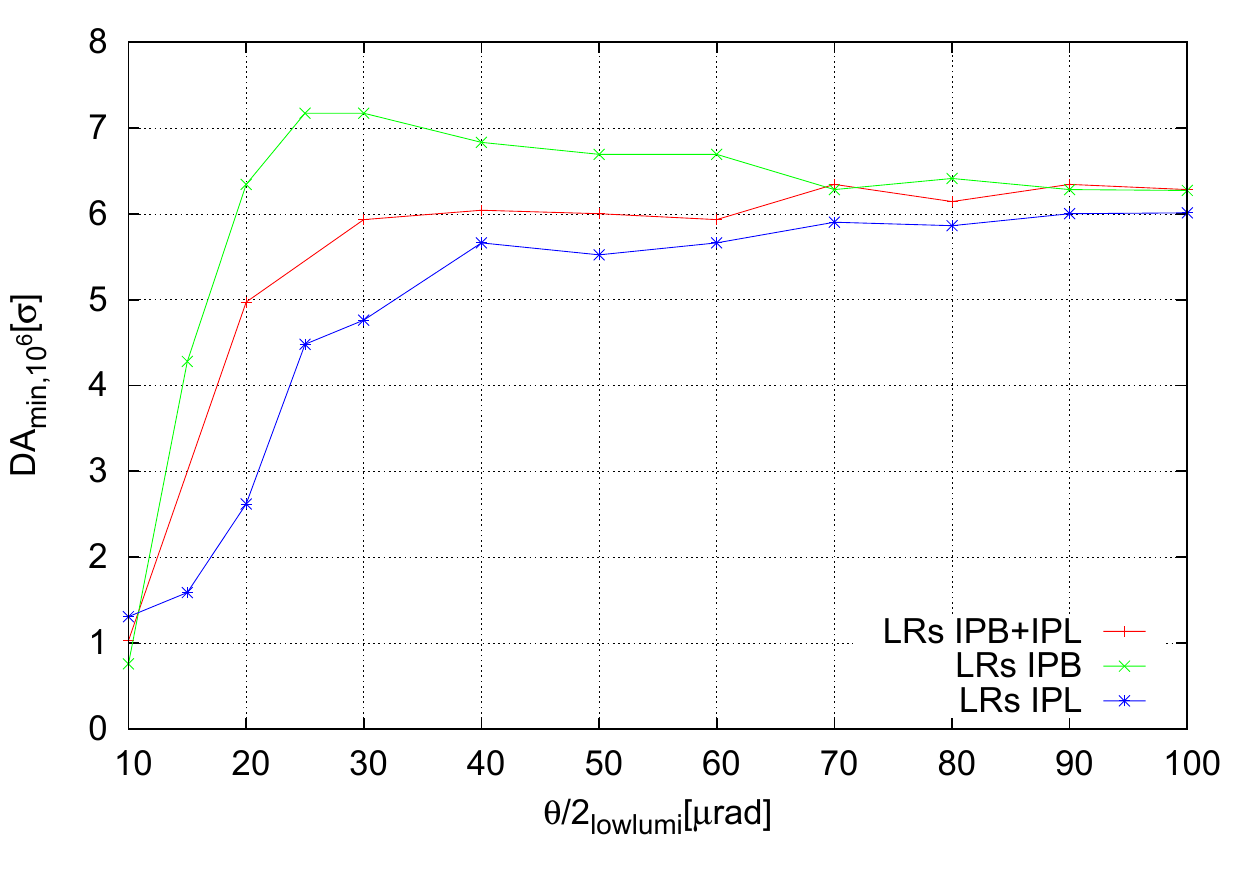}	
    \caption{Minimum Dynamic Aperture, evaluated over $10^6$ turns, as a function of the half crossing angle  at IPB and IPL, including beam-beam long range interactions in IPB only (the green line), in IPL only (the blue line) and in both IPB and IPL (the red line)~\cite{garciaIpac}.}
     \label{fig:DA_IPLB}
\end{figure}

In order to keep the impact of the two low luminosity experiments IPB and IPL  in the shadow of the high luminosity ones, IPA and IPG, a crossing angle of $\SI{180}{\micro rad}$ is required for the $\bstar=\SI{3}{\m}$ optics at these two experiments. As visible in Fig.~\ref{fig:DA_IPLB} the DA does not depend anymore on the crossing angle because it is defined by the dynamics of IPA and IPG. In this configuration the long-range effects of IPB and IPL are negligible and the impact of the long-range beam-beam effects coming from these two experiments can be neglected leaving more margin to push the performances of IPA and IPG. 

\begin{figure}
    \centering
    \includegraphics*[width=0.65\textwidth]{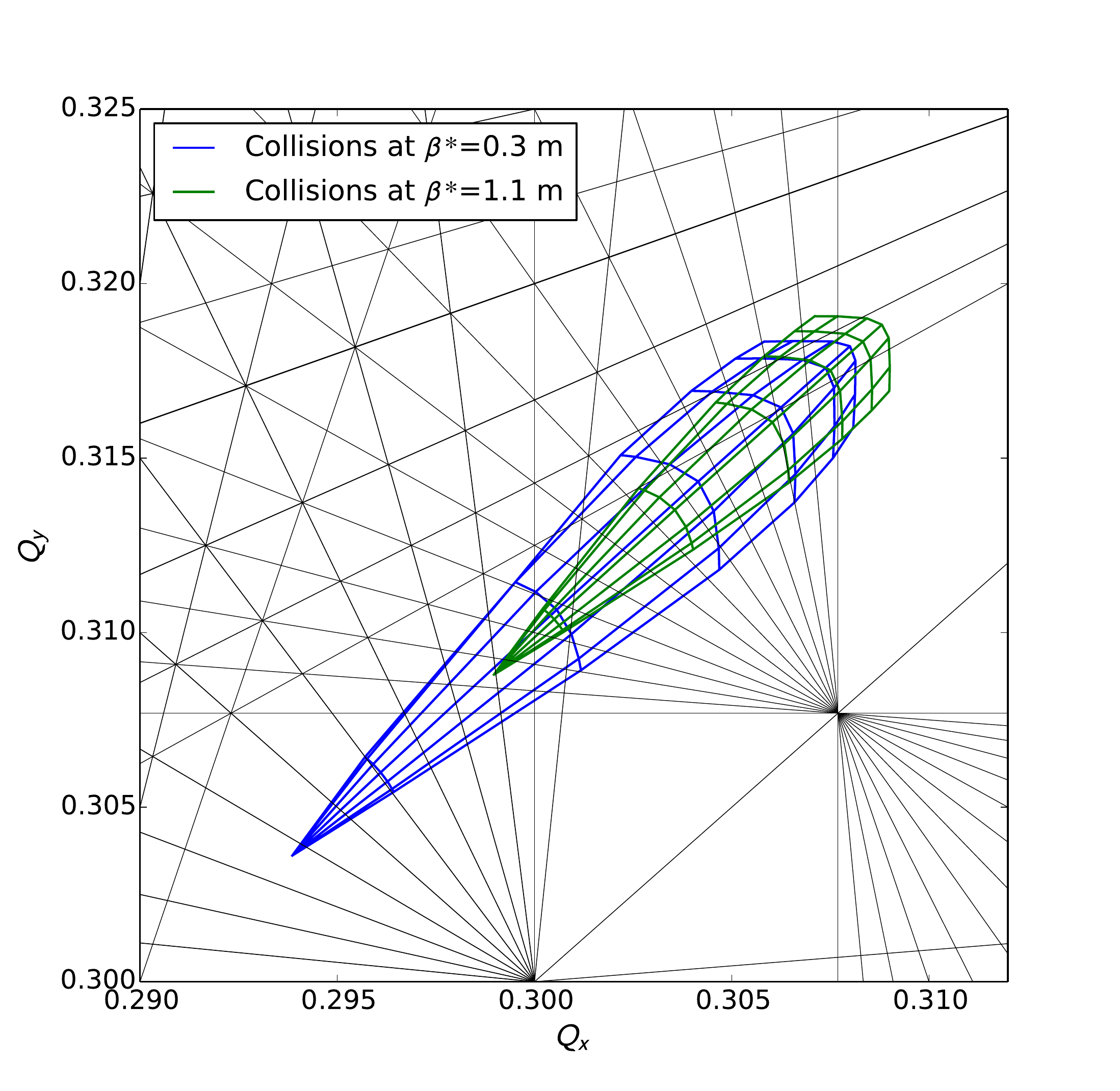}
    \caption{Two-dimensional tune footprints with head-on collisions in IPA and IPG for the Baseline scenario with $\bstar = \SI{1.1}{m}$ (green color) and for the Ultimate scenario $\bstar = \SI{0.3}{m}$ with a reduced normalized emittance of $\epsilon_n=\SI{1.5}{\micro m}$ rad (blue color).}
    \label{fig:footcoll}
\end{figure}

The total  beam-beam tune shift for two head-on collisions in IPA and IPG  will be $\Delta Q_{bbho}=0.011$,  as also visible in Fig.~\ref{fig:footcoll}, where the two dimensional tune footprints with two head-on collisions in IPA and IPG are shown  for the Baseline scenario with $\bstar = \SI{1.1}{m}$ (the green color) and for the Ultimate scenario with a reduced normalized emittance $\epsilon_n=\SI{1.5}{\micro m}$ (the blue color) in order to take into account the effect of the synchrotron radiation shown by the emittance reduction in Fig.~\ref{collideandsqueeze} at the end of the "Collide \& Squeeze".
For the case with the reduced normalized emittance ($\epsilon_n=\SI{1.5}{\micro m}$) the total head-on beam-beam tune shift increases up to approximately 0.016 (the blue color in Fig.~\ref{fig:footcoll}). Since the total beam-beam tune shift $\Delta Q_{bbho}$ is limited to be less than 0.03~\cite{Herr:604005, pieloniHB2012, Buffat:2261037, vikbuffat} the two low luminosity experiments IPB and IPL are required to operate with a transverse offset resulting in a maximum tune shift of $1-2\times 10^{-3}$. However if the total beam-beam tune shift approaches the value of 0.03 and it is not tolerated, mitigations can be applied such as the blow-up of the transverse emittances with controlled noise.

\section{Energy Deposition from Collision Debris \label{sec:rma:IR_radiation}}

 This section evaluates energy deposition from collision debris in the IR  for the nominal design, alternative triplet and low luminosity experimental insertions described in Sec.~\ref{sec:optics}.

\subsection{Nominal Design}
\begin{figure}
  \begin{center}
    \includegraphics*[trim=1cm 8cm 1cm 8cm, width=0.8\textwidth]{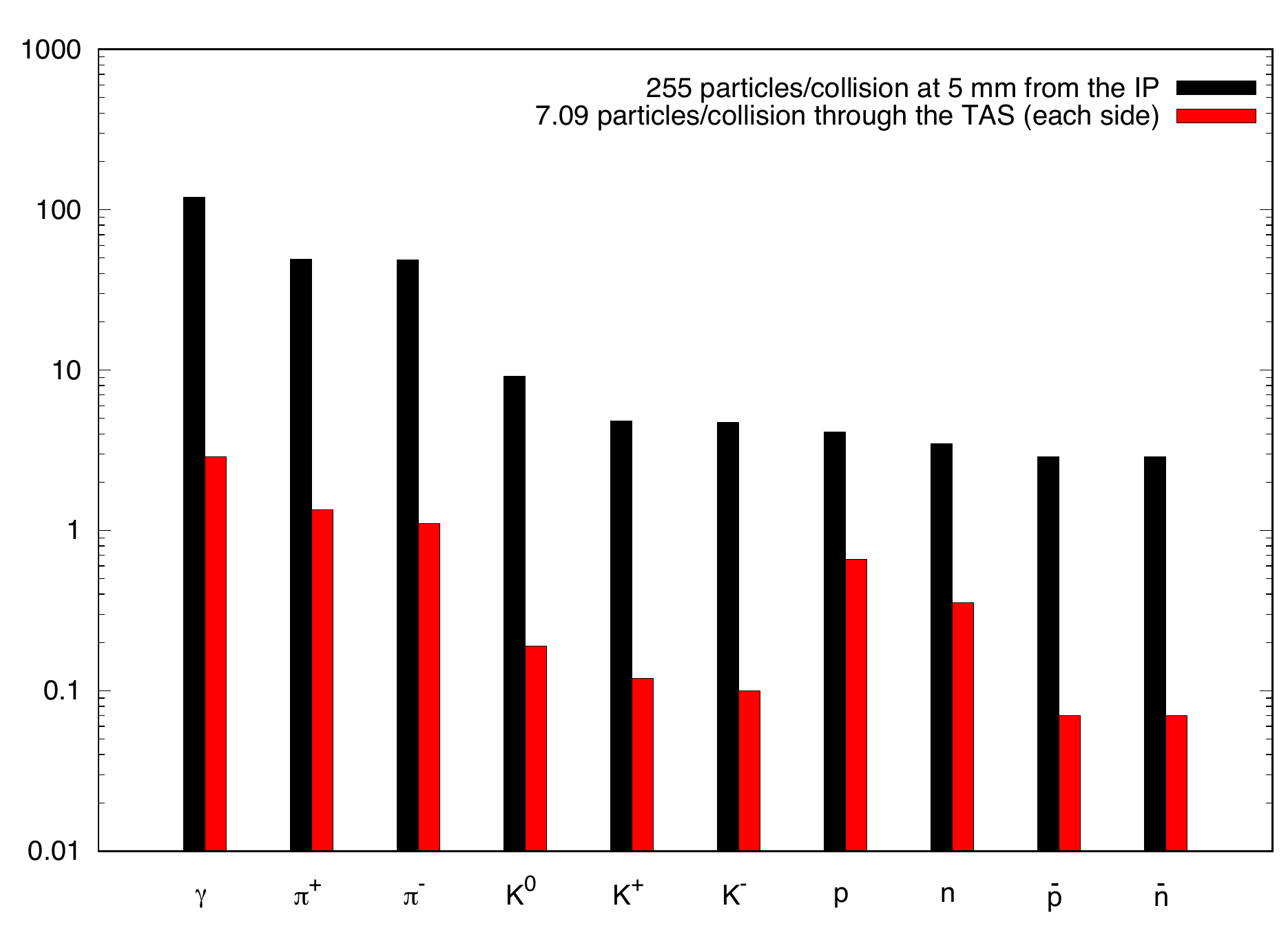}
  \end{center}
  \caption{Collision debris from a single \SI{100}{TeV} c.m. proton-proton inelastic reaction at 5 mm from the interaction point (black) and at the exit of the TAS (red) at \SI{35}{m} from the IP.}
  \label{fig:inf:colldeb}
\end{figure}

Proton-Proton inelastic collisions taking place in the FCC-hh, particularly in the two high luminosity detectors, generate a large number of secondary particles. Moving away from the IP, this multiform population evolves even before interacting with the surrounding materials due to the decay of unstable particles, such as neutral pions which decay into photon pairs. Most of these particles are intercepted by the detector and release their energy within the cavern. However, the most energetic particles, emitted at small angles with respect to the beam direction, travel farther inside the vacuum chamber and reach the accelerator elements, causing a significant impact on the magnets along the EIR, particularly in the final focusing quadrupoles and the separation dipoles. Fig.~\ref{fig:inf:colldeb} shows the particle population close to the interaction point and at the exit of the TAS: the average multiplicity of a single \SI{100}{TeV} c.m. proton-proton inelastic interaction is $\sim$255. At ultimate instantaneous luminosity conditions ($30 \times 10^{34} cm^{-2} s^{-1}$) the power released toward each side of the IP is $\SI{260}{kW}$, that is impacting upon the FCC-hh elements and consequently dissipated in the machine, the nearby equipment (e.g. electronics racks), and the tunnel walls. It is important to study how these particles are lost in order to implement the necessary protections for shielding sensitive parts of the machine and in particular of the magnets. 

\begin{figure}
  \begin{center}
    \includegraphics*[trim=1cm 2cm 1cm 2cm, width=0.9\textwidth]{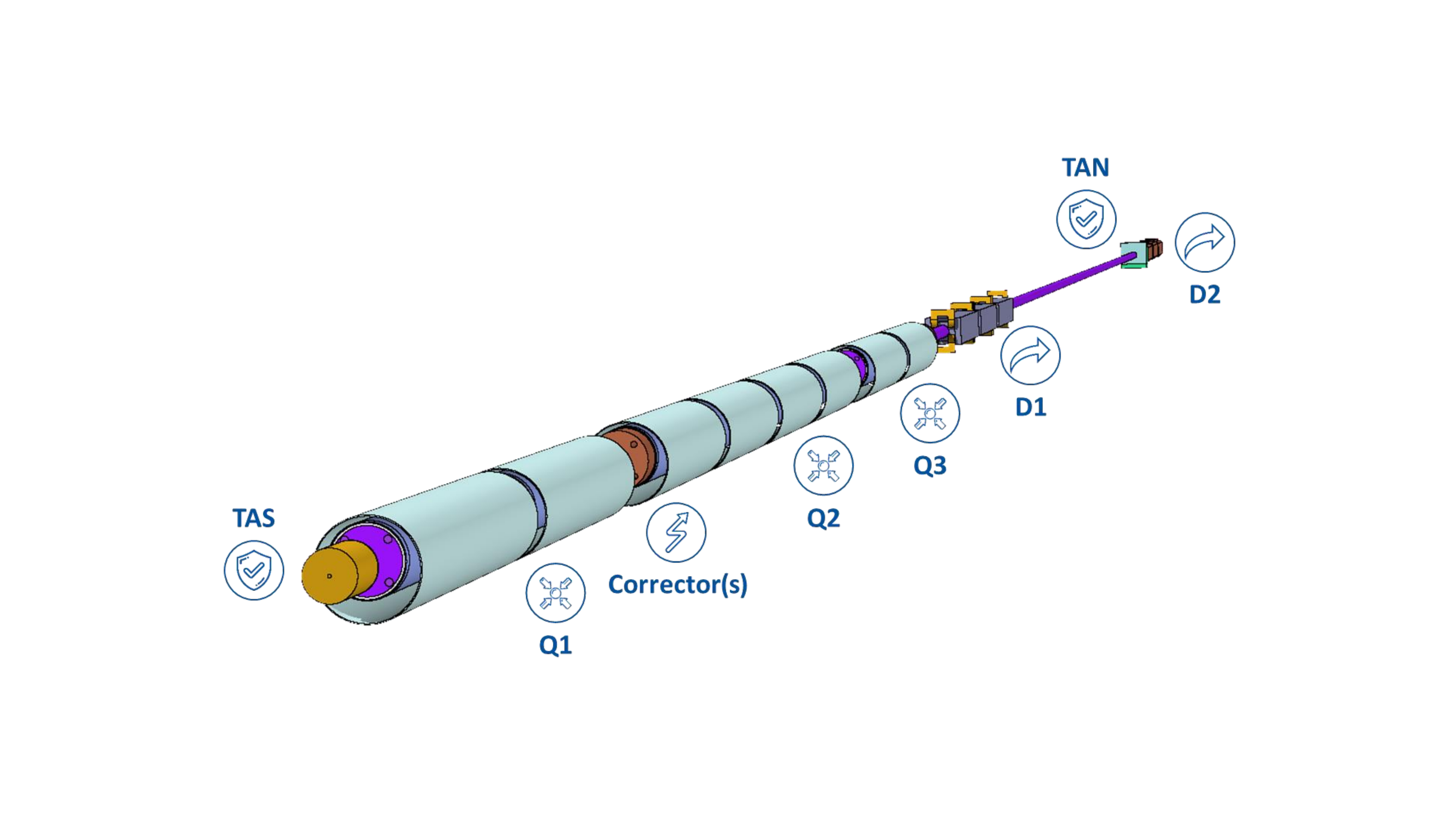}
  \end{center}
  \caption{3D rendering of the FLUKA geometry of the EIR, including $\sim$\SI{700}{m} of beam line. The picture shows the first $\sim$\SI{500}{m} including the TAS, the inner triplet, the TAN, the separation and recombination dipoles D1 and D2.  }
  \label{fig:inf:EIRFLUKAgeo3D}
\end{figure}

In this context, Monte Carlo simulation of particle interaction with matter plays a crucial role relying on a detailed implementation of physics model and an accurate 3D description of the region of interest. The FLUKA code \cite{fluka1, fluka14} is extensively used in this conceptual design study, based on the experience collected in the LHC and HL-LHC design \cite{Apollinari:2284929} as well as the benchmarks already available in literature for these machines \cite{bib:inf:Skordis2015}. Fig.~\ref{fig:inf:EIRFLUKAgeo3D} shows part of the FLUKA model of the EIR, for the latest layout available at the time of the simulation ($\lstar = \SI{40}{m}$) including $\SI{700}{m}$ of accelerator line with the inner triplet, the separation and recombination dipoles (D1 and D2), the TAS, the Target Absorber Neutrals (TAN), and the matching section (Q4-Q7). The following coil apertures (in diameter) were implemented in the model: Q1 (MQXC) $\SI{164}{mm}$, Q2 (MQXD) and Q3 (MQXE) $\SI{210}{mm}$, orbit correctors $\SI{210}{mm}$, Q4 (MQY) $\SI{70}{mm}$, Q5 (MQYL) and Q6 (MQYL) $\SI{60}{mm}$, Q7 (MQM) $\SI{50}{mm}$. The matching section quadrupoles include a rectellipse beam-screen modelled according to optics constraints. To protect the inner quadrupoles coils, a $\SI{35}{mm}$ thick tungsten shielding was implemented in the mechanical design of the triplet magnets and the orbit correctors: the shielding thickness reported in this study is the maximum allowed in order to comply with optics requirements. The first separation dipole, D1 (MBXW), is a single aperture warm dipole, with a pole tip aperture of \SI{170}{mm}. The TAN, made of a \SI{4}{m} long tungsten absorber, includes twin diverging apertures of \SI{52}{mm}. D2 (MBRW) is a twin aperture warm dipole: each module has been modelled with two parallel bores centered at a separation distance varying from the first to the last, in order to reach on the non-IP side the arc value of \SI{250}{mm}. Proton-proton collisions at 100 TeV c.m. with a vertical half crossing angle of \SI{100}{$\mu$rad} have been simulated and the particle shower was tracked all along the accelerator elements \cite{bib:abe:Cerutti:FCC2018, bib:inf:Infantino:EURCIR2018}. The study of the matching section requires an extreme computational effort to achieve a statistically meaningful outcome and therefore it is planned to be finalized at a later stage. Results concerning the triplet-D2 area are presented in the following. 

The total power deposited in the cold magnets (Tab.~\ref{tab:inf:TotalPower}) is shared between the cold mass and the massive tungsten shielding, with a $\sim$15-85 ratio. In particular, the Q1B (MQXC.B1RA) turns out to be the most impacted element of the triplet with a total power of about \SI{2}{kW} in the cold mass and \SI{13}{kW} in the shielding.

\begin{table}
  \begin{center}
  \caption{\label{tab:inf:TotalPower} Total power distribution in the EIR elements.}
  \begin{tabular}{ |l|S|S|S| } 
	\hline
	\textbf{Element} & \multicolumn{3}{c|}{\textbf{Total Power [\si{kW}]}} \\
	  & \bf Cold Shielding & \bf Cold Mass & \bf Warm Mass \\ \hline
	TAS       &  &  & 26.5 \\ \hline
	Q1a    & 4.6 & 0.78 &  \\ \hline
	Q1b    & 13 & 1.92 &  \\ \hline
	C1     & 0.06 & 0.06 &  \\ \hline
    Q2a    & 1.53 & 0.32 &  \\ \hline
	Q2b    & 0.7 & 0.09 &  \\ \hline
	Q2c    & 4.6 & 0.63 &  \\ \hline
	Q2d    & 5.93 & 0.81 &  \\ \hline
    C2     & 0.51 & 0.05 &  \\ \hline
	Q3a    & 6.02 & 0.77 &  \\ \hline
	Q3b    & 7.8 & 0.95 &  \\ \hline
	C3     & 0.94 & 0.17 &  \\ \hline
    D1a    &  &  & 4.99  \\ \hline
	D1b    &  &  & 3.57 \\ \hline
	D1c    &  &  & 3.57\\ \hline
	D1d    &  &  & 3.96 \\ \hline
	TAN      &  &  & 107 \\ \hline
	D2a    &  &  & 0.07 \\ \hline
	D2b    &  &  & 0.01 \\ \hline
	D2c    &  &  & 0.003 \\ \hline
	D2d    &  &  & 0.002 \\ \hline
  \end{tabular}
  \end{center}
\end{table}

\begin{figure}
  \begin{center}
    \includegraphics*[trim=6cm 6cm 6cm 6cm,clip, width=1.0\textwidth]{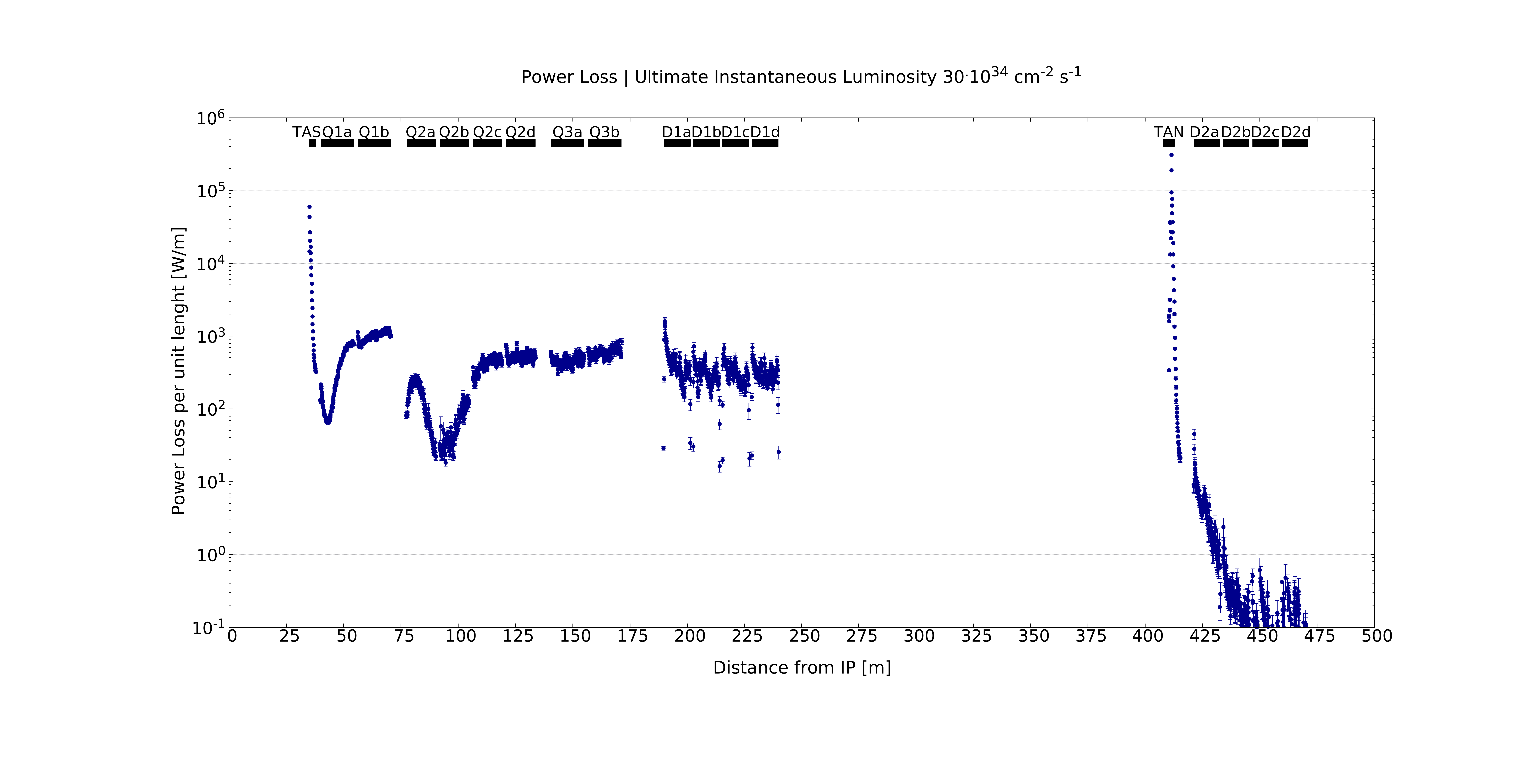}
  \end{center}
  \caption{Absorbed power profile in the elements of the TAS-D2 region at the ultimate instantaneous luminosity of {$30 \times 10^{34} cm^{-2} s^{-1}$}.}
  \label{fig:inf:PowerLoss_TAS_D2}
\end{figure}

Fig.~\ref{fig:inf:PowerLoss_TAS_D2} shows the absorbed power profile (in \si{W/m}) along the TAS-D2 region, where triplet cold mass and shielding contributions are added up. In the Q1B cold mass a linear power loss of $\sim$\SI{150}{W/m} is reached. On the other hand, a preliminary evaluation, based on the cooling capabilities of the beam-screen of the \SI{16}{T} main dipoles, indicates that four helium tubes of \SI{15}{mm} diameter, placed in a \SI{45}{\degree} pattern with respect to the mid planes, would allow to dissipate the \SI{13}{kW} ($\sim$\SI{0.8}{kW/m}) deposited in the Q1B shielding \cite{bib:inf:Infantino:EURCIR2018}. The possibility of a shielding mechanical design embedding such a cooling circuit is currently under investigation.  

\begin{figure}
  \begin{center}
    \includegraphics*[trim=4cm 4cm 4cm 4cm,clip, width=1.0\textwidth]{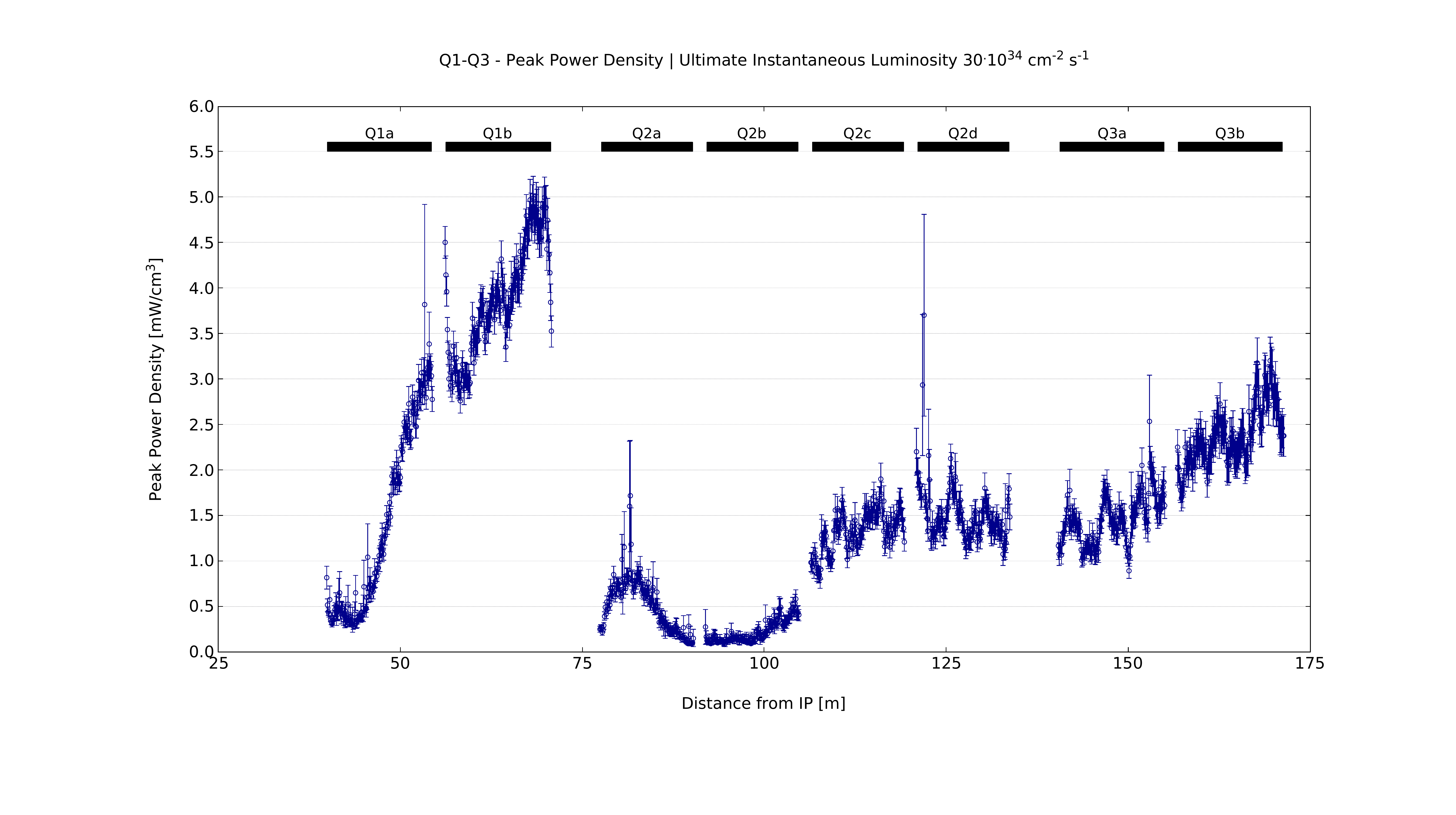}
  \end{center}
  \caption{Peak power density profile in the triplet quadrupole coils at the ultimate instantaneous luminosity of {$30 \times 10^{34} cm^{-2} s^{-1}$}. Values are averaged over the radial cable thickness, with an azimuthal resolution of \SI{2}{\degree}. Vertical bars indicate the statistical error.}
  \label{fig:inf:PeakPowerTriplet}
\end{figure}

Figure~\ref{fig:inf:PeakPowerTriplet} shows the peak power density profile in the triplet quadrupole coils, reaching a maximum of \SI{5}{mW/cm^{3}} at the end of the Q1B that matches with no margin a first conservative estimate of the quench limit. Studies to better determine the latter are currently ongoing \cite{bib:inf:Shoerling}. Recently, \cite{bib:inf:Bottura} showed that the \SI{11}{T} Nb$_{3}$Sn HL-LHC dipoles are expected to withstand steady state loads ten times higher.

\begin{figure}
  \begin{center}
    \includegraphics*[trim=4cm 4cm 4cm 3cm,clip, width=1.0\textwidth]{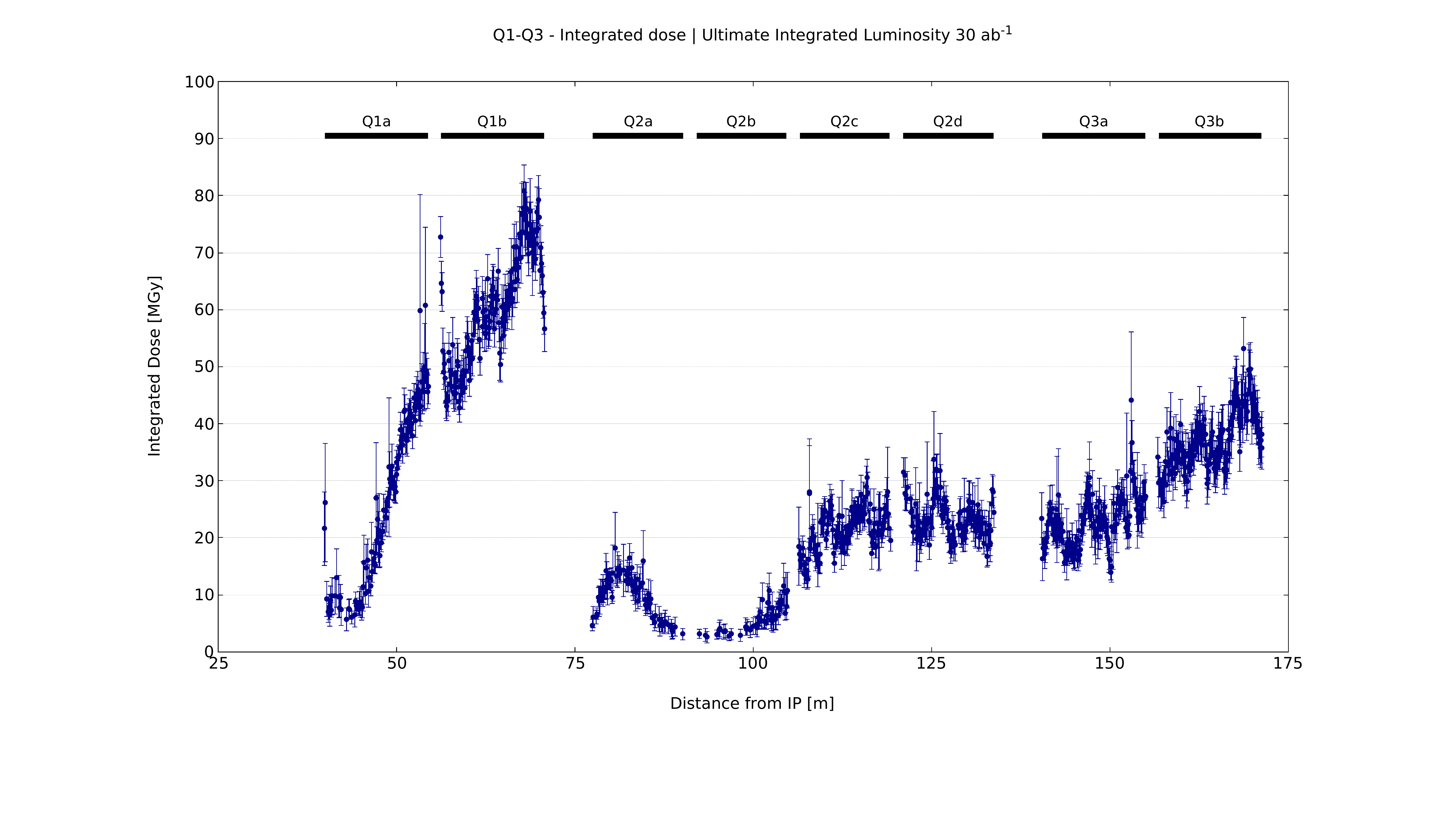}
  \end{center}
  \caption{Peak dose profile in the triplet quadrupole coils for the ultimate integrated luminosity target (\SI{30}{ab^{-1}}). Values refer to a radial and azimuthal resolution of \SI{3}{mm} and \SI{2}{\degree}, respectively. Vertical bars indicate the statistical error.}
  \label{fig:inf:PeakDoseTriplet}
\end{figure}

To estimate the integrated luminosity reach with respect to the insulator lifetime, the absorbed dose in the magnet coils was calculated. Figure~\ref{fig:inf:PeakDoseTriplet} shows the peak dose profile for the ultimate integrated luminosity goal (\SI{30}{ab^{-1}}). Assuming an operational limit of \SI{30}{MGy} for conventional radiation resistant insulator materials, the most critical element (Q1B) exceeds it by a factor \SI{2.5}{}. As previously mentioned, the model is already containing the maximum shielding thickness allowed by beam aperture requirements (\SI{35}{mm}). Nevertheless, a Q1 split featuring a larger Q1B aperture at the price of a lower gradient, to be compensated in the Q1A, would allow a respective increase of the shielding thickness reducing the maximum dose. Moreover, crossing angle polarity and plane alternation are known to significantly reduce the maximum dose, by  more equally distributing the radiation load in the coils~\cite{bib:abe:Cerutti:FCC2018}. Finally, the dose limit might be increased by using more radiation hard insulator, e.g. epoxy/cyanate-ester blends \cite{bib:inf:Shoerling}. Alternatively, the replacement of the inner triplet once during the FCC-hh era might be considered.

\begin{figure}
  \begin{center}
    \includegraphics*[trim=4cm 4cm 4cm 4cm,clip, width=1.0\textwidth]{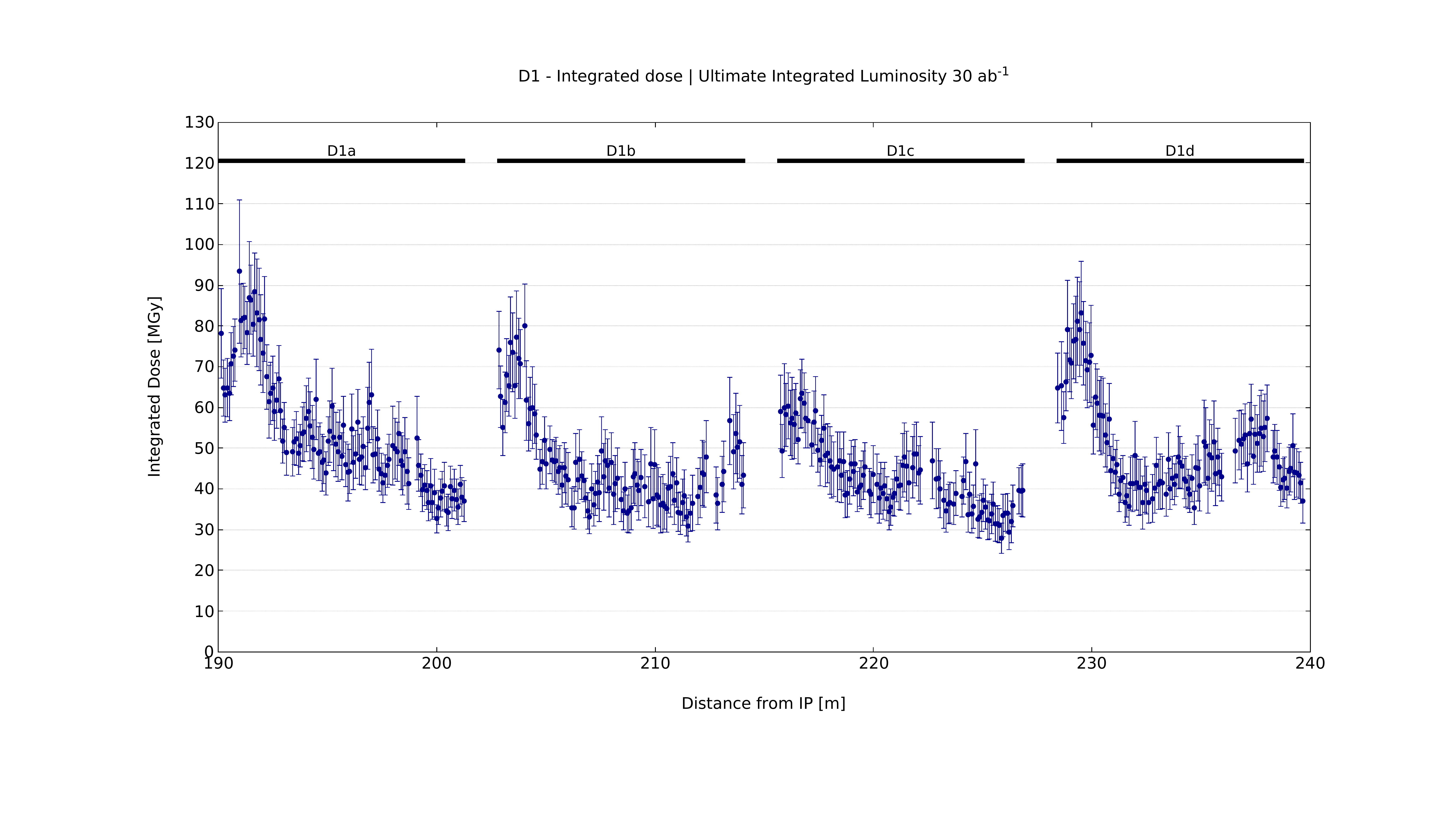}
  \end{center}
  \caption{Peak dose profile in the D1 warm separation dipole coils for the ultimate integrated luminosity target (\SI{30}{ab^{-1}}). Values are averaged over a \SI{3}{mm}x\SI{3}{mm} transverse area. Vertical bars indicate the statistical error.}
  \label{fig:inf:PeakDoseD1}
\end{figure}

With regard to the warm dipoles, the peak dose profile in D1 is reported in Fig.~\ref{fig:inf:PeakDoseD1}. Both D1 and D2 have been modelled without embedding any shielding but with a design moving the return coils far from the beam pipe, to reduce their exposure to the collision debris. This solution allows already for an important gain. However, the picture is rather similar to what was shown above for the triplet quadrupoles, with a peak dose in the first module (D1A) evaluated to be three times the \SI{30}{MGy} limit. In this case, a front mask as well as internal shielding can offer a substantial benefit. With regard to D2, the highest value expected at the D2A IP end is below \SI{10}{MGy}, while peak doses ten times lower have been found in the D2C and D2D. Thanks to the protection provided by the TAN, the D2 presents no major concern for the coil insulator lifetime, even for the ultimate integrated luminosity target. 

\subsection{Energy Deposition in Alternative Triplet}

\begin{figure}
   \centering
	\includegraphics[width=1.0\textwidth]{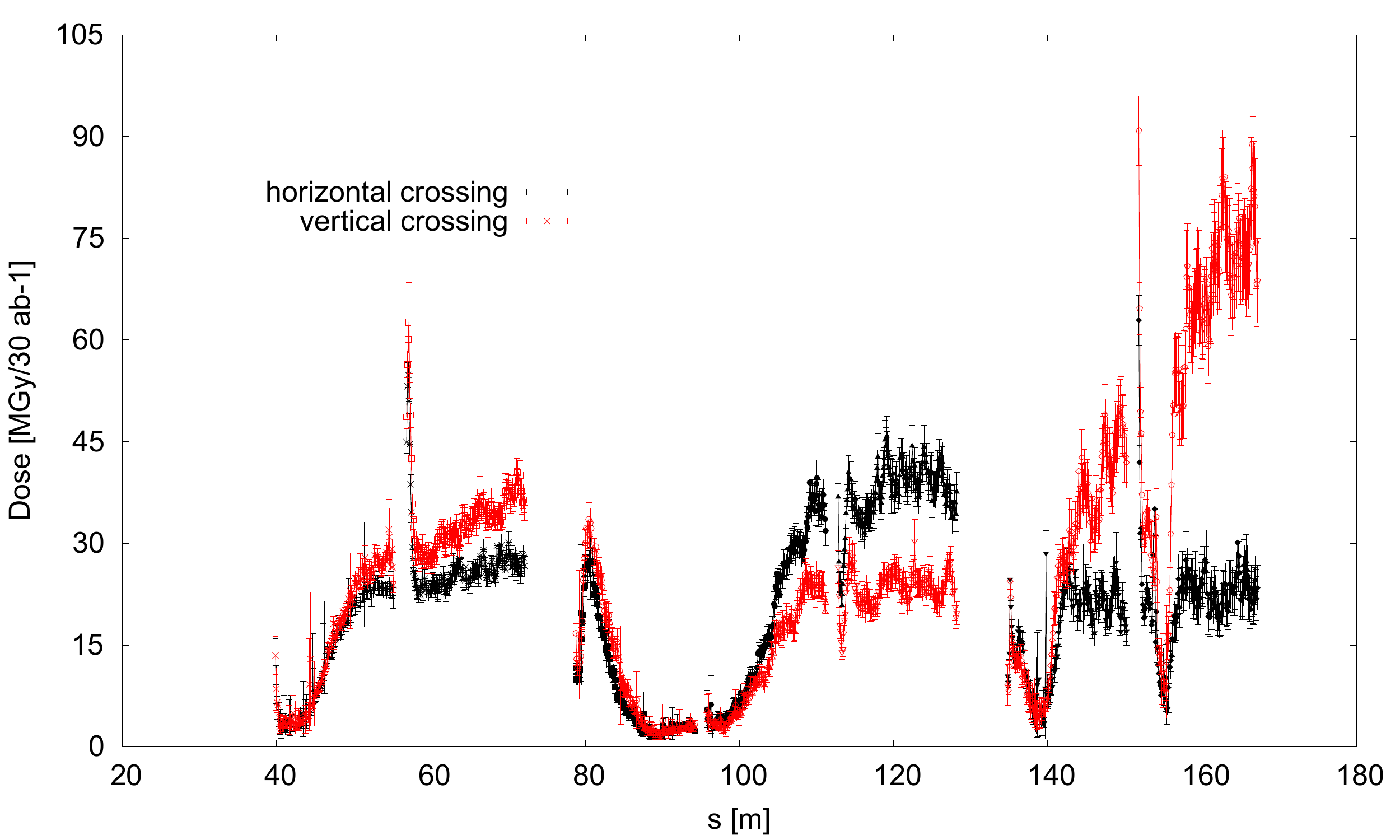}
   \caption{Peak dose profile for alternative triplet (round optics) for the ultimate integrated luminosity target (\SI{30}{ab^{-1}}).}
   \label{fig:abe:energy_round}
   \centering
	\includegraphics[width=1.0\textwidth]{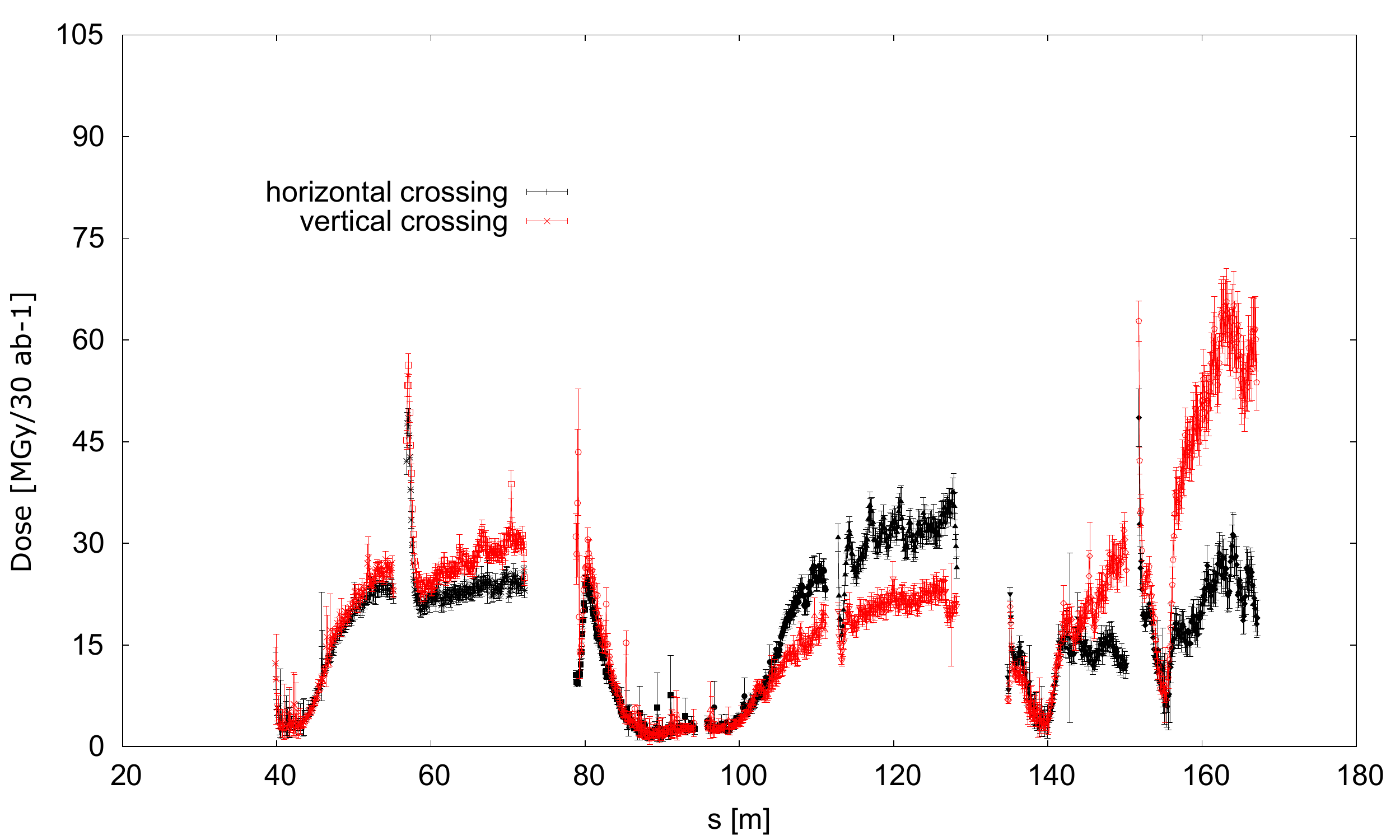}
   \caption{Peak dose profile for the alternative triplet (flat optics) for the ultimate integrated luminosity target (\SI{30}{ab^{-1}}).}
   \label{fig:abe:energy_flat}
\end{figure}

Figure~\ref{fig:abe:energy_round} shows the peak dose in the triplet magnets along the longitudinal axis.  
The maximum dose is found at Q3, with a maximum of 30~MGy/10 ab$^{-1}$ (excluding the peak at the beginning, that can be reduced by reducing the missing gap for the interconnects).
 This means 65~MGy for the entire life of the magnet, assumed to resist, at least, an integrated luminosity of 18.5~ab$^{-1}$.
The dose can also be reduced when using the alternate crossing scheme,
in a similar ways as the baseline triplet~\cite{bib:rma:PhysRevAccelBeams.20.081005,bib:abe:Cerutti:FCC2018}.
 On the other hand, the peak dose profile for the alternative flat beam option is shown in Fig.~\ref{fig:abe:energy_flat}. 
  The peak dose is reduced to 
 from 55 to 42~MGy for 18.5~ab$^{-1}$, due to the lower crossing angle allowed by the flat beam optics.

\subsection{Energy Deposition in the Low Luminosity EIR}
\label{subsec::LLeir_enDep}
The energy deposition in the low luminosity EIR has been assessed with FLUKA simulations, for both vertical and horizontal crossing.

\begin{figure}
   \centering
	\includegraphics[width=0.7\textwidth]{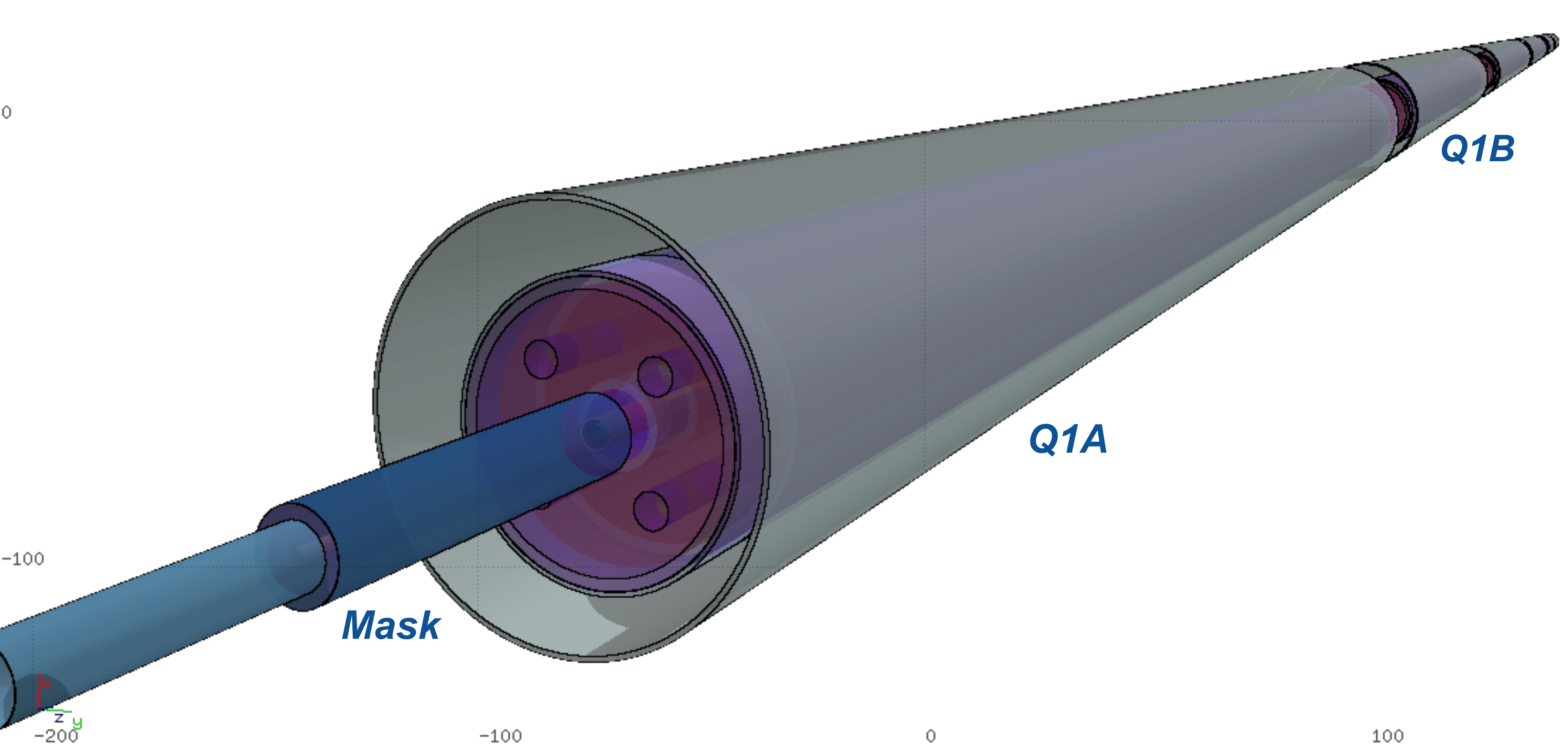}
   \caption{Low luminosity triplet geometry as modelled in FLUKA.}
   \label{fig:LLtriplet}
\end{figure}

For this purpose, the insertion region has been modelled, as shown in Fig.~\ref{fig:LLtriplet}. The quadrupole design is very similar to the one for the main EIR described in Section ~\ref{sec:Final_focus_triplet}, but the model has been scaled down  to cope with the smaller coil radius of 32~mm. 
In order to better protect the superconducting magnets, the 10~mm thick tungsten shielding is prolonged in the interconnect cold bore, with tentative gaps of 70~cm. In addition to this, a mask has been put in front of the Q1A, to shield its entrance. The mask is clearly visible in Fig.~\ref{fig:LLtriplet} and  it is modelled as a 76~cm long tungsten (INERMET180) cylinder with an external radius of 81~mm and a free radial aperture of 13.26~mm.  

The power impacting on the quadrupoles and the orbit correctors C1 and C2 is reported in Table~\ref{tab:LLtotPow} for the shielding and the cold mass separately, assuming an instantaneous luminosity of 5$\times$10$^{33}$~cm$^{-2}$~s$^{-1}$. Even if the mask in front of Q1A intercepts about 280~W, this magnet remains the most exposed and the total power on the cold mass is about 150~W for both crossing schemes.

\begin{table}
  \begin{center}
  \caption{Total power in the magnets of the inner triplet for vertical and horizontal crossing, assuming an instantaneous luminosity of 5$\times$10$^{33}$~cm$^{-2}$~s$^{-1}$. The contribution to the shielding and the cold mass are quoted separately. \label{tab:LLtotPow}}
  \begin{tabular}{ |l|c|c|c|c|c|c|} 
	\hline
	\bf Magnet & \multicolumn{3}{|c|}{\bf Vertical Crossing [\si{W}] } & \multicolumn{3}{|c|}{\bf Horizontal Crossing [\si{W}] } \\
	\hline
	& \bf Total & \bf Shielding & \bf Cold & \bf Total & \bf Shielding & \bf Cold \\ 
	& &  & \bf Mass  &  &  & \bf Mass \\ \hline
	Q1A & 249 & 101 & 147.9 &251 & 102 & 149\\
	\hline
	Q1B & 268 & 183 & 85 & 269 & 184 & 85\\
	\hline
	C1 & 27 & 19 & 8 & 28 & 19 & 8 \\
	\hline
	Q2A & 118 & 82 & 36 & 119 & 83 & 36\\
	\hline
	Q2B & 204 & 147 & 57 & 191 & 137 & 54 \\
	\hline
	Q3A & 111  & 77 & 34 & 113 & 80 & 33\\
	\hline
	Q3B & 113 & 81 & 31 & 132 & 95 & 37\\
	\hline
	C2 & 15 & 11 & 4 & 18 & 13 & 5\\
	\hline
	\bf Total & \bf 1105 & \bf 701 & \bf 404 & \bf 1121&\bf  714 & \bf 407\\
	\hline
  \end{tabular}
  \end{center}
\end{table}

Less than 40\% of the total power generated in the collision is deposited in the inner triplet. The remaining 63\%, which corresponds to about 2.7~kW for an instantaneous luminosity of 5$\times$10$^{33}$~cm$^{-2}$~s$^{-1}$, escapes downstream on both sides of the IP and it will be deposited elsewhere in the accelerator.  
\begin{figure}
 \centering
  \includegraphics[trim={2.4cm 2cm 4.1cm 1cm},clip,width=0.49\textwidth]{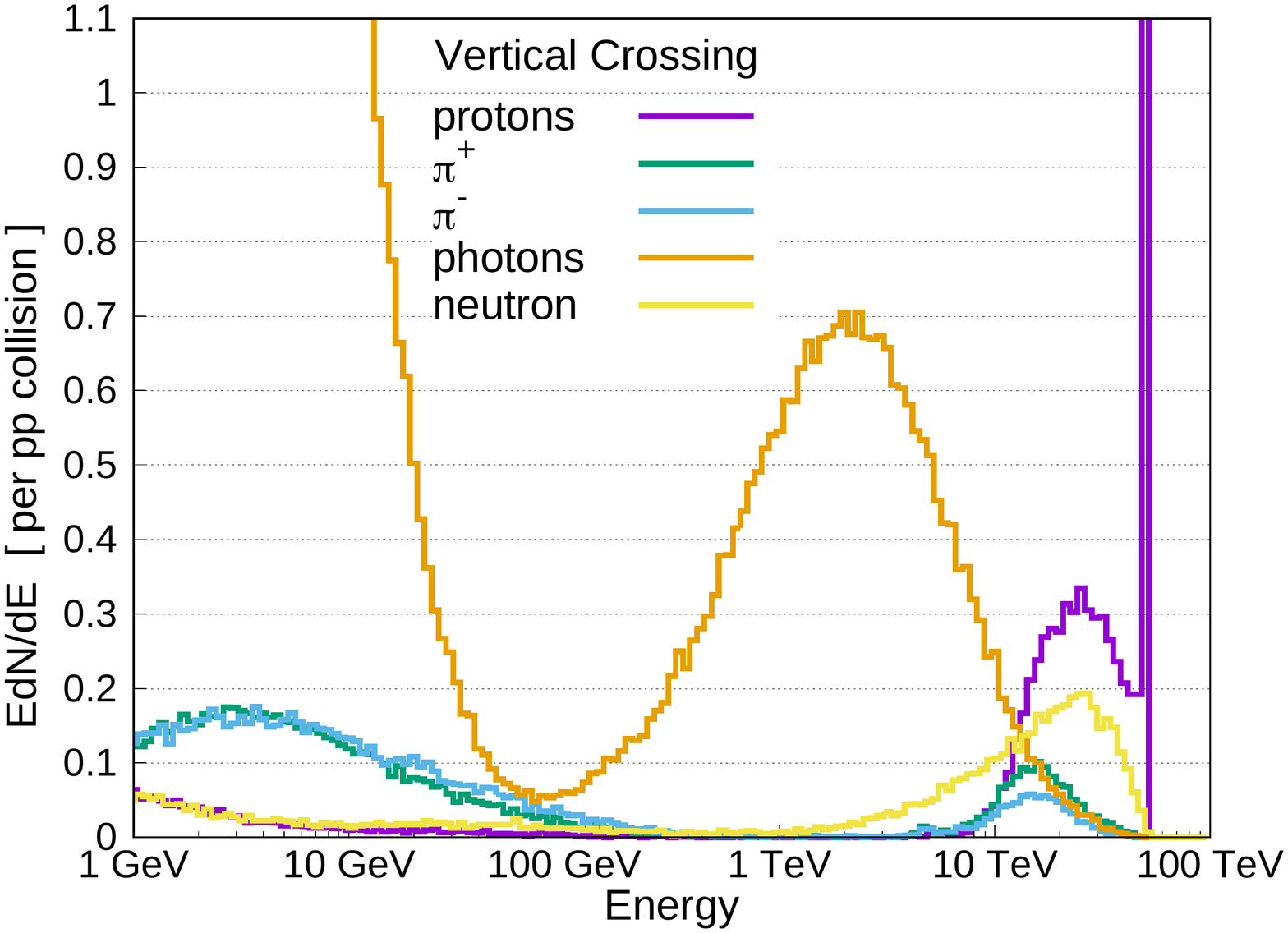}
 \includegraphics[trim={2.4cm 2cm 4.1cm 1cm},clip,width=0.49\textwidth]{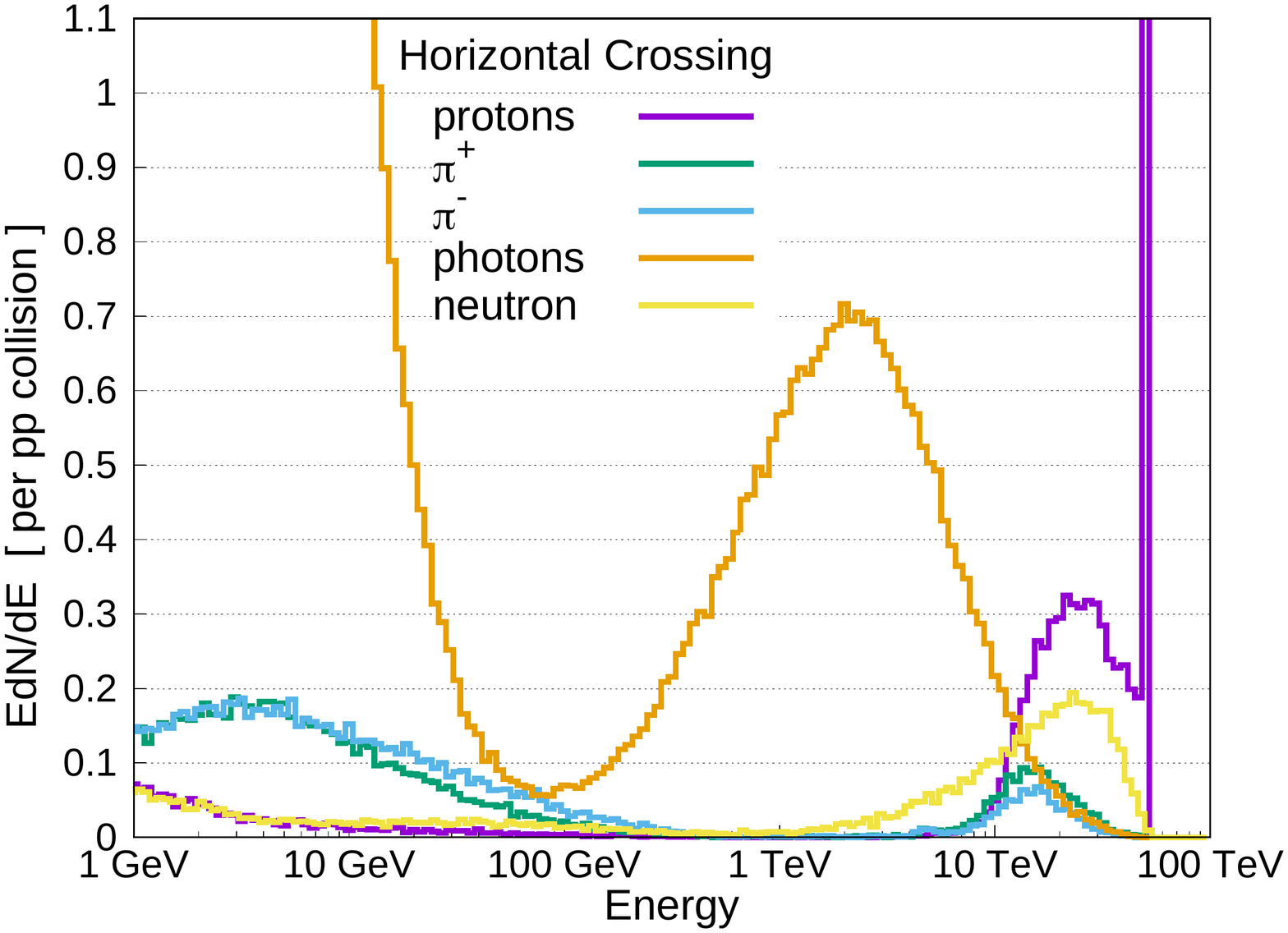}
\caption{Spectra of collision debris particles in the vacuum chamber at the exit of Q3B for vertical crossing (left) and horizontal crossing (right) schemes. The distributions are normalised to one proton-proton collision. }
\label{fig::LLspectra}
\end{figure}

Figure~\ref{fig::LLspectra} shows the spectra of particles at the exit of Q3B for vertical crossing on the left and horizontal crossing on the right. The peak at 50~TeV is due to protons produced in single diffractive events. These protons travel much further in the accelerator and are expected to impact in the dispersion suppression region. The charged pions and the protons of few tens of TeV will instead be lost on D1, on the TAN or in the matching section. Many high energy photons and neutrons escape as well downstream Q3 and they will be captured by the TAN or even at longer distances from the IP, depending on on their angle. 
In order to precisely assess the impact of these particles in the accelerator components, new calculations are foreseen, which will extend the simulation to the matching section, as it has been done for the high luminosity EIR.

\begin{figure}
   \centering
	\includegraphics[angle=-90,width=0.8\textwidth]{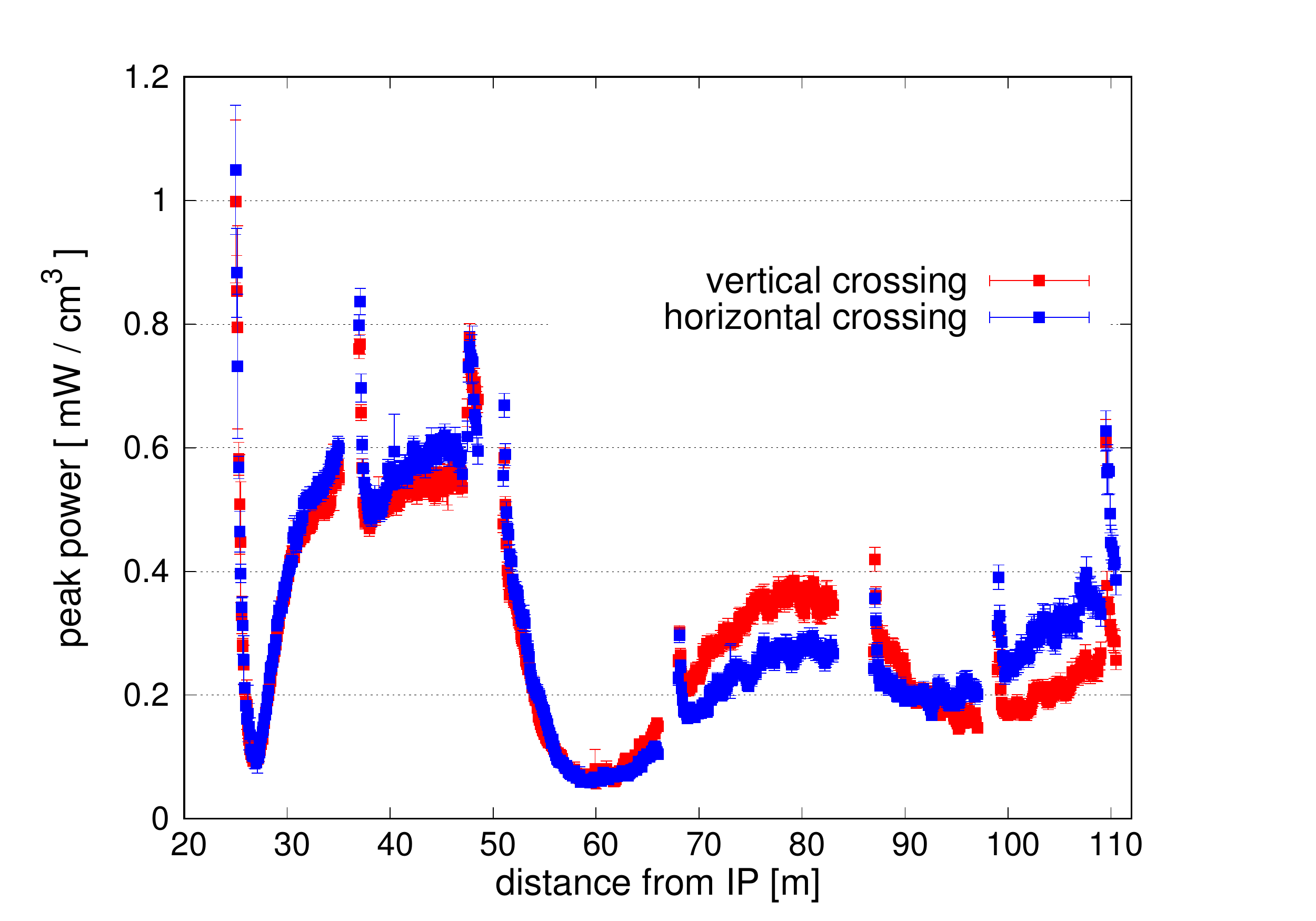}
   \caption{Peak power density in the innermost cable of the inner triplet magnets as a function of the distance from the IP, for an instantaneous luminosity of 5$\times$10$^{33}$~cm$^{-2}$~s$^{-1}$. The resolution along the z-axis is 10~cm and the resolution on the azimuthal direction is 2~deg. A radial average is considered along the cable thickness (18.6~mm for quadrupoles and 5~mm for correctors) and the maximum over the azimuthal direction is considered.}
   \label{fig::LLpeakpow}
\end{figure}

The peak power density in the magnet inner cable is presented in Figure~\ref{fig::LLpeakpow} as a function of the distance from the interaction point for both 
the vertical and the horizontal crossing. 
For both cases, the values are safely below the quench limit for the superconducting coils. The maximum is indeed 1~mWcm$^{-3}$ and it is reached at the entrance of Q1A. Without the presence of the mask, this value would be more than 30 times higher and would significantly exceed quench limits. The presence of a peak at the entrance of each magnet is due to the shielding gaps in the interconnects. 

For what concerns the dose, the maximum is reached as well at the entrance of Q1A for both schemes and, for an integrated luminosity of 500~fb$^{-1}$, it remains below 30~MGy, which is the limit presently assumed for the damage of insulators and organic materials. 

\begin{figure}
 \centering
  \includegraphics[width=0.8\textwidth]{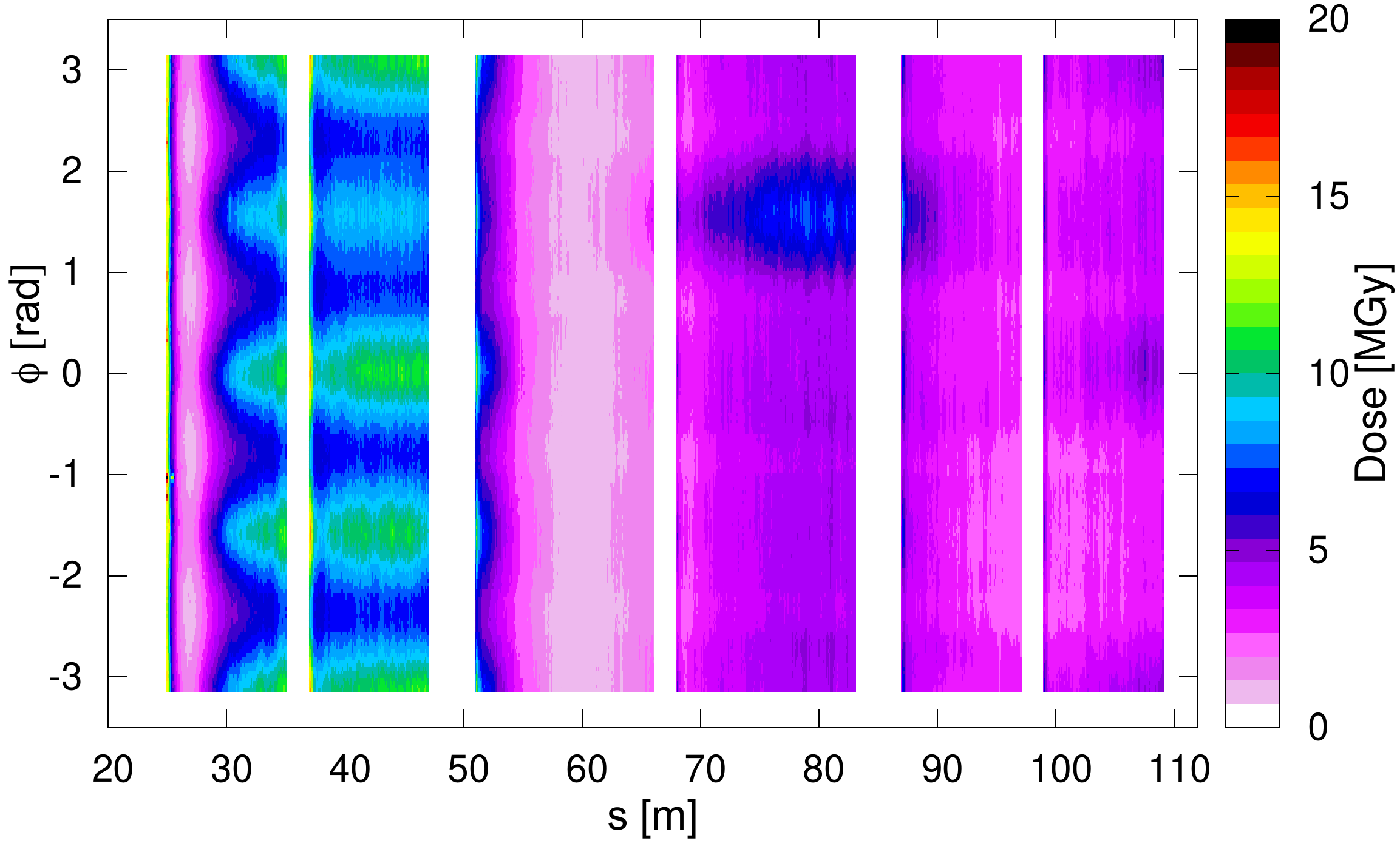}
  \includegraphics[width=0.8\textwidth]{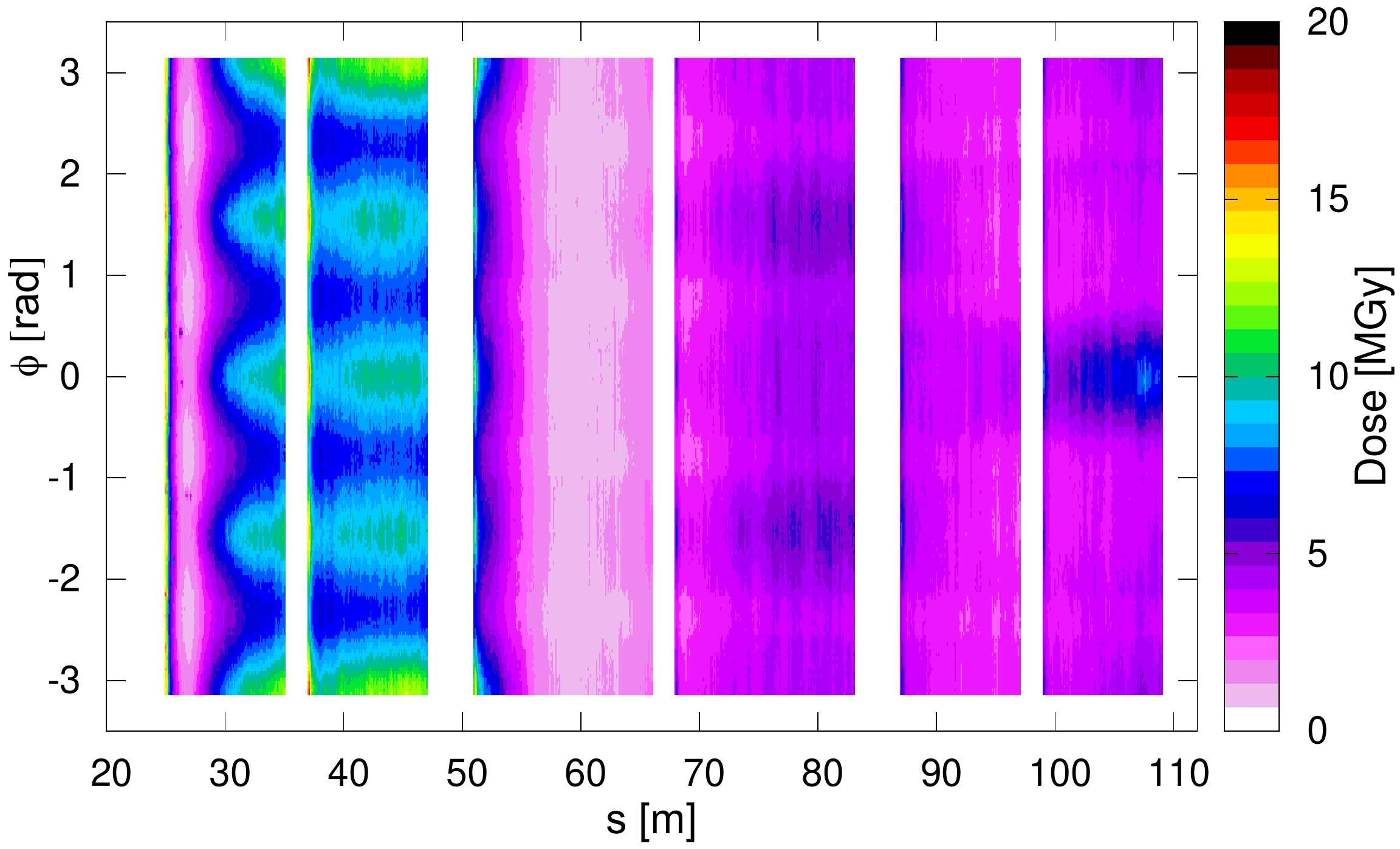}
  \caption{Dose distribution averaged over the innermost \SI{3}{mm} of the magnet coils, as a function of z and of the azimuthal angle for vertical (top) and horizontal (bottom) crossing, normalised to an integrated luminosity of \SI{500}{fb^{-1}}. }
  \label{fig::LLdose}
\end{figure}

The cause of the shape difference between the red and the blue curves in Fig.~\ref{fig::LLpeakpow} is the different crossing scheme of the collisions. The crossing choice influences as well the azimuthal position of the peaks. This can be seen from Fig.~\ref{fig::LLdose}, which shows the dose distribution in the innermost strands of the magnet coils as a function of the distance from the IP and of the azimuthal angle $\Phi$, for vertical crossing on top and horizontal crossing on bottom. The observed asymmetric pattern is due to the combined effects of the crossing angle and plane and of the focusing/defocusing action of the quadrupoles, which sweep low energy secondary particles into the magnets, preferentially along the vertical and horizontal planes.

The triplet polarity is DFD in the horizontal plane for positive particles, which are more abundant in p-p collisions. In Q1 positive particles are therefore deflected in the horizontal plane and, in the case of vertical crossing, hit the coil symmetrically at 0 and $\pi$. On the vertical plane, negative particles impact at $-\frac{\pi}{2}$\footnote{The position of the peak is -$\frac{\pi}{2}$, because the crossing angle is negative. In the case of positive crossing, the peak would have been at $\frac{\pi}{2}$.}, because of the crossing angle offset. In the case of horizontal crossing, positive particles impact mainly at $\pi$, because of the initial crossing angle. For this reason both peak power density and peak dose values in Q1A and Q1B are higher for this case. The lower peaks at $\pm\frac{\pi}{2}$ are due to negative particles symmetrically deflected in the vertical plane. 
Due to the polarity change in Q2, positive particles are deflected on the vertical plane and impact at $\frac{\pi}{2}$ for the vertical case, because of the initial crossing angle.  In the case of h-crossing two symmetric and lower peaks are present at $\pm\frac{\pi}{2}$. 
Finally in Q3, where the polarity is inverted again, positive particles are deflected in the horizontal plane and are collected at 0 for h-crossing, while 
for v-crossing they hit symmetrically at 0 and at $\pi$. Higher power density and dose values are therefore observed in Q3B for the horizontal crossing.

\section{Photon Background From Synchrotron Radiation \label{sec:synchrotron_radiation}}
The amount of power radiated by Synchrotron Radiation (SR) strongly depends on the relativistic $\gamma$ Lorentz factor of the moving particle, and thus on its energy to mass ratio: $P\propto \gamma^4 \rightarrow P \propto (E/m)^{4}$ \cite{bib:obg:cernradwalker}.\\
Due to their mass,  SR emitted by protons is usually a very small source of backgrounds in the experiments, even in very high energy proton beams such as LHC.
However, in the case of FCC, in which beams are planned to reach 50 TeV of energy, also this possible source of background should be carefully evaluated.\par
The critical energy of the emitted SR scales with Lorentz factor $\gamma$ and bending radius $\rho$ according to $\gamma^3/\rho$.
While the increase of FCC-hh center of mass energy with respect to the LHC is about a factor 7, the critical energy of emitted photons increases by a factor 100, shifting the energy spectrum from hard ultraviolet for LHC (which is easily absorbed) into soft X-rays of several keV for FCC. Since the Beryllium of the inner beam pipe can start to become transparent at these energies, some of these photons could traverse the beam pipe and may potentially constitute a background in the detectors.\par 

To address this study, a dedicated software tool has been developed, validated and used. MDISim~\cite{bib:obg:MDISim} is a toolkit that combines existing standard tools MAD-X~\cite{bib:obg:madx}, ROOT~\cite{bib:obg:Root} and GEANT4~\cite{bib:obg:Geant4}. It reads the MAD-X  optics files, and it uses its Twiss (and optionally Survey) output file to export the geometry and the magnetic field information in a format which can be directly imported in GEANT4 to perform particle tracking, including the generation of secondaries and detailed modelling of the relevant absorption processes.\par

\begin{figure}[hb!]
  \begin{center}
  \includegraphics*[width=\textwidth]{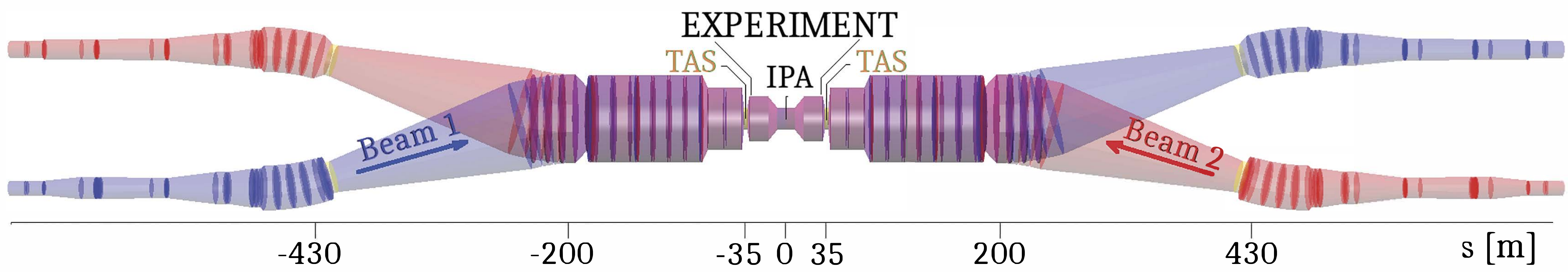}
  \caption{\label{fig:obg:fcchhExpArea} FCC-hh interaction region top view as 
  resulting from MDISim, from -700~m to 700~m.
  The beam pipe is in blue for beam~1 and in red for beam~2. The transverse dimensions have been scaled up 500 times for visualization.}
  \end{center}
\end{figure}

MDISim has been used to reconstruct the region from -700~m to 700~m of the EIR around IPA, see Figure~\ref{fig:obg:fcchhExpArea} for a top view. 
The study also applies to the second interaction point named IPG, that has the same design and is located approximately 50~km away from IPA.\par
The beam pipe apertures upstream and downstream IPA are symmetrical. 
As described in Section~\ref{sec:optics}  the beam pipe is made of 0.8~mm thick Beryllium with an inner radius of 20~mm from IPA to $\pm$8~m, representing a critical region due to its proximity to the vertex detector. The following 8~m are covered by a Beryllium cone with an opening angle of 2.5~mrad. From $\pm16$~m to $\pm$35~m respectively, the material is Aluminum and the beam pipe radius is 40~mm. The entire detector layout occupies the region between $\pm25$~m, followed by a forward shielding section from $\pm25$~m to $\pm35$~m. 
At $\pm$35~m from the IPA, outside the detector and shield area, the TAS is placed  
as absorber  to protect the insertion quadrupoles from collision debris and 
its aperture radius is 20~mm.
The aperture radii outside the region between the two TAS are larger than 56~mm, until the collimator TAN at $\pm$412~m where it is reduced to 29~mm.\par
\begin{figure}
  \begin{center}
  \includegraphics*[width=\textwidth]{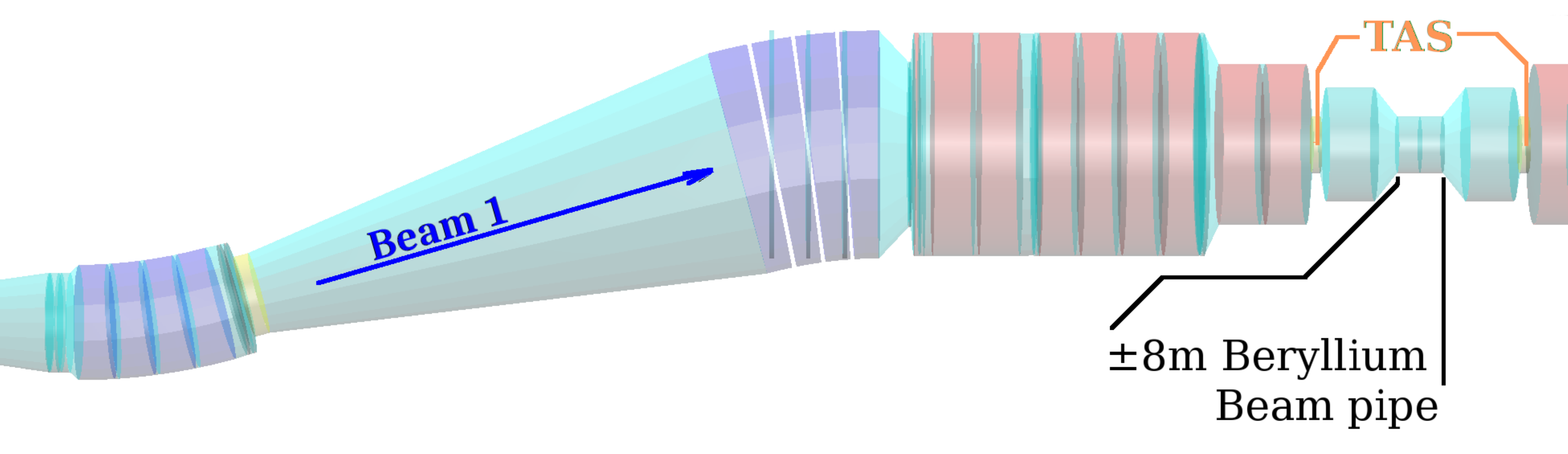}
  \caption{\label{fig:obg:fcchhbeam1} Top view of the  beam pipe 3D-model  obtained with MDISim. Dipoles are in violet, drifts in light blue, quadrupoles in orange, collimators in yellow.}
  \end{center}
\end{figure}

The dipoles in the experimental region shown in Figure~\ref{fig:obg:fcchhExpArea}
are 2~T magnets 11.3~m long, differently from the nominal arc dipoles which are 16~T.
These low field dipoles are located at about 200~m and at 430~m upstream and downstream IPA providing a bending angle of 135~$\mu$rad each. 
The aim of our study was to determine SR photons  coming from these dipoles 
and entering the TAS that might impact the detector, also possibly crossing the Beryllium beam pipe from -8~m to 8~m around IPA. Figure~\ref{fig:obg:fcchhbeam1} shows a zoom
of this region.

As reference position we have chosen the TAS. We give the power of the photons passing through the TAS and look in detail where these photons were generated and calculate the energy spectra and hit positions downstream of the TAS and determine the fraction of photons hitting the inner Beryllium pipe.\par

Table~\ref{tab:obg:dipolestable} gives the exact longitudinal position
of the last eight dipoles 470~m upstream the IPA. The critical energy in the dipoles is 0.536~keV, the mean energy 0.165~keV and the emitted power by each dipole is 6.4~W. In total, their power emitted is about 50~W. However, only a fraction of these photons will reach the experimental area, due to geometrical reasons and to the TAS presence, and even fewer of them will hit the $\pm8$~m Beryllium beam pipe.

\begin{table}
  \begin{center}
  \caption{\label{tab:obg:dipolestable} 
  Longitudinal position of the the last eight dipoles 470~m  upstream the IPA, shown in Fig.~\ref{fig:inf:PowerLoss_TAS_D2}. They are all
  2~T and 11.3~m, 
  their critical energy is 0.536~keV,
  mean energy is 0.165~keV and each has an the emitted power of
  6.4~W. In total the emitted power is about ~50~W.}\vspace*{5mm}


  \begin{tabular}{|c|c|}\hline
  \bf Dipole name&\bf Distance from IPA [m]\\\hline
 D1a 	&	190.0  \\
 D1b 	&	202.8  \\
 D1c	&	215.6  \\
 D1d	&	228.4 \\
 D2a 	&	421.3  \\
 D2b 	&	434.1  \\
 D2c 	&	446.9  \\
 D2d 	&	459.7  \\\hline
  \end{tabular}
  \end{center}
\end{table}

As described above we used MDISim to produce the geometry and the magnetic field description. GEANT4 has been used to perform the detailed simulation, starting at approximately $-700$~m from IPA with a Gaussian beam  with the expected size and emittance. 
These protons were tracked with the Monte Carlo,  taking into account  the production of SR photons and their subsequent propagation.
For this study the baseline and ultimate optics discussed in Section~\ref{sec:optics}  were used, with and without the horizontal crossing angle.
The results of this simulation, summarized in Table~\ref{tab:obg:sr_results}, suggest that about 10~W are expected to enter the experiment area with no crossing angle. The power of the photon flux at the inner Beryllium pipe located from -8~m  to 8~m around IPA remains below 1~W. 
A possible 10~Tm detector spectrometer placed in the experimental area
would increase this value by one Watt.
Even if the power deposited on the Beryllium pipe remains small, the simulations show that the number of photons is significant and merits a closer investigation. To evaluate the amount of particles that can pass through the Beryllium, we also used a more local GEANT4 simulation, in which the photons with the energy spectra obtained from the beams were directly impacting at a  $200~\mu$rad grazing angle on the 0.8~mm Beryllium pipe, to provide an upper limit estimate for the ultimate crossing angle.\par
\begin{table}
\begin{center}
    \caption{\label{tab:obg:sr_results}  Summary of the SR power emitted per beam in the last 700~m upstream IPA that reaches the experimental area~P$_{TAS}$, and the fraction that impacts the inner Be beam pipe~P$_{Be}$, for the baseline and ultimate configurations with and without crossing angle. The number of photons hitting the Be~N$_{\gamma Be}$ and their mean energy~E$_{mBe}$ are also shown.}
    \vspace*{5mm}
 \begin{tabular}{|c|c|r|r|r|r|}\hline
        \bf Lattice & \bf Half Crossing&\bf P$\boldsymbol{_{TAS}}$&\bf P$\boldsymbol{_{Be}}$&\bf N$\boldsymbol{_{\gamma Be}}$ & \bf E$\boldsymbol{_{mBe}}$\\
        & \bf Angle  $\boldsymbol{[\mu}$rad]&\bf [W] & \bf [W] & [$\boldsymbol{10^9}$] & \bf [keV]\\\hline
        
         Baseline &    0  & 8.5 & 0.74 & 1.1 & 0.166\\
         Ultimate &    0  & 8.7 & 0.73& 1.1 & 0.163\\\hline
         Baseline &   52  & 26.5 &1.17& 1.8 & 0.163\\
         Ultimate &  100  & 46.4 & 12.86 & 16.0 & 0.198\\\hline
    \end{tabular}
\end{center}
\end{table}

\begin{figure}
  \begin{center}
  \includegraphics*[width=0.45\textwidth]{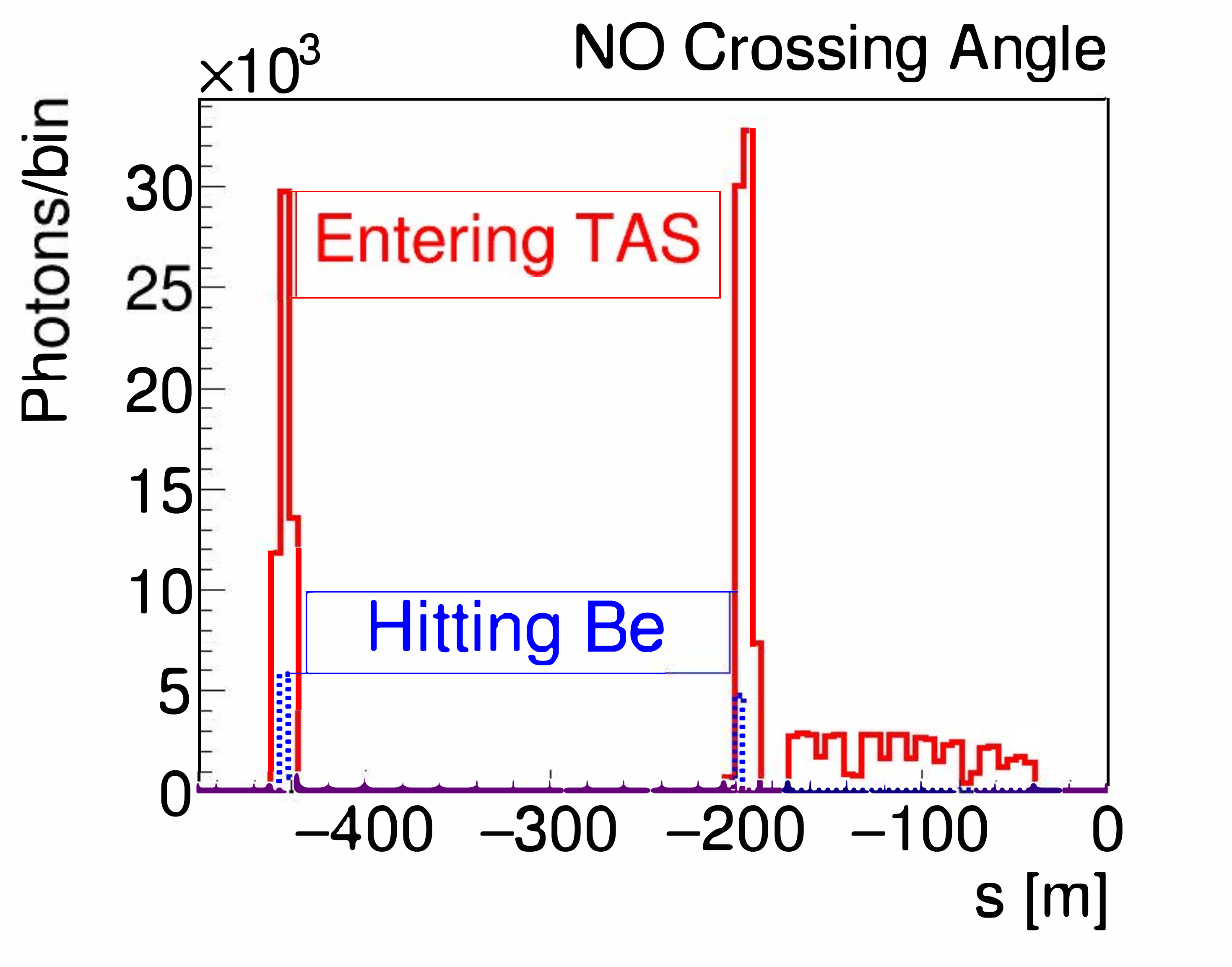}
  \includegraphics*[width=0.45\textwidth]{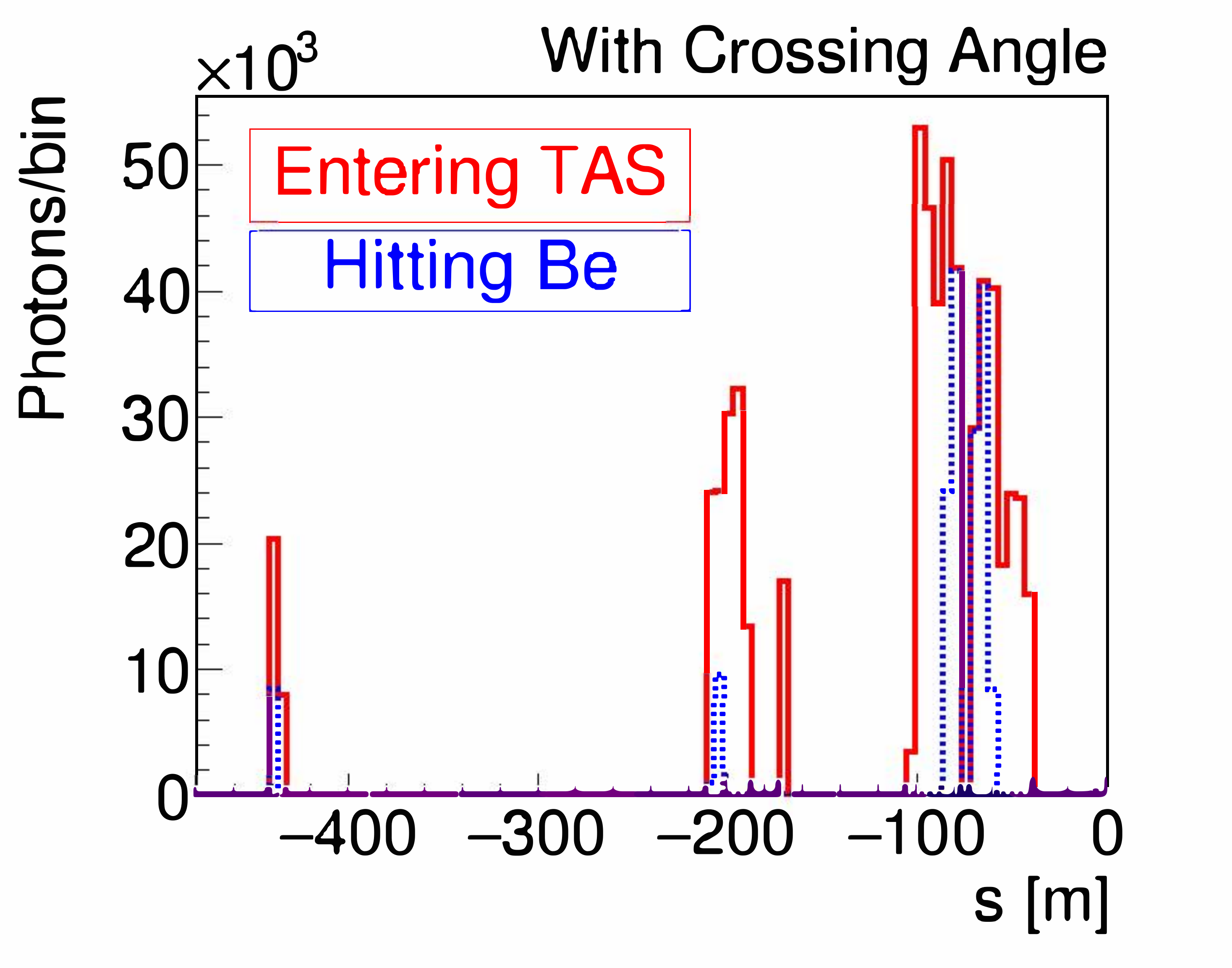}
  \caption{\label{fig:fco:fcchhphotonsVSs} Histogram of SR photon generation upstream IPA for the ultimate optics, where IPA is at z=0. {\bf Left plot:}~The two peaks at -450 and -200~m are photons produced by 2~T dipoles that reach the two TAS regions (in red), but few of them hit the Be pipe (in blue). In addition, few photons at are generated by quadrupole magnets downstream -200~m, and none of them hit the Be beam pipe. {\bf Right plot:}~With crossing angle, additional radiation comes from quadrupole magnets in the last 100~m. Radiation coming from MQXD.A2LA.H at -77.6~m and MQXC.B1LA.H at -56.3~m hits the Be pipe.
  }
  \end{center}
\end{figure}

Table~\ref{tab:obg:sr_results} summarizes the study of the SR
impacting the experimental area. Without crossing angle, the SR emitted by beam protons in the last magnetic elements upstream IPA is small and only a very minor source of backgrounds to the experiments.
With crossing angle, we get a small increase in power and a more significant increase in the number of photons reaching the experimental area. A small part of the increase is due to the extra SR generated by the field of the corrector magnets that generate the crossing angle (shown in Table~\ref{tab:obg:bumps}).
\begin{table}
\begin{center}
    \caption{\label{tab:obg:bumps} Correctors upstream IPA used in the optics lattices with crossing angle.}\vspace*{5mm}
    \begin{tabular}{|l|c|c|S|S|}\hline
 \bf Corrector name&\bf Distance&\bf Length &\multicolumn{2}{|c|}{\bf B field [T]} \\ 
& \bf to IPA [m] & \bf [m] & \bf Baseline & \bf Ultimate \\\hline
 MCBXDHV.A2LA.H	& 75.8 & 1.3 & -0.168 & -0.562\\
 MCBXCHV.3LA.H & 174.2 & 1.3 & 1.226 & 1.957\\
 MCBRDH.4LA.H1&	474.0 & 3.0 & -0.821 & -1.536\\\hline
    \end{tabular}
\end{center}
\end{table}
We find that the power from the SR of the corrector magnets is 0.6~W for the baseline and 1.8~W for the ultimate lattices, or a rather small increase compared to the power produced by the 11.3~m long 2~T dipoles.
Figure~\ref{fig:fco:fcchhphotonsVSs} shows the z-position at the origin of the photons reaching the TAS. We can see, that without crossing angle (left plot of Figure~\ref{fig:fco:fcchhphotonsVSs}), almost all photons entering the experimental originate in the 2~T dipoles located -450 and -200~m upstream of IPA, and that only few of these hit the Beryllium beam pipe.
The simulation includes SR from quadrupoles which increases significantly with crossing angle over the last 100\,m as can be seen in  Fig.~\ref{fig:fco:fcchhphotonsVSs}). With crossing angles many of these photons and in particular those generated by the quadrupole magnets  MQXD.A2LA.H at -77.6~m and MQXC.B1LA.H at -56.3~m will hit the Beryllium pipe.
The magnetic field gradient of these quadrupoles in the last 100~m does not vary much between the different optics configurations. Much of the increase with crossing angle
is from the last quadrupoles.\par
\begin{figure}
  \begin{center}
  \includegraphics*[width=0.45\textwidth]{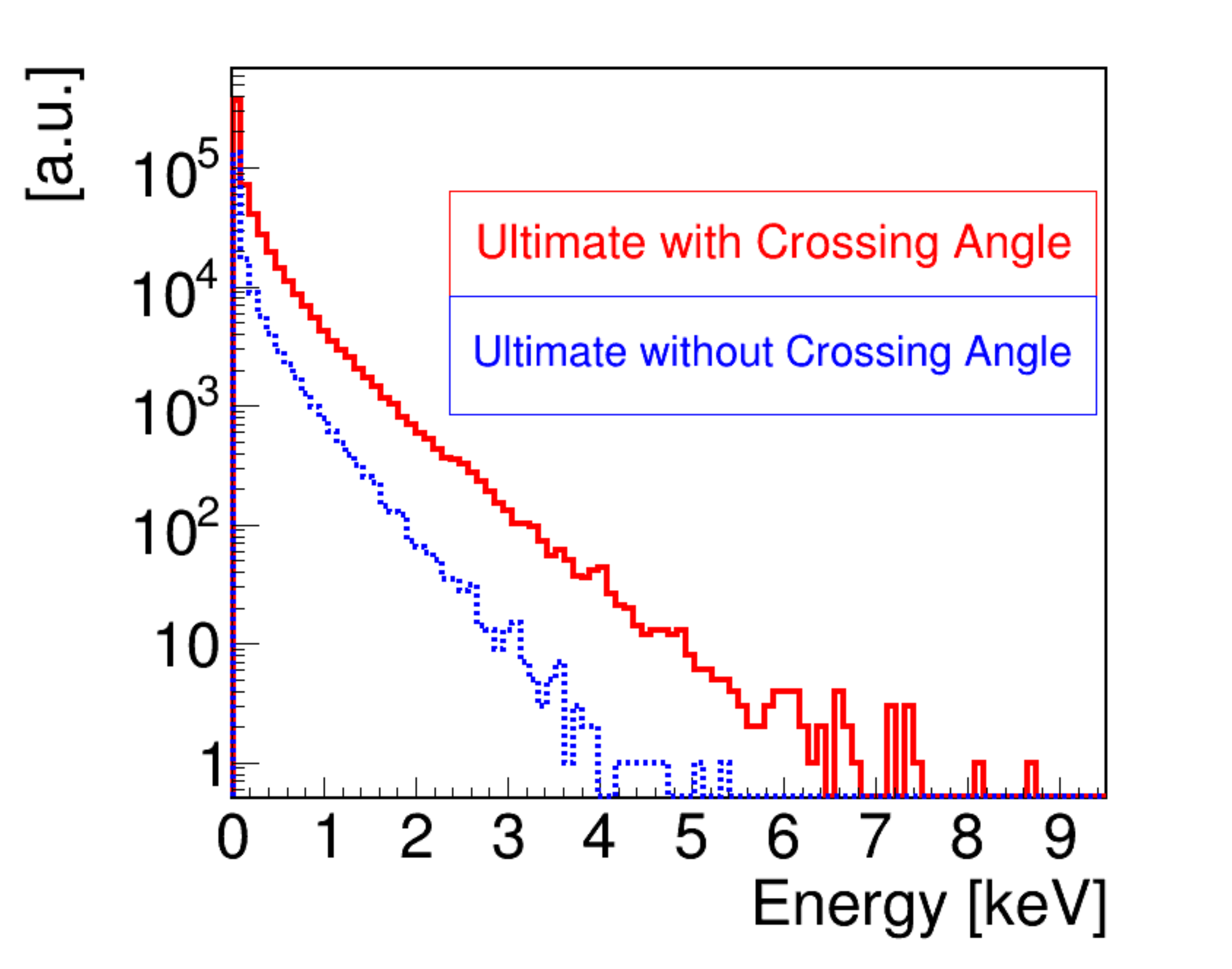}
  \includegraphics*[width=0.45\textwidth]{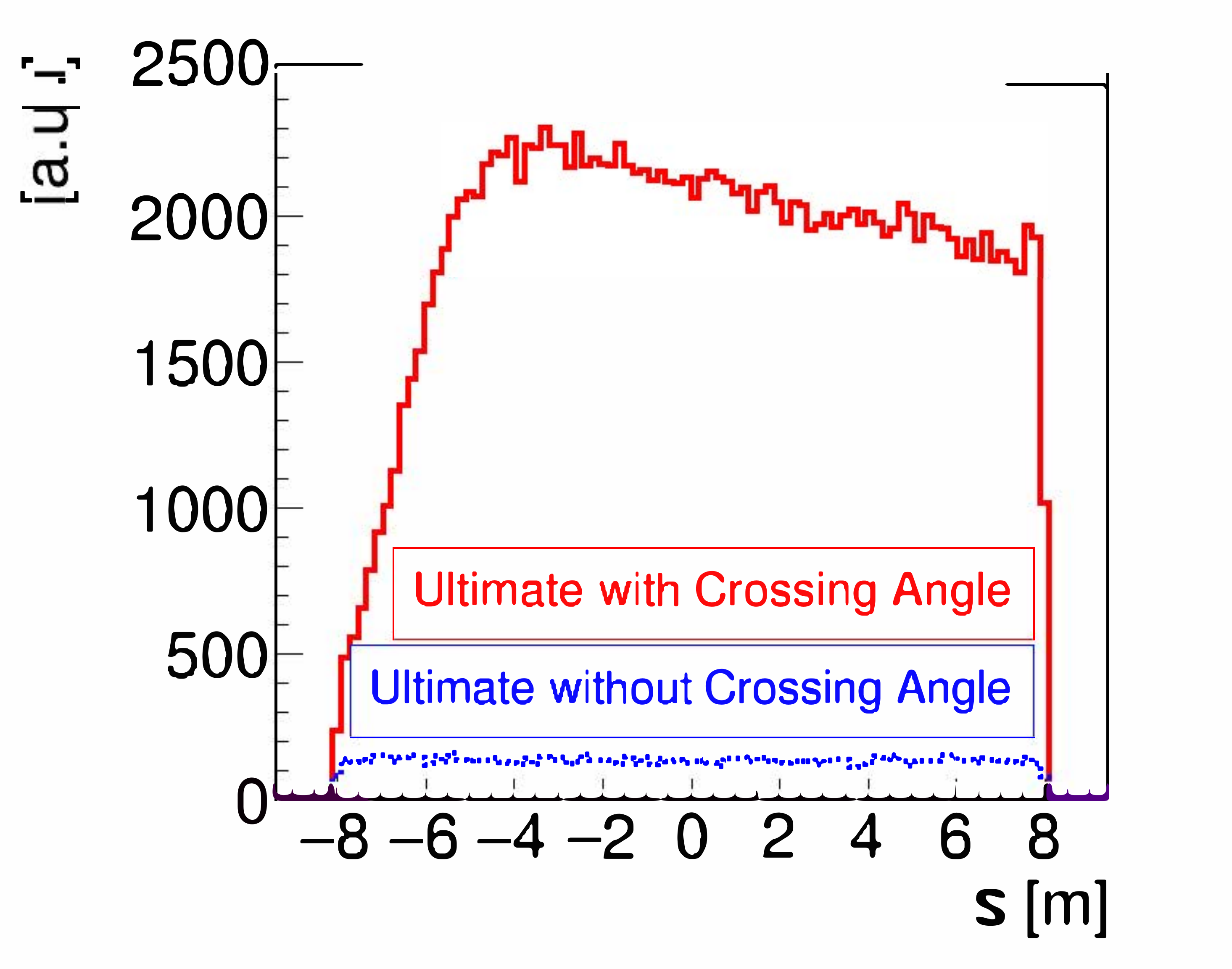}
  \caption{\label{fig:fco:fcchhphotonsspectrum}  {\bf Left:} Photon energy distribution entering the TAS with crossing angle (in red), and without crossing angle (in blue) for the ultimate optics. {\bf Right:} Photons hitting point in the last -8~m to 8~m Be beam pipe around IPA located at z=0. Without crossing angle (in blue) the photons are uniformly distributed, while With crossing angle (in red) they accumulate towards the center.}
  \end{center}
\end{figure}
\par
In conclusion, the contribution of synchrotron radiation photons into the experimental area has been addressed by means of a detailed GEANT4 simulation using MDISim. Similar studies were also carried out for other lattice versions~\cite{bib:obg:FrancescosRepport} and the results have been benchmarked with SYNRAD~\cite{bib:obg:SYNRAD}, finding a very good agreement.
We expect less than 1 photon per bunch with an energy of the order of 1~keV to traverse the Beryllium beam pipe towards the experiments.
So we conclude that the photon background from synchrotron radiation does not generate significant noise in the detector, the impact on the detector performance is  expected to be minimal.

\section{Cross Talk Between Experimental Insertion Regions}

Proton collisions at the interaction points of the FCC-hh may contribute to background in the subsequent detector, and losses between the detectors. As the proton luminosity is high, this may be of concern.

\begin{figure}
	\centering
	\includegraphics*[width=0.8\textwidth]{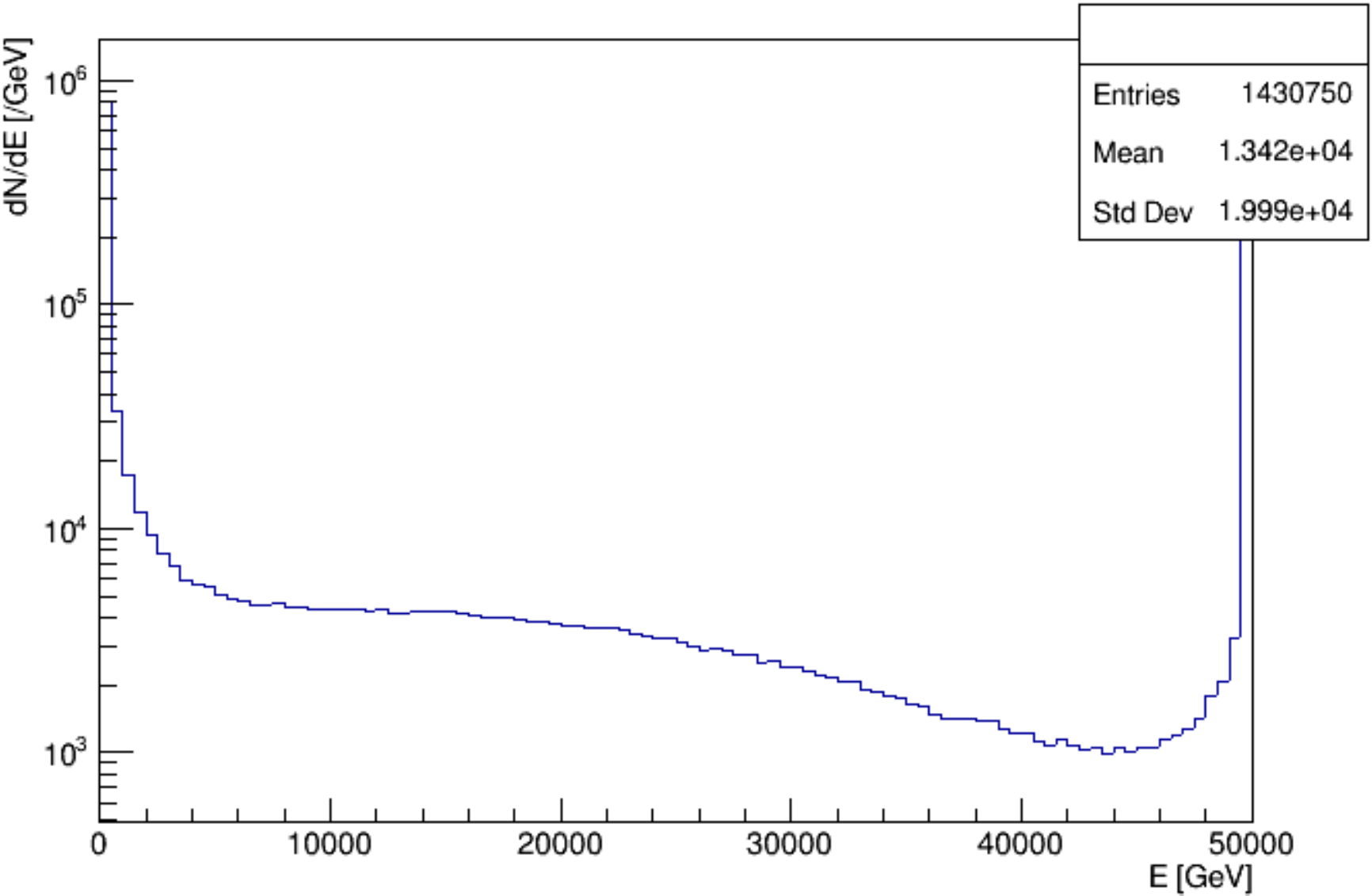}
	\caption{\label{fig:hr:e_distn} Proton energy distribution \SI{3}{ m} downstream of IPA for $10^6$ $pp$ collision events.}
\end{figure}

Using the upgraded version of the DPMJET-III event generator \cite{bib:hra:Fedynitch_PHD} inside FLUKA~\cite{fluka1} we generate the debris from the \SI{50}{TeV} proton-proton ($pp$) collisions with vertical crossing. Due to the rigidity of the charged particles, only protons are transported by the accelerator. Muons are the only other major concern, and are treated separately. The energy distribution of protons is large, this is shown in Fig.~\ref{fig:hr:e_distn}.

For proton cross-talk, we use the energy of \SI{49.95}{TeV} as a threshold. Protons from the collision with energy greater than this are defined as `elastic', and those with energy below this are defined as `inelastic' protons. PTC~\cite{bib:hra:PTC} and MERLIN~\cite{bib:hra:IPAC18_MERLIN} are used to perform tracking of both elastic and inelastic protons to determine the cross-talk. That is, the number of collision debris protons that will reach the next detector. We generate the debris at IPA and track to IPB. 

For the case of elastic protons, nearly all reach IPB with a spot size similar to that of the beam. This is likely to lead to an emittance growth, but should pose no major concern. Around 2 inelastic protons per bunch crossing will arrive at IPB under nominal settings, this rises to $\approx$ 9 under ultimate settings. The mean energy of these few protons is \SI{49.89}{TeV}, and they are unlikely to be of concern in terms of cross-talk.

What is of greater concern is the loss of inelastic protons between the two points. Most are lost in the short straight section and dispersion suppressor (DS) regions post IPA.  A detailed study of losses in the detector, inner triplet, and separation and recombination dipoles has been conducted in detail with FLUKA. This is documented in section~\ref{sec:rma:IR_radiation}. Therefore we focus on losses after these elements, of which, the DS losses are of great concern as they are bottlenecks for off-momentum particles and the proton energy is high.

It was decided to mitigate these losses using HL-LHC style `TCLD' collimators in the DS. Two \SI{1}{m} long TCLD collimators were placed before the first quadrupole in cells 8 and 10, at the points where the dispersion rises rapidly. With these collimators in place, the DS losses are minimised, as all particles are intercepted by the TCLDs. In these simulations all apertures are treated as black absorbers.

\begin{figure}
	\centering
	\includegraphics*[width=\textwidth]{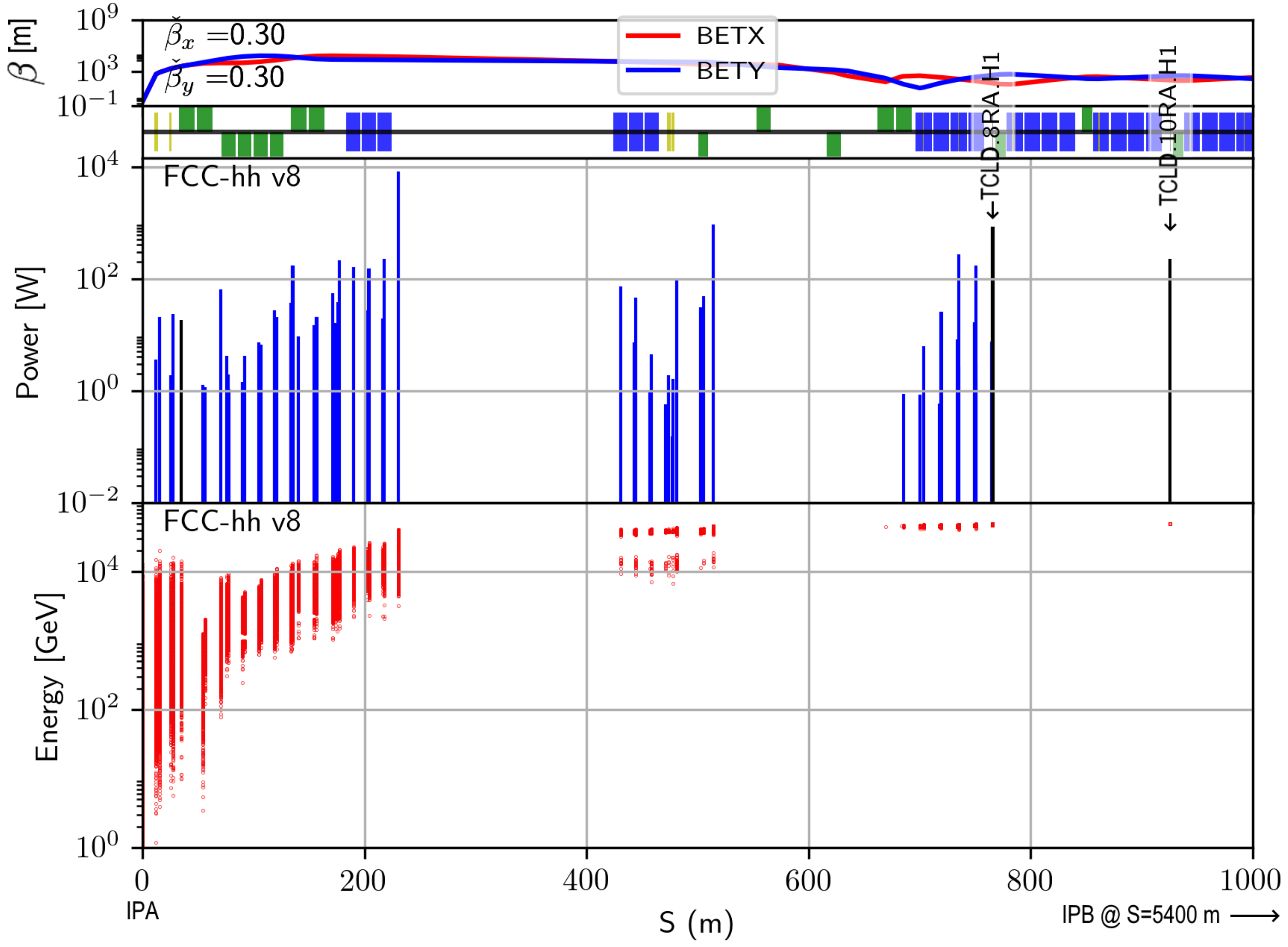}
	\caption{\label{fig:hr:loss_energy} Proton losses in the first 1 km post IPA (S = 0 m, IPB at S = 5400 m). The top plot shows power deposition per element, the bottom plot shows the energy distribution of the losses per element, both with the TCLD collimators included in the lattice.}
\end{figure}

These collimators are placed in regions of relatively low $\beta$ function, thus allowing larger jaw gaps so as not to violate the collimation hierarchy. A jaw half-gap of 35~$\sigma$ was found to be sufficient. Using MERLIN the power and energy of lost particles in the short straight section are shown, in the presence of the two TCLD collimators, in Fig.~\ref{fig:hr:loss_energy}. We note that all losses are shown per element, therefore the largest peak in the power plot corresponds to the loss over the full length of a $\approx$ \SI{200}{m} long drift.

\begin{figure}
	\centering
	\includegraphics*[width=0.7\textwidth]{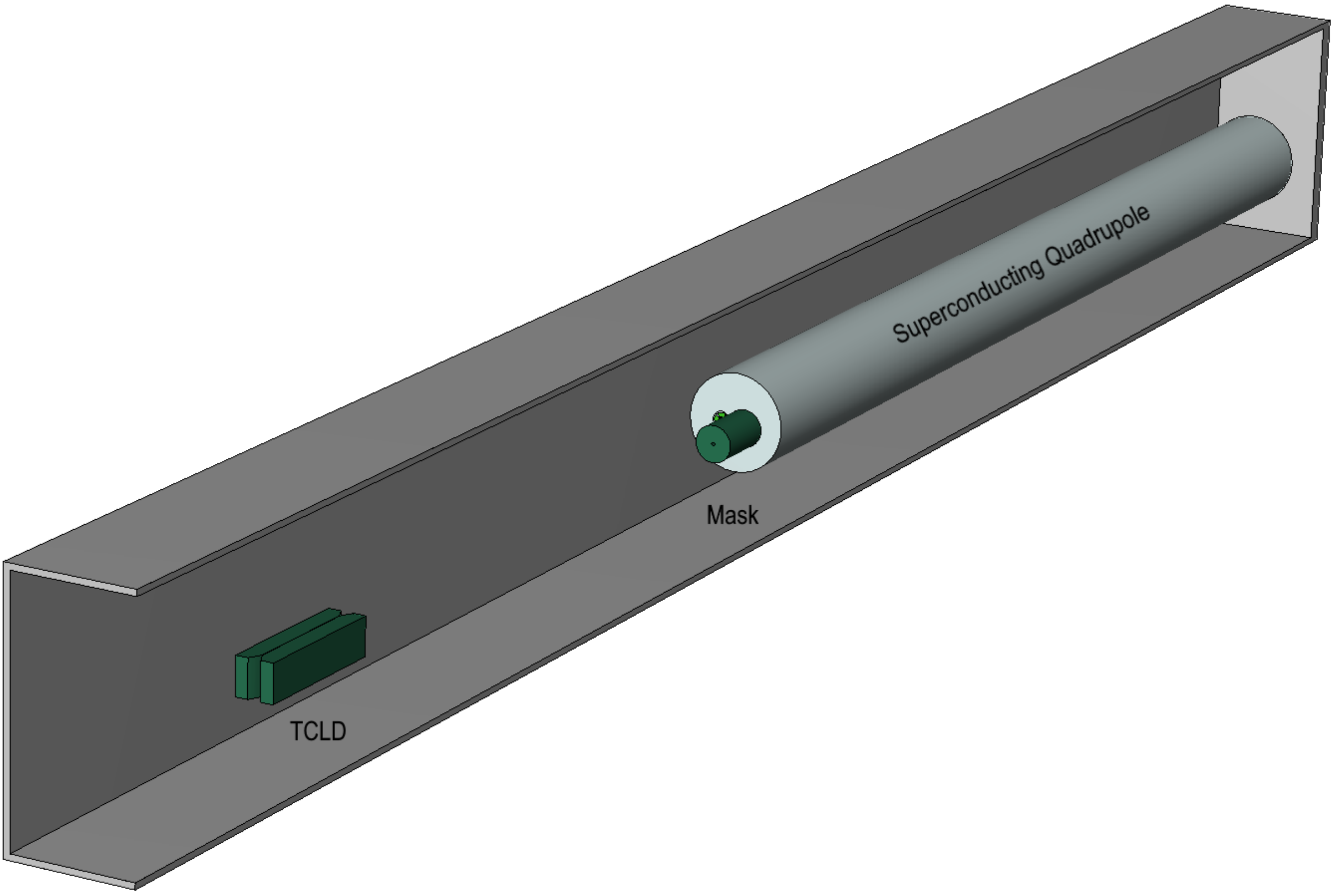}
	\caption{\label{fig:hr:TCLD_FLUKA} FLUKA model used for shower simulations in the dispersion suppressor. The green objects are first the TCLD collimator, followed by a \SI{50}{cm} long mask, both made of INERMET180. Following this is the first quadrupole in the DS. The particles are loaded \SI{63}{cm} before the collimator.}
\end{figure}

Collimators clean the `primary' halo, but produce a `secondary' halo in turn. In order to verify that this secondary shower would not exceed the maximum energy deposition allowed on the subsequent superconducting quadrupole, a two step simulation was used. Firstly the inelastic protons were transported from IPA to the TCLDs using MERLIN, to generate hits on the collimator jaws. These hits were fed into a FLUKA model, shown in Fig.~\ref{fig:hr:TCLD_FLUKA}, which consists of the first TCLD collimator, a drift space, followed by a \SI{50}{cm} long mask prior to the superconducting quadrupole. The quadrupole coils are simulated as a mixture of 50\% Nb$_{3}$Sn and 50\% copper. INERMET180 has been chosen as the material for the TCLD jaw and quadrupole mask, as in the current LHC absorbers. The distance between the collimator and the mask and quadrupole gives space for the shower to spread, thus minimising the load on the quadrupole coils.

As the first collimator (in cell 8) has the higher load, it was used for shower simulations. A jaw half-gap of 35~$\sigma$ was shown to intercept all inelastic protons whilst not violating the betatron collimation hierarchy. There is the possibility that this could still interfere with the momentum cleaning hierarchy - as the momentum cleaning was not defined at the time of this investigation.

\begin{figure}
	\centering
	\includegraphics*[width=0.7\textwidth]{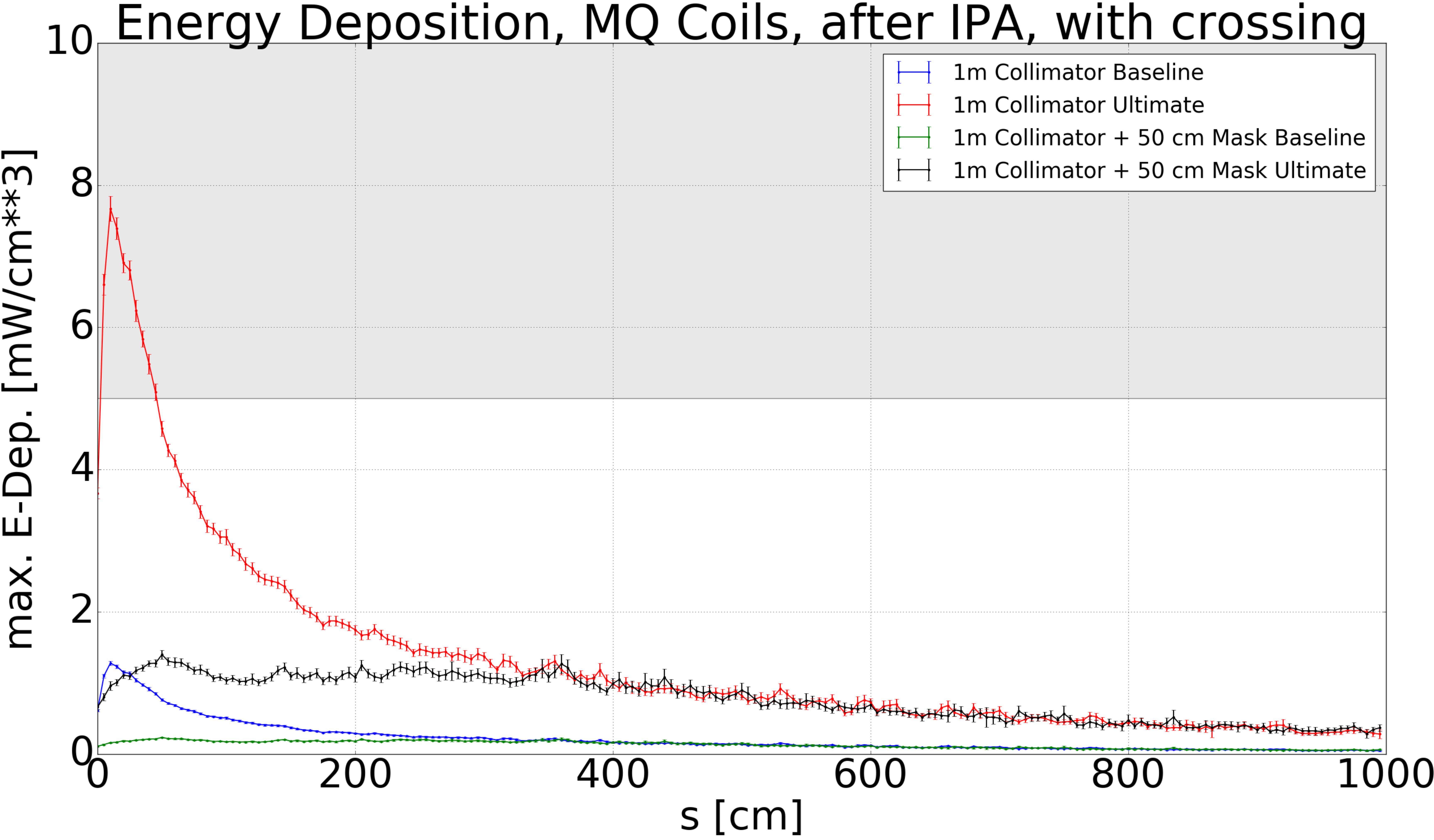}
	\caption{\label{fig:hr:TCLD_deposition} Maximum energy deposition per \si{cm^3} in \SI{5}{cm} bins along the first quadrupole in cell 8 after IPA for baseline and ultimate configurations.}
\end{figure}

It is evident from Fig.~\ref{fig:hr:TCLD_deposition}, which shows the maximum energy deposition in the first quadrupole post-TCLD, that the \SI{50}{cm} mask is required in order to stay below the limit of $\approx$ 5 - \SI{10}{mW~cm^3}~\cite{bib:hra:Todesco} at ultimate parameters. For baseline parameters the mask is not required. Thus we may mitigate the DS losses due to inelastic protons from collision debris using the two \SI{1}{m} long INERMET180 TCLDs, placed in cells 8 and 10, before the first quadrupole in the cell.

\begin{figure}
	\centering
	\includegraphics*[width=0.6\textwidth]{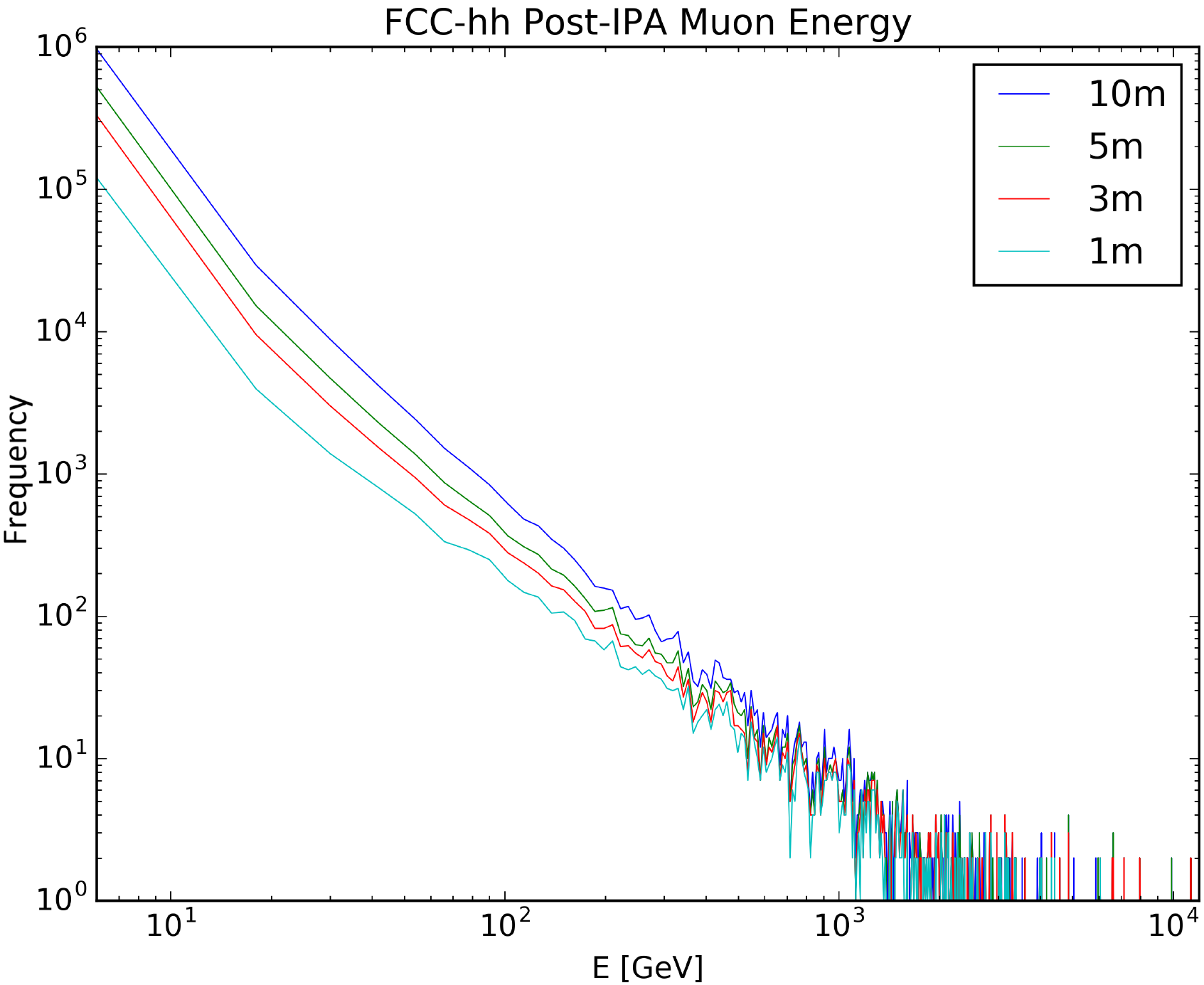}
	\caption{\label{fig:hr:muon_energy} Muon energy distribution at intervals downstream of IPA, generated using DPMJET-III in FLUKA, with no detector or accelerator model.}
\end{figure}

As muons have a large mean free path, they can travel kilometres in dense materials, therefore muon cross-talk may be a concern. The muon energy distribution generated from 10$^6$ \SI{50}{TeV} $pp$ collisions using DPMJET-III inside FLUKA is shown in Fig.~\ref{fig:hr:muon_energy}. Low energy muons are produced from a multitude of particle physics processes, this results in an increase in the number of low energy muons as the observation point is moved further from the collision point, within a small range. High energy muon production is rare, the highest energy muon produced is around \SI{20}{TeV}.

We may separate muon energy loss into; ionisation, bremsstrahlung, production of electron-positron pairs, and through photo-nuclear and photo-nucleon interactions. This approach is summarised in Eq.~(\ref{eqn:hr:muondedx})~\cite{bib:hra:Muon_Range}:
\begin{equation}\label{eqn:hr:muondedx}
\left< \frac{-dE}{dx} \right> = a(E) + b(E)E\ ,
\end{equation}
where $a(E)$ is the ionisation contribution, and $b(E) = b_b(E) + b_p(E) + b_n(E)$ is the sum of the contributions of bremsstrahlung, pair production, and photo-nuclear/nucleon interactions~\cite{bib:hra:PDG}. In the continuous slowing down approximation the range is given by
\begin{equation}\label{eqn:hr:muonrange1}
R(E) = \int_{E_0}^{E} (a(E') + b(E')E')^{-1} dE'.
\end{equation}

\begin{figure}
	\centering
	\includegraphics*[width=0.5\textwidth]{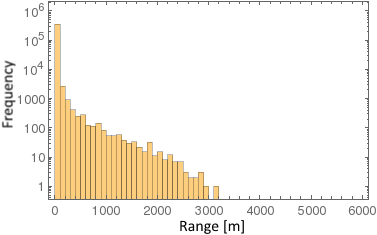}
	\caption{\label{fig:hr:muon_range} Theoretical range of collision debris muons in rock.}
\end{figure}

At high energy $a$ and $b$ are constant, and this becomes
\begin{equation}\label{eqn:hr:muonrange2}
R(E) \approx \frac{1}{b} ln \left( 1 + \frac{E}{E_c}\right)\ ,
\end{equation}
where the electronic and radiative losses are equal at the critical energy $E_c$. We use this approach to calculate the theoretical range of muons in standard rock, which has a specific gravity of \SI{2.65}{g \centi m^{-3}} and $\left< \frac{Z}{A}\right> = 0.5$, and in which the muon critical energy is \SI{693}{GeV}. The result of this calculation is shown in Fig.~\ref{fig:hr:muon_range}, which gives a maximum range of \SI{3.3}{km} for collision debris muons in the FCC-hh through standard rock. This analytical calculation does not include the interaction of collision debris with the detector.

\begin{figure}
	\centering
	\includegraphics*[width=\textwidth]{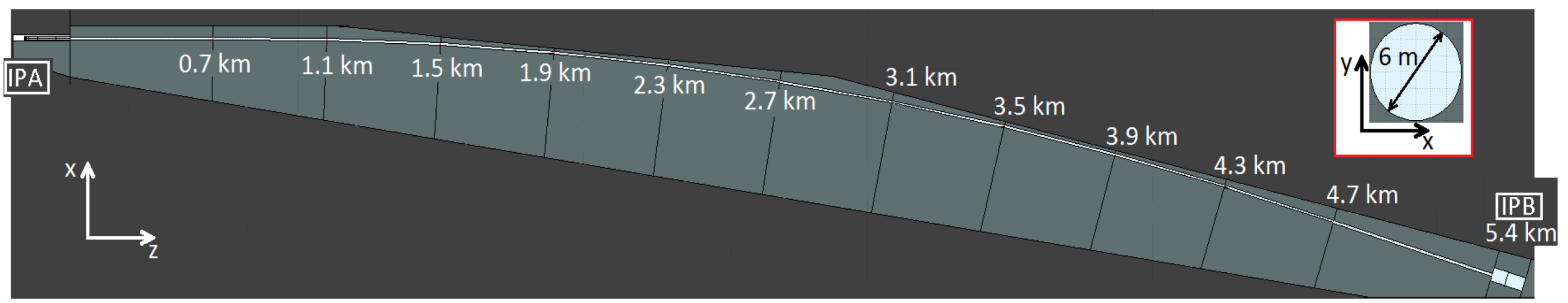}
	\caption{\label{fig:hr:tunnel_FLUKA} Cross section of the FLUKA FCC-hh tunnel model. The distance from IPA to each point along the tunnel central line, where muons are observed, is indicated. IPB is located \SI{5.4}{km} away from IPA. The lighter grey area is modelled as standard rock. Note that this model was based on an older version of the FCC-hh lattice.}
\end{figure}

In order to verify the analytical expectation, $5 \times 10^4$ $pp$ collisions were generated using DPMJET-III inside FLUKA with a complete model of the detector~\cite{bib:hra:Besana} in order to generate the initial muon distribution. These muons were then tracked in FLUKA using a total of 10$^9$ histories through the tunnel model shown in Fig.~\ref{fig:hr:tunnel_FLUKA}. 

\begin{figure}
	\centering
	\includegraphics*[width=0.9\textwidth]{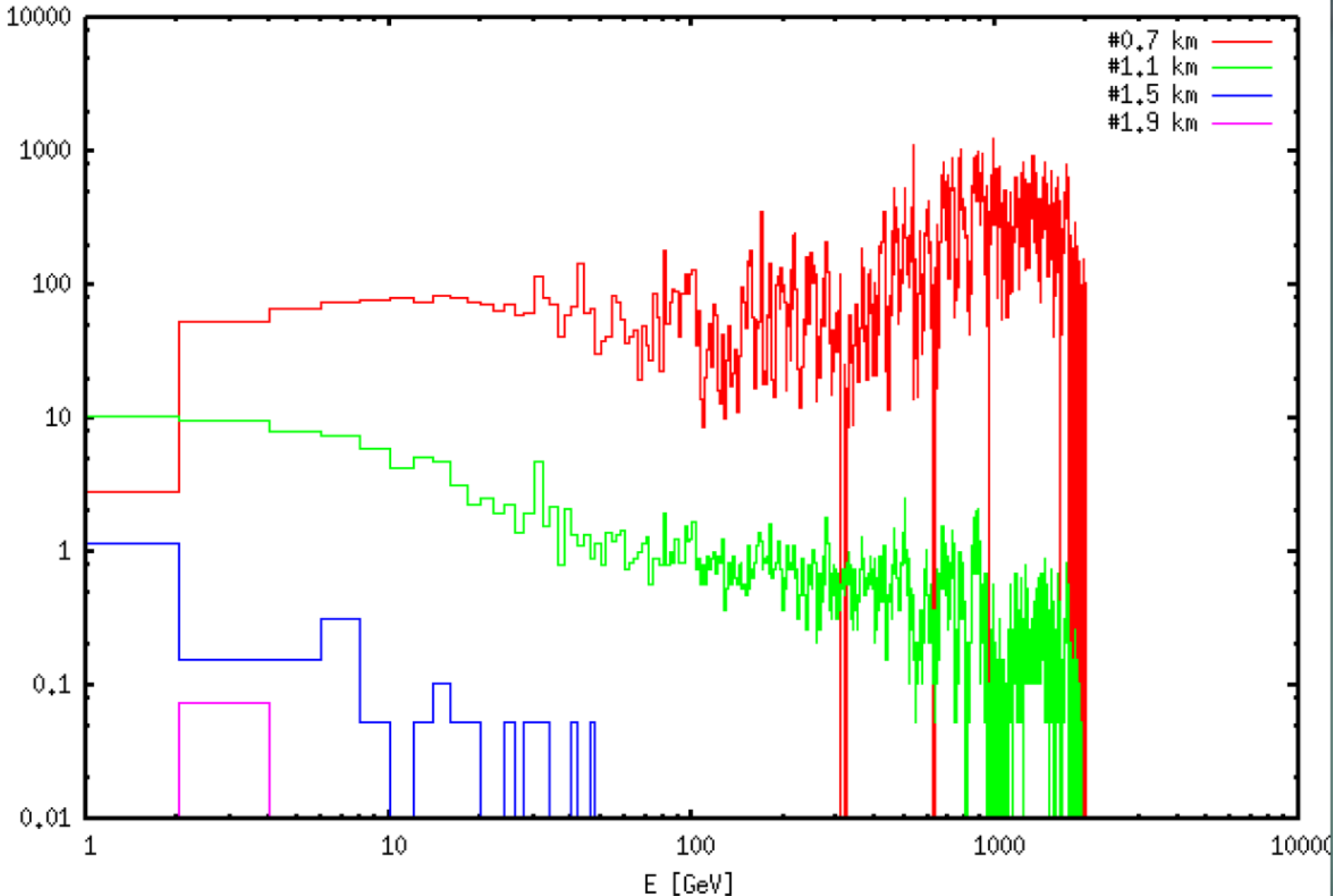}
	\caption{\label{fig:hr:muon_results} Muon distribution at different positions along the tunnel, as indicated in Fig.~\ref{fig:hr:tunnel_FLUKA}.}
\end{figure}

The muon energy distributions along the tunnel model are shown in Fig.~\ref{fig:hr:muon_results}. From this it is clear that few muons travel \SI{1.9}{km}, no muons travel beyond \SI{2.7}{km}, thus we may conclude that muon cross-talk should not be an issue at the FCC-hh.

In summary, elastic protons  with an energy greater than \SI{49.95}{TeV} from collisions at IPA nearly all reach IPB with a spot size similar to the beam. This should result in an emittance growth of the beam. Inelastic protons, with an energy less than \SI{49.95}{TeV} pose a greater threat. Only 2 - 9 inelastic protons per bunch crossing are foreseen to reach IPB, this is deemed negligible. The losses from inelastic protons in the short straight section and dispersion suppressor regions post IPA are of concern. By using two \SI{1}{m} long TCLD collimators with INERMET180 jaws, the losses in the DS region post IPA was mitigated. Shower simulations of the inelastic proton impacts upon the first TCLD in cell 8 were performed in order to gauge secondary shower damage on the next superconducting element, the first quadrupole in the cell. Results show that for baseline parameters the energy deposition is below the suggested limit, and for the ultimate parameters the addition of a \SI{50}{cm} long INERMET180 mask would be required to protect the quadrupole coils~\cite{bib:hra:IPAC18_Proton_Cross_Talk}. 

Despite high energy muons of up to \SI{20}{TeV} being created in the $pp$ collisions, by analytical and Monte Carlo methods we have shown that muons should not travel far enough through rock or the accelerator tunnel to reach the subsequent detector~\cite{bib:hra:IPAC18_Muon_Cross_Talk}.

Photons and other charged hadrons in the collision debris are ignored as their rigidity means that they will not be accepted in the accelerator, and thus cannot be transported to IPB.

\section{Hardware Specifications}
Table~\ref{tab:rma:hl_eir_magnet_specifications} lists the specifications for the magnets of the high luminosity EIRs. The cryostats of the triplet quadrupoles will have to be designed so they can support thick and consequently heavy shielding inside the coil apertures. The field quality specifications of the triplet magnets using the same notation as in~\cite{hl-lhc-tech-design}  are given in Tables~\ref{tab:errorsQ1} and~\ref{tab:errorsQ2}.
\begin{table}
  \begin{center}
  \caption{\label{tab:rma:hl_eir_magnet_specifications} Magnet parameters of the high luminosity EIRs.}
  \begin{tabular}{ |l|c|c|c|c| } 
	\hline
	 \bf Magnet & \bf Length & \bf Field & \bf Coil aperture& \bf Number 	\\
	  & \bf [\si{m}] & \bf strength & \bf diameter [\si{mm}] & \bf per IP \\
	  \hline
	\multicolumn{5}{|l|}{\bf Triplet quadrupoles} \\ \hline
	Q1 & 14.3 & \SI{130}{\tesla /\metre} & 164 &	4	\\ \hline
	Q2 & 12.5 & \SI{105}{\tesla /\metre} & 210 &	8 	\\ \hline
	Q3 & 12.5 & \SI{105}{\tesla /\metre} & 210 & 4	\\ \hline
	\multicolumn{5}{|l|}{\bf Separation and recombination dipoles} \\ \hline
	D1 & 11.3 & \SI{2}{\tesla} & 170 &	8	\\ \hline
	D2 & 11.3 & \SI{2}{\tesla} & 91 &	8 	\\ \hline
	\multicolumn{5}{|l|}{\bf Matching quadrupoles} \\ \hline
	Q4 & 9.1 & \SI{200}{\tesla /\metre} & 70 &	2	\\ \hline
	Q5/6 & 12.8 & \SI{260}{\tesla /\metre} & 60 &	4	\\ \hline
	Q7 & 14.3 & \SI{400}{\tesla /\metre} & 50 &	4	\\ \hline
	\multicolumn{5}{|l|}{\bf Non-linear correctors} \\ \hline
	Sextupole  & \multirow{2}{*}{0.07}  & \multirow{2}{*}{\SI{460}{\tesla \per \metre^2}} & \multirow{2}{*}{210} & \multirow{2}{*}{2/2} \\ 
	(normal/skew) & & & & \\ \hline
	Octupole & \multirow{2}{*}{0.21} & \multirow{2}{*}{\SI{4000}{\tesla \per \metre^3}} & \multirow{2}{*}{210} & \multirow{2}{*}{2/2} \\
	(normal/skew) & & & & \\ \hline
	\multicolumn{5}{|l|}{\bf Orbit correctors} \\ \hline
	MCBX & 1.3 & \SI{3}{\tesla} & 210 & 6	\\ \hline
	MCBRD & 3.0 & \SI{4}{\tesla} & 70 & 4	\\ \hline
	MCBYM & 1.5 & \SI{4}{\tesla} & 60 & 4	\\ \hline	
	MCB & 1.2 & \SI{4}{\tesla} & 50 & 4	\\ \hline
  \end{tabular}
  \end{center}
\end{table}
\begin{table}
\centering
\caption{\label{tab:errorsQ1} Field error components of Q1 and Q3 with $R_{\text{ref}} = \SI{55}{mm}$ for Q1 and $R_{\text{ref}} = \SI{70}{mm}$ for Q3.}
\begin{tabular}{|c|cc|cc|cc|}
		\hline
		 & \multicolumn{2}{c|}{\bf Systematic} & \multicolumn{2}{c|}{\bf Uncertainty} & \multicolumn{2}{c|}{\bf Random} \\
		\bf Normal & \bf Injection & \bf High & \bf Injection & \bf High & \bf Injection & \bf High \\ 
		& & \bf Field & & \bf Field & & \bf Field \\ \hline
		$b_{1}$ & 0.000 & 0.000 & 0.000 & 0.000 & 0.000 & 0.000\\
		$b_{2}$ & 0.000 & 0.000 & 0.000 & 0.000 & (10) & (10)\\
		$b_{3}$ & 0.000 & 0.000 & 0.82 & 0.82 & 0.82 & 0.82\\
		$b_{4}$ & 0.000 & 0.000 & 0.57 & 0.57 & 0.57 & 0.57\\
		$b_{5}$ & 0.000 & 0.000 & 0.42 & 0.42 & 0.42 & 0.42\\
		$b_{6}$ & -19.947 & -0.357 & 1.1 & 1.1 & 1.1 & 1.1\\
		$b_{7}$ & 0.000 & 0.000 & 0.19 & 0.19 & 0.19 & 0.19\\
		$b_{8}$ & 0.000 & 0.000 & 0.13 & 0.13 & 0.13 & 0.13\\
		$b_{9}$ & 0.000 & 0.000 & 0.07 & 0.07 & 0.07 & 0.07\\
		$b_{10}$ & 3.664 & -0.129 & 0.2 & 0.2 & 0.2 & 0.2\\
		$b_{11}$ & 0.000 & 0.000 & 0.26 & 0.26 & 0.26 & 0.26\\
		$b_{12}$ & 0.000 & 0.000 & 0.18 & 0.18 & 0.18 & 0.18\\
		$b_{13}$ & 0.000 & 0.000 & 0.009 & 0.009 & 0.009 & 0.009\\
		$b_{14}$ & 0.158 & -0.866 & 0.023 & 0.023 & 0.023 & 0.023\\
		$b_{15}$ & 0.000& 0.000 & 0.000 & 0.000 & 0.000 & 0.000\\ \hline
		\bf Skew & \multicolumn{6}{c|}{}\\ \hline
		$a_{1}$ & 0.000 & 0.000 & 0.000 & 0.000 & 0.000 & 0.000\\
		$a_{2}$ & -0.877 & -0.877 & 0.000 & 0.000 & (10) & (10)\\
		$a_{3}$ & 0.000 & 0.000 & 0.65 & 0.65 & 0.65 & 0.65\\
		$a_{4}$ & 0.000 & 0.000 & 0.65 & 0.65 & 0.65 & 0.65\\
		$a_{5}$ & 0.000 & 0.000 & 0.43 & 0.43 & 0.43 & 0.43\\
		$a_{6}$ & 0.062 & 0.062 & 0.31 & 0.31 & 0.31 & 0.31\\
		$a_{7}$ & 0.000 & 0.000 & 0.19 & 0.19 & 0.19 & 0.19\\
		$a_{8}$ & 0.000 & 0.000 & 0.11 & 0.11 & 0.11 & 0.11\\
		$a_{9}$ & 0.000 & 0.000 & 0.08 & 0.08 & 0.08 & 0.08\\
		$a_{10}$ & 0.002 & 0.002 & 0.04 & 0.04 & 0.04 & 0.04\\
		$a_{11}$ & 0.000 & 0.000 & 0.026 & 0.026 & 0.026 & 0.026\\
		$a_{12}$ & 0.000 & 0.000 & 0.014 & 0.014 & 0.014 & 0.014\\
		$a_{13}$ & 0.000 & 0.000 & 0.01 & 0.01 & 0.01 & 0.01\\
		$a_{14}$ & -0.004 & -0.004 & 0.005 & 0.005 & 0.005 & 0.005\\
		$a_{15}$ & 0.000 & 0.000 & 0.000 & 0.000 & 0.000 & 0.000\\ \hline
	  \end{tabular}
\end{table}
\begin{table}
\centering
\caption{\label{tab:errorsQ2} Field error components of Q2 with $R_{\text{ref}} = \SI{70}{mm}$.}
\begin{tabular}{|c|cc|cc|cc|}
		\hline
		 & \multicolumn{2}{c|}{\bf Systematic} & \multicolumn{2}{c|}{\bf Uncertainty} & \multicolumn{2}{c|}{\bf Random} \\
		\bf Normal & \bf Injection & \bf High & \bf Injection & \bf High & \bf Injection & \bf High \\ 
		& & \bf Field & & \bf Field & & \bf Field \\ \hline
		$b_{1}$ & 0.000 & 0.000 & 0.000 & 0.000 & 0.000 & 0.000\\
		$b_{2}$ & 0.000 & 0.000 & 0.000 & 0.000 & (10) & (10)\\
		$b_{3}$ & 0.000 & 0.000 & 0.82 & 0.82 & 0.82 & 0.82\\
		$b_{4}$ & 0.000 & 0.000 & 0.57 & 0.57 & 0.57 & 0.57\\
		$b_{5}$ & 0.000 & 0.000 & 0.42 & 0.42 & 0.42 & 0.42\\
		$b_{6}$ & -19.752 & -0.317 & 1.1 & 1.1 & 1.1 & 1.1\\
		$b_{7}$ & 0.000 & 0.000 & 0.19 & 0.19 & 0.19 & 0.19\\
		$b_{8}$ & 0.000 & 0.000 & 0.13 & 0.13 & 0.13 & 0.13\\
		$b_{9}$ & 0.000 & 0.000 & 0.07 & 0.07 & 0.07 & 0.07\\
		$b_{10}$ & 3.631 & -0.132 & 0.2 & 0.2 & 0.2 & 0.2\\
		$b_{11}$ & 0.000 & 0.000 & 0.26 & 0.26 & 0.26 & 0.26\\
		$b_{12}$ & 0.000 & 0.000 & 0.18 & 0.18 & 0.18 & 0.18\\
		$b_{13}$ & 0.000 & 0.000 & 0.009 & 0.009 & 0.009 & 0.009\\
		$b_{14}$ & 0.151 & -0.865 & 0.023 & 0.023 & 0.023 & 0.023\\
		$b_{15}$ & 0.000& 0.000 & 0.000 & 0.000 & 0.000 & 0.000\\ \hline
		\bf Skew & \multicolumn{6}{c|}{}\\ \hline
		$a_{1}$ & 0.000 & 0.000 & 0.000 & 0.000 & 0.000 & 0.000\\
		$a_{2}$ & -1.003 & -1.003 & 0.000 & 0.000 & (10) & (10)\\
		$a_{3}$ & 0.000 & 0.000 & 0.65 & 0.65 & 0.65 & 0.65\\
		$a_{4}$ & 0.000 & 0.000 & 0.65 & 0.65 & 0.65 & 0.65\\
		$a_{5}$ & 0.000 & 0.000 & 0.43 & 0.43 & 0.43 & 0.43\\
		$a_{6}$ & 0.071 & 0.071 & 0.31 & 0.31 & 0.31 & 0.31\\
		$a_{7}$ & 0.000 & 0.000 & 0.19 & 0.19 & 0.19 & 0.19\\
		$a_{8}$ & 0.000 & 0.000 & 0.11 & 0.11 & 0.11 & 0.11\\
		$a_{9}$ & 0.000 & 0.000 & 0.08 & 0.08 & 0.08 & 0.08\\
		$a_{10}$ & 0.002 & 0.002 & 0.04 & 0.04 & 0.04 & 0.04\\
		$a_{11}$ & 0.000 & 0.000 & 0.026 & 0.026 & 0.026 & 0.026\\
		$a_{12}$ & 0.000 & 0.000 & 0.014 & 0.014 & 0.014 & 0.014\\
		$a_{13}$ & 0.000 & 0.000 & 0.01 & 0.01 & 0.01 & 0.01\\
		$a_{14}$ & -0.007 & -0.007 & 0.005 & 0.005 & 0.005 & 0.005\\
		$a_{15}$ & 0.000 & 0.000 & 0.000 & 0.000 & 0.000 & 0.000\\ \hline
	  \end{tabular}
\end{table}

For the separation and recombination dipoles D1 and D2 in the high luminosity EIRs, normal conducting dipoles, similar to the MBXW and MBW designs of the LHC, were chosen because of the radiative environment and because they can provide better field quality. The field quality specification using the same notation as in~\cite{hl-lhc-tech-design} are listed in Tables~\ref{tab:errorsD1} and \ref{tab:errorsD2}.

The required strengths of the non-linear corrector package behind the triplet were obtained from the dynamic aperture studies. The coil apertures are the same as in the triplet quadrupoles Q2 and Q3 in order to avoid exposure to collision debris. With this the possible field strengths could be determined~\cite{bib:Louzguiti:private} and the lengths requirements calculated. The sextupole and octupole correctors require only lengths of a few centimetres. Thus it is possible to increase the coil apertures of the sextupole and octupole correctors further in order to reduce energy deposition if necessary.

Table~\ref{tab:rma:hl_eir_magnet_specifications} also lists the hardware specifications of the high luminosity IR orbit correctors. The single aperture MCBX magnets have nested coils, allowing them to deflect the beam in both planes. Each matching section quadrupole is equipped with one orbit corrector of the same aperture, hence the need for 3 classes. The MCB class is identical to the arc orbit correctors and two units are placed next to Q7 in order to provide enough strength. No strengths requirements for the low luminosity EIR orbit correctors have been established so far, but we expect to require 4 single aperture correctors per IP in the triplet region, as well as 5 double aperture correctors with 70~mm coil aperture per IP and 5 double aperture correctors with 50~mm coil aperture per IP.
\begin{table}
\centering
\caption{\label{tab:errorsD1} Field error components of D1 with $R_{\text{ref}} = \SI{46}{mm}$. The values are based on the MBXW magnet design for LHC.}
\begin{tabular}{|c|cc|cc|cc|}
		\hline
		 & \multicolumn{2}{c|}{\bf Systematic} & \multicolumn{2}{c|}{\bf Uncertainty} & \multicolumn{2}{c|}{\bf Random} \\
		\bf Normal & \bf Injection & \bf High & \bf Injection & \bf High & \bf Injection & \bf High \\ 
		& & \bf Field & & \bf Field & & \bf Field \\ \hline
		$b_{1}$ & 0.000 & 0.000 & 0.000 & 0.000 & 0.000 & 0.000\\
		$b_{2}$ & -0.200 & -0.300 & 0.000 & 0.000 & 0.100 & 0.200 \\
		$b_{3}$ & 0.100 & -0.900 & 0.000 & 0.000 & 0.300 & 0.000\\
		$b_{4}$ & 0.000 & 0.000 & 0.000 & 0.000 & 0.000 & 0.000\\
		$b_{5}$ & -0.100 & -0.100 & 0.000 & 0.000 & 0.200 & 0.000\\ \hline
		\bf Skew & \multicolumn{6}{c|}{}\\ \hline
		$a_{1}$ & 0.000 & 0.000 & 0.000 & 0.000 & 0.000 & 0.000\\
		$a_{2}$ & -0.200 & -0.100 & 0.000 & 0.000 & 0.200 & 0.100\\
		$a_{3}$ & 0.000 & 0.000 & 0.000 & 0.000 & 0.000 & 0.100\\
		$a_{4}$ & 0.000 & 0.000 & 0.000 & 0.000 & 0.000 & 0.000\\
		$a_{5}$ & 0.000 & 0.000 & 0.000 & 0.000 & 0.000 & 0.100\\ \hline
	  \end{tabular}
\end{table}
\begin{table}
\centering
\caption{\label{tab:errorsD2} Field error components of D2 with $R_{\text{ref}} = \SI{28}{mm}$. The values are based on the MBW magnet design for LHC.}
\begin{tabular}{|c|cc|cc|cc|}
		\hline
		 & \multicolumn{2}{c|}{\bf Systematic} & \multicolumn{2}{c|}{\bf Uncertainty} & \multicolumn{2}{c|}{\bf Random} \\
		\bf Normal & \bf Injection & \bf High & \bf Injection & \bf High & \bf Injection & \bf High \\ 
		& & \bf Field & & \bf Field & & \bf Field \\ \hline
		$b_{1}$ & 0.000 & 0.000 & 0.000 & 0.000 & 0.000 & 0.000\\
		$b_{2}$ & 0.300 & -1.400 & 0.000 & 0.000 & 1.800 & 1.100 \\
		$b_{3}$ & 1.500 & -0.400 & 0.000 & 0.000 & 0.400 & 0.800\\
		$b_{4}$ & 0.000 & 0.300 & 0.000 & 0.000 & 0.400 & 0.800\\
		$b_{5}$ & -0.400 & -0.500 & 0.000 & 0.000 & 0.300 & 0.200\\
		$b_{6}$ & 0.000 & 0.000 & 0.000 & 0.000 & 0.400 & 0.300\\
		$b_{7}$ & -0.300 & -0.200 & 0.000 & 0.000 & 0.200 & 0.200\\
		$b_{8}$ & 0.000 & 0.100 & 0.000 & 0.000 & 0.200 & 0.200\\
		$b_{9}$ & -0.100 & 0.000 & 0.000 & 0.000 & 0.000 & 0.200\\
		$b_{10}$ & 0.000 & 0.100 & 0.000 & 0.000 & 0.000 & 0.200\\
		$b_{11}$ & 0.000 & 0.100 & 0.000 & 0.000 & 0.000 & 0.100\\ \hline
		Skew & \multicolumn{6}{c|}{}\\ \hline
		$a_{1}$ & 0.000 & 0.000 & 0.000 & 0.000 & 0.000 & 0.000\\
		$a_{2}$ & 0.100 & 0.200 & 0.000 & 0.000 & 0.100 & 0.200\\
		$a_{3}$ & 0.000 & -0.100 & 0.000 & 0.000 & 0.100 & 0.300\\
		$a_{4}$ & 0.000 & 0.100 & 0.000 & 0.000 & 0.000 & 0.200\\
		$a_{5}$ & 0.000 & -0.100 & 0.000 & 0.000 & 0.000 & 0.100\\
		$a_{6}$ & 0.000 & 0.000 & 0.000 & 0.000 & 0.100 & 0.200\\
		$a_{7}$ & 0.000 & -0.100 & 0.000 & 0.000 & 0.000 & 0.100\\
		$a_{8}$ & 0.000 & 0.000 & 0.000 & 0.000 & 0.000 & 0.100\\ \hline
	  \end{tabular}
\end{table}

Orbit correction studies have shown that the alignment tolerances for the high luminosity EIR elements listed in Table~\ref{tab:alignment_tolerances} result in a residual orbit below \SI{1}{mm} (90th percentile). It should be noted that the residual orbit is very sensitive to misalignments of the strong Q7 quadrupoles. Thus those elements need to be aligned more precisely than the other matching quadrupoles, possibly requiring a remote alignment system as proposed for the HL-LHC \cite{MainaudDurand:remoteAlignment}.
\begin{table}
\centering
\caption{\label{tab:alignment_tolerances} Alignment specifications for the high luminosity EIR elements.}
\begin{tabular}{|c|c|c|c|}
		\hline
		\textbf{Element} & \textbf{Error} & \textbf{Value} & \textbf{Comments}\\ \hline
		Separation dipole D1 & roll angle $\sigma(\phi)$  & \SI{1.0}{\milli rad} & \\
		Recombination dipole D2 & roll angle $\sigma(\phi)$  & \SI{1.0}{\milli rad} & \\ \hline
		Triplet quadrupoles Q1-Q3 & $\sigma(x), \sigma(y)$ & \SI{0.2}{\milli \metre} & \\
		Matching quadrupoles Q4-Q6 & $\sigma(x), \sigma(y)$ & \SI{0.5}{\milli \metre} & \\
		\multirow{2}{*}{Matching quadrupole Q7} & \multirow{2}{*}{$\sigma(x), \sigma(y)$} & \multirow{2}{*}{\SI{0.2}{\milli \metre}} & remote \\ 
		& & & alignment \\ \hline
		\multirow{2}{*}{BPM} & $\sigma(x), \sigma(y)$ & \SI{0.3}{\milli \metre} &  \\
							  & $\sigma(\text{read})$ & \SI{0.05}{\milli \metre} & accuracy \\ \hline
	  \end{tabular}
\end{table}

For collision optics beyond ultimate parameters (with $\bstar$ down to \SI{0.2}{m}) a crab voltage of \SI{18.1}{MV} per beam on either side of each high luminosity IP was sufficient to provide full crabbing. This corresponds to \SI{145}{MV} in total. Following a direct scaling from the HL-LHC lattice, \SI{20}{m} of space were allocated for the crab cavities on each side of the two main IPs. No detailed studies on number of cavities or cryostat design were done yet. It should be noted that a radiation mitigation strategy to protect the triplet is to change the crossing plane on the two main IPs at least once during the lifetime. This will also require an exchange of the crab cavities (horizontally/vertically deflecting). Since IPA and IPG will always run with different crossing planes, it should be possible to simply exchange the hardware between the two main IPs during a shutdown. This should be taken into account when designing the cryostats and RF connections.

The specifications for the triplet quadrupoles in the low luminosity experimental insertions are listed in Table~\ref{tab:rma:ll_eir_magnet_specifications}. Contrary to the high luminosity EIRs, the separation and recombination dipoles in the low luminosity EIRs are chosen to be superconducting. This allows a significantly shorter separation, providing more space for both experiment and the injection hardware.

\begin{table}
  \begin{center}
  \caption{\label{tab:rma:ll_eir_magnet_specifications} Magnet parameters of the low luminosity EIRs.}
  \begin{tabular}{ |l|c|c|c|c| } 
	\hline
	 \bf Magnet & \bf Length [\si{m}] & \bf Field & \bf Coil aperture & \bf Number	\\  
	  & & \bf strength & \bf diameter [\si{mm}] & \bf per IP \\ \hline
	\multicolumn{5}{|l|}{\bf Triplet quadrupoles} \\ \hline
	Q1 & 10 & \SI{270}{\tesla / \metre} & 64 & 4	\\ \hline
	Q2 & 15 & \SI{270}{\tesla / \metre} & 64 & 4 	\\ \hline
	Q3 & 10 & \SI{270}{\tesla / \metre} & 64 & 4	\\ \hline
	\multicolumn{5}{|l|}{\bf Separation and recombination dipoles} \\ \hline
	D1 & 12.5 & \SI{12}{\tesla} & 100 &	4	\\ \hline
	D2 & 15 & \SI{10}{\tesla} & 60 & 4 	\\ \hline
	\multicolumn{5}{|l|}{\bf Matching quadrupoles} \\ \hline
	Short type & 9.1 & \SI{200}{\tesla / \metre} & 70 & 6	\\ \hline
	Long type & 12.8 & \SI{300}{\tesla / \metre} & 50 & 5 	\\ \hline
  \end{tabular}
  \end{center}
\end{table}



\section{Summary and Outlook}
We have presented the first complete design of the FCC-hh IR, which demonstrates to meet the
ambitious project goals in terms of energy and luminosity.
The main unknowns reside in the assumed Nb$_{3}$Sn magnet technology,
with the largest peak magnetic field of about 11~T for the high luminosity EIR magnets.
This technology will
profit from developments and experimental demonstrations within the HL-LHC project.
The current design  addresses the severe limitations coming from the 
high power luminosity debris. The current layout can withstand the  synchrotron radiation 
from the beam in the IR. The various detectors are placed sufficiently far apart to 
avoid any significant exchange of radiation.

Further luminosity upgrades beyond the Ultimate scenario are conceivable.
The IP beam size is first limited by the strength of the arc sextupoles followed 
by the triplet quadrupole aperture. Both limitations could be mitigated by 
developing magnets with High Temperature Superconducting (HTS) materials.

\section*{Acknowledgements}
The European Circular Energy-Frontier Collider Study (EuroCirCol) project has
received funding from the European Union’s Horizon 2020 research and innovation
programme under grant No 654305. The information herein only reflects the views of its
authors and the European Commission is not responsible for any use that may be made
of the information.
These studies are also supported by the Swiss Institute of Accelerator Physics and Technology (CHART).

\bibliographystyle{unsrt}

\bibliography{bibliography_file_updated}

\end{document}